\documentclass[aps,prd,amsmath,amssymb,nofootinbib,notitlepage,superscriptaddress,11pt]{revtex4}
\usepackage[colorlinks=true, pdfstartview=FitV, linkcolor=red, citecolor=blue, urlcolor=blue, pdftitle={},pdfsubject={}, pdfkeywords={}]{hyperref}

\usepackage{stmaryrd}
\usepackage{comment}
\usepackage{color}
\usepackage{amsmath}
\usepackage{amssymb}
\usepackage{amsthm}
\usepackage{mathrsfs}
\usepackage{graphicx}
\usepackage{marvosym}
\usepackage{amsmath,bm}
\usepackage{array}
\usepackage{latexsym}
\usepackage{enumerate}
\usepackage{slashed}
\usepackage{hyperref}
\usepackage{color}

\allowdisplaybreaks[1]

\usepackage[normalem]{ulem}

\begin{document}

\title{Spin alignment of vector mesons by glasma fields}
\author{Avdhesh Kumar}
\affiliation{
	Institute of Physics, Academia Sinica, Taipei 11529, Taiwan\\
}
\author{Berndt M\"uller}
\affiliation{
	Department of Physics, Duke University, Durham, North Carolina 27708-0305, USA\\
}
\author{Di-Lun Yang}
\affiliation{
	Institute of Physics, Academia Sinica, Taipei 11529, Taiwan\\
}
\date{\today}

\begin{abstract}
We explain how spin alignment of vector mesons can be induced by background fields, such as electromagnetic fields or soft gluon fields. Our study is based on the quantum kinetic theory of spinning quarks and antiquarks and incorporates the relaxation of the dynamically generated spin polarization. The spin density matrix of vector mesons is obtained by quark coalescence via the Wigner function and kinetic equation. Our approach predicts a local spin correlation that is distinct from the non-local expressions previously obtained in phenomenological derivations. We estimate the magnitude of such local correlations in the glasma model of the preequilibrium phase of relativistic heavy ion collisions. It is found that the resulting spin alignment could be greatly enhanced and may be comparable to the experimental measurement in order of magnitude. We further propose new phenomenological scenarios to qualitatively explain the transverse-momentum and centrality dependence of spin alignment in a self-consistent framework.    
\end{abstract}
\maketitle
\section{Introduction}
Strongly interacting matter produced in the peripheral collisions of two heavy nuclei at the relativistic energies  carries a huge orbital angular momentum transferred by the two colliding nuclei. Due to spin-orbit coupling a part of such an initial orbital angular momentum can be transformed into the spin part which may lead to the spin polarization of emitted particles~\cite{Liang:2004xn, Liang:2004ph,Voloshin:2004ha,Voloshin:2017kqp}. Indeed,  a non-zero global and local  spin polarization of hadrons has been measured by the STAR Collaboration~\cite{STAR:2017ckg,STAR:2019erd} at BNL, ALICE Collaboration at CERN~\cite{ALICE:2019onw}, and HADES Collaboration~\cite{Kornas:2020qzi}. 
Theoretically, relativistic hydrodynamic predictions  based on global thermodynamic equilibrium formula, which connects the mean spin pseudo-vector of a fermion with four-momentum to the thermal vorticity~\cite{Becattini2013a,Becattini:2016gvu,Karpenko:2016jyx,Becattini:2017gcx,Fang:2016vpj}, can successfully explain the experimentally measured global polarization of $\Lambda$ hyperons   ~\cite{Karpenko:2016jyx,Xie:2017upb,Pang:2016igs,Becattini:2017gcx,Li:2017slc,Wei:2018zfb,Ryu:2021lnx}. 

However, predictions for the {\it local} spin polarization, i.~e. the momentum dependence of the longitudinal spin polarization\cite{Becattini:2017gcx,Xia:2018tes}, disagree with the measured values~\cite{STAR:2019erd}. This result has triggered further developments in the theoretical studies related to proper understanding of the origin of spin polarization and spin transport in relativistic heavy ion collisions \cite{Hidaka:2017auj,Liu:2020dxg,Liu:2021uhn,Becattini:2021suc,Yi:2021ryh,Hidaka:2017auj,Hidaka:2018ekt,Yang:2018lew,Shi:2020htn,Fang:2022ttm,Wang:2022yli,Lin:2022tma,Weickgenannt:2022zxs,Weickgenannt:2022qvh,Bhadury:2020puc,Bhadury:2020cop,Buzzegoli:2022kqx,Bhadury:2022ulr}. These investigations mainly explore the possible role of symmetric gradients of hydrodynamic variables known as the thermal shear~\cite{Hidaka:2017auj,Liu:2021uhn,Becattini:2021suc} and of gradients of chemical-potential ~\cite{Hidaka:2017auj,Liu:2020dxg}, spin potential~\cite{Buzzegoli:2021wlg}. See recent reviews ~\cite{Becattini:2022zvf,Hidaka:2022dmn} for further references about spin polarization. More recently, several studies performed with thermal shear corrections in local equilibrium indicated that the agreement with the local spin polarization data can only be achieved if the temperature gradients in thermal vorticity and shear are neglected~\cite{Becattini:2021iol} or if the mass of the $\Lambda$ hyperon is replaced with the constituent strange quark mass~\cite{Fu:2021pok,Florkowski:2021xvy}. 

In addition to spin polarization measurements, experimental studies of the spin alignment of vector mesons have been performed~\cite{ALICE:2019aid,ALICE:2022sli,Mohanty:2021vbt,STAR:2022fan}. The spin alignment is characterized by the deviations of the ($00$)-component of the spin density matrix $\rho_{00}$ from its equilibrium value $1/3$~\cite{Schilling:1969um,Park:2022ayr}.  Measurements indicate that the spin alignment is much larger than predictions based on the assumption of thermal equilibrium~\cite{Becattini:2007sr,Becattini:2007nd} and  the spin coalescence model \cite{Liang:2004xn,Yang:2017sdk}. Furthermore, spin alignment values strongly vary with collision energy and with the flavors of the quark and anti-quark that form the vector mesons. At LHC energies~\cite{ALICE:2019aid} values $\rho_{00}<1/3$ for global spin alignment is observed for both $\phi$ and $K^{*0}$ mesons at small transverse momenta, while at RHIC energies~\cite{STAR:2022fan}, $\rho_{00}>1/3$ for $\phi$ and $\rho_{00}\approx 1/3$ for $K^{*0}$ were found. There have been also recent measurements associated with the spin alignment of $J/\psi$ \cite{ALICE:2022sli}. This puzzling behavior has motivated the development of alternative mechanisms for the formation of spin alignment. In spite of substantial theoretical  efforts~\cite{Sheng:2019kmk,Sheng:2020ghv,Xia:2020tyd,Muller:2021hpe,Yang:2021fea,Goncalves:2021ziy,Sheng:2022wsy,Li:2022neh,Sheng:2022ffb,Li:2022vmb,Wagner:2022gza}, this issue remains an open question. 

In one of the approaches~\cite{Muller:2021hpe,Yang:2021fea} based on the quantum kinetic theory (QKT) for the spin-$1/2$ fermions~\cite{Son:2012wh,Stephanov:2012ki,Chen:2012ca,Hidaka:2016yjf,Gao:2019znl,Weickgenannt:2019dks,Hattori:2019ahi,Wang:2019moi,Yang:2020hri,Wang:2020pej,Weickgenannt:2020aaf} (see also a recent review ~\cite{Hidaka:2022dmn} and references therein) with the inclusion of color degrees of freedom, it was shown that the turbulent color fields occurring in weakly coupled anisotropic quark-gluon plasmas (QGP) could dynamically generate spin polarization of quarks and lead to $\rho_{00}<1/3$ for spin alignment of $\phi$ mesons at small transverse momentum. (A similar mechanism~\cite{Hauksson:2023tze} could also induce a jet polarization in anisotropic QGP.) In QKT, such a dynamical source term expressed in terms of coherent color fields could capture early-time effects and result in spin polarization at freeze-out, whereas collisions at late time could lead to suppression of such early-time effects by means of relaxation or enhancement by quantum corrections from gradient terms such as vorticity \cite{Kapusta:2019sad,Li:2019qkf,Yang:2020hri,Weickgenannt:2020aaf,Wang:2020pej,Wang:2021qnt,Fang:2022ttm,Hongo:2022izs,Wang:2022yli}.      

Although Weibel-type instabilities \cite{Mrowczynski:1988dz,Mrowczynski:1993qm,Romatschke:2003ms} can be one of the sources for generation of the color fields in an expanding QGP, our focus here is on the color fields~\cite{Guerrero-Rodriguez:2021ask} arising from the glasma phase~\cite{Lappi:2006fp,Lappi:2006hq} that is thought to precede the formation of QGP. The glasma phase is commonly described by the color glass condensate (CGC) effective theory \cite{McLerran:1993ni,McLerran:1993ka,McLerran:1994vd,Gelis:2010nm,Albacete:2014fwa}. Notably, such color fields are not effective in creating a nonzero spin polarization due to their fluctuating properties \cite{Kumar:2022ylt}, but they can contribute to  spin correlations of quarks and antiquarks that lead to spin alignment of vector mesons~\cite{Kumar:2022ylt}. 

In this paper, we re-examine the spin alignment of vector mesons arising from the color fields in the glasma using newly derived equation for the $\rho_{00}$-component of spin density matrix from the vector-meson kinetic equation in the quark-coalescence scenario. The new expression of the $\rho_{00}$-component of a spin density matrix, unlike the phenomenological one adopted in our previous work~\cite{Kumar:2022ylt}, involves the contributions from spin correlations of both the color-singlet and color-octet components of the axial-charge current densities for quarks and antiquarks that are dynamically generated by the fluctuating color fields. We also calculate the spin correlation due to the $U(1)$ magnetic field generated by the colliding nuclei and discuss the momentum dependence of the spin alignment. 

The paper is structured as follows: In Sec.~\ref{sec:DSP}, we show how spin polarization is generated by the background electromagnetic fields in the framework of QKT, followed by a discussion of the contribution from color fields. In Sec.~\ref{sec:SPD}, we  derive a new equation for the $\rho_{00}$-component of the spin density matrix from the vector-meson kinetic equation in the quark coalescence scenario and obtain a simplified expression in the non-relativistic approximation. In Sec.~\ref{sec:spin_alignment_glasma}, we estimate the contribution from color fields in the glasma phase. We also estimate the contribution from the $U(1)$ magnetic fields generated by the colliding nuclei. In Sec.~\ref{sec:SAV:QA}, we qualitatively analyze the momentum dependence of the spin alignment of vector mesons from the glasma effect and from an effective potential. Finally, we present conclusions and an outlook in Sec.~\ref{sec:conc}. Various technical details have been relegated to the appendices.

Throughout this paper we use 
the mostly minus signature of the Minkowski metric $\eta^{\mu\nu} = {\rm diag} (1, -1,-1,-1)  $ 
and the completely antisymmetric tensor $ \epsilon^{\mu\nu\rho\lambda} $ with $ \epsilon^{0123} = 1 $. 
We introduce the notations $A^{(\mu}B^{\nu)}\equiv A^{\mu}B^{\nu}+A^{\nu}B^{\mu}$,  $A^{[\mu}B^{\nu]}\equiv A^{\mu}B^{\nu}-A^{\nu}B^{\mu}$, and $\tilde{F}^{\mu\nu}\equiv\epsilon^{\mu\nu\alpha\beta}F_{\alpha\beta}/2$. Greek and roman indices are used for space-time and spatial components, respectively, unless otherwise specified. 

\section{Dynamical spin polarization}
\label{sec:DSP}

To track the dynamical spin polarization for non-equilibrium fermions created in early times of heavy ion collisions in the presence of strong (chromo-) electromagnetic fields led by colliding nuclei, the QKT developed in recent years is one of the most suitable theoretical frameworks. In this section, we review the so-called axial kinetic theory (AKT) constructed in Refs.~\cite{Hattori:2019ahi,Yang:2020hri} with further inclusion of color degrees of freedom \cite{Muller:2021hpe,Yang:2021fea}, which incorporate a scalar kinetic equation (SKE) and an axial-vector kinetic equation (AKE) to delineate the intertwined dynamics between charge and spin evolution, and further derive the terms associated with dynamical spin polarization with approximated spin relaxation from collisions. We shall begin with the case with $U(1)$ electromagnetic fields and then discuss the scenario for quarks influenced by color fields.

\subsection{Background electromagnetic fields}

In order to study the spectra of spin polarization and spin correlation of massive fermions, we will focus on the vector and axial-vector components of the Wigner function, which are given by
\begin{eqnarray}
	\mathcal{V}^{\mu}(p,x) = \frac{1}{4}{\rm tr}\left(\gamma^{\mu}S^{<}(p,x)\right),
	\quad
	\mathcal{A}^{\mu}(p,x) = \frac{1}{4}{\rm tr}\left(\gamma^{\mu}\gamma^{5}S^{<}(p,x)\right),
\end{eqnarray}
respectively. Here
\begin{eqnarray}\label{eq:WT}
	S^{<}(p,x)=\int d^4Ye^{ip\cdot Y/\hbar}\langle\bar{\psi}(x_2)U(x_2,x_1)\psi(x_1)\rangle
\end{eqnarray}
represents the Wigner function of massive fermions,
where $x=(x_1+x_2)/2$, $Y=x_1-x_2$. Also, $U(x_2,x_1)$ denotes the gauge link and $p_{\mu}$ represents the kinetic momentum, which ensure the gauge invariance of $S^{<}(p,X)$. One may obtain perturbative solutions of $\mathcal{V}^{\mu}(p,x)$ and $\mathcal{A}^{\mu}(p,x)$ and corresponding kinetic equations from the Kadanoff-Baym equation by utilizing the $\hbar$ expansion as the gradient expansion of Wigner functions in phase space. Due to the quantum nature of spin, we may adopt the power counting,  
$\mathcal{V}^{\mu}\sim \mathcal{O}(\hbar^0)$ and $\mathcal{A}^{\mu}\sim \mathcal{O}(\hbar)$ and focus on the leading-order contribution. In such a case, we have 
\begin{eqnarray}
\mathcal{V}^{\mu}(p,x)=2\pi\delta(p^2-m^2)f_{V},
\end{eqnarray}  
where $f_{V}(p,x)$ and $m$ denote the distribution function and mass of the fermions, respectively. The dynamics of $f_V$ is dictated by the SKE as a standard Vlasov equation, $p\cdot\Delta f_V=\mathcal{C}[f_V]$, with the on-shell condition $p^2=m^2$. Here $\Delta_{\mu}=\partial_{\mu}+eF_{\nu\mu}\partial^{\nu}_{p}$ with $F_{\nu\mu}$ being the field strength of electromagnetic fields and $\mathcal{C}[f_V]$ corresponds to the collision term depending on the details of interaction. Our focus will be instead $\mathcal{A}^{\mu}$ delineating the spin polarization through quantum corrections of $\mathcal{O}(\hbar)$. See Ref.~\cite{Hidaka:2022dmn} for a comprehensive review and technical details. 

In the particle rest frame with a frame vector $n^{\mu}_r=p^{\mu}/m$, the magnetization-current term in $\mathcal{A}^{\mu}$ vanishes and the $\mathcal{A}^{\mu}$ reduces to
\begin{eqnarray}\label{eq:Amu_WF_QED}
	\mathcal{A}^{\mu}(p,x)=2\pi\Big[\delta(p^2-m^2)\tilde{a}^{\mu}
	+\hbar e\tilde{F}^{\mu\nu}p_{\nu}\delta'(p^2-m^2)f_V
	\Big],
\end{eqnarray}
where  $\delta'(x)\equiv \partial\delta(x)/\partial x$ and $\tilde{a}^{\mu}(p,x)$ represents an effective spin four-vector. For practical applications to the spin polarization in heavy ion collisions, one usually evaluates the spin-polarization or correlation spectra near chemical equilibrium with $f_V$ in local thermal equilibrium, while $\tilde{a}^{\mu}$ need not reach thermal equilibrium and thus could carry early-time effects. Consequently, we will refer the contribution of $\tilde{a}^{\mu}$ to spin polarization or correlation as the dynamical one and which from the second term in $\mathcal{A}^{\mu}$ carrying only the information at chemical freeze-out as the non-dynamical one. The phase-space evolution of $\tilde{a}^{\mu}(p,x)$ is governed by the AKE,
\begin{eqnarray}\label{AKE_nr}
	\Box^{(n_r)}\mathcal{A}^{\mu}=\hat{\mathcal{C}}^{(n_r)\mu}_{1}+\hbar\hat{\mathcal{C}}^{(n_r)\mu}_{2},
\end{eqnarray}
where 
\begin{eqnarray}\notag
	\Box^{(n_r)}\mathcal{A}^{\mu}&=&\delta(p^2-m^2)
	\Big(p\cdot\Delta\tilde{a}^{\mu} +eF^{\nu\mu}\tilde{a}_{\nu}-\frac{e}{2}{\hbar\epsilon^{\mu\nu\rho\sigma}p_{\rho}(\partial_{\sigma}F_{\beta\nu})
		\partial_{p}^{\beta}f_{V}}\Big)
	\\
	&&+\hbar e\tilde{F}^{\mu\nu}p_{\nu}\delta'(p^2-m^2)p\cdot\Delta f_V.
\end{eqnarray}
For simplicity, we may neglect the terms, $p_{\alpha}F^{\alpha\beta}\partial_{p\beta}\tilde{a}^{\mu}$ and $F^{\nu\mu}\tilde{a}_{\nu}$, which are suppressed in the weak-field limit when $\tilde{a}^{\mu}$ is dynamically generated by $F^{\mu\nu}$. Also, we adopt the relaxation-time approximation for the collision term by postulating $\hat{\mathcal{C}}^{(n_r)\mu}_{1}+\hbar\hat{\mathcal{C}}^{(n_r)\mu}_{2}=-\delta(p^2-m^2)p_0(\tilde{a}^{\mu}-\tilde{a}^{\mu}_{\rm eq})/\tau_{\rm R}$, where $\tilde{a}^{\mu}_{\rm eq}(p,x)$ denotes the equilibrium value of $\tilde{a}^{\mu}$ and $\tau_{\rm R}$ represents a constant spin relaxation time. The practicability of this simplification will be further discussed later. Accordingly, the off-shell AKE reduces to
\begin{eqnarray}\label{eq:AKE}
p\cdot\partial\tilde{a}^{\mu}-\frac{e}{2}{\hbar\epsilon^{\mu\nu\rho\sigma}p_{\rho}(\partial_{\sigma}F_{\beta\nu})
	\partial_{p}^{\beta}f_{V}}=-\frac{p_0(\tilde{a}^{\mu}-\tilde{a}^{\mu}_{\rm eq})}{\tau_{\rm R}},
\end{eqnarray}
which yields
\begin{eqnarray}\nonumber\label{eq:sol_AKE}
	\tilde{a}^{\mu}(p,x)&=&\frac{1}{2p_0}\int^{\infty}_{-\infty} dx'_0\Theta(x_0-x'_0)e^{-(x_0-x'_0)/\tau_{\rm R}}\bigg[\hbar \epsilon^{\mu\nu\rho\sigma}p_{\rho}\Theta(x_0')\big(\partial_{x'\sigma}eF_{\beta\nu}(x')\big)
	\partial_{p}^{\beta}f_{V}(p,x')
	\\
	&&+\frac{2p_0\tilde{a}^{\mu}_{\rm eq}(p,x')}{\tau_{\rm R}}\bigg]\bigg|_{\text{c}},
\end{eqnarray}
where $|_{\text{c}}=\{x^{\prime i}_{\rm T}=x^{i}_{\rm T},x^{\prime i}_{\parallel}=x^{i}_{\parallel}-p^{i}(x_0-x'_0)/p_0\}$ and $\Theta(x)$ denotes a unit-step function of $x$. Here $V^{i}_{\rm T}$ and $V^{i}_{\parallel}$ represent the perpendicular and parallel components with respect to the spatial momentum $p^{i}$ for an arbitrary spatial vector $V^{i}$, respectively. We also assume $\partial_{x'\sigma}F_{\beta\nu}(x')\neq 0$ starting at $x'_0=0$ as the initial time. We will further assume $\tilde{a}^{\mu}_{\rm eq}$ is a constant, whereby Eq.~(\ref{eq:sol_AKE}) reduces to
\begin{align}
	\tilde{a}^{\mu}(p,x)=\tilde{a}^{\mu}_{\rm eq}+\frac{\hbar e}{2p_0}\int^{\infty}_{-\infty} dx'_0\Theta(x_0-x'_0)\Theta(x_0')e^{-(x_0-x'_0)/\tau_{\rm R}}\epsilon^{\mu\nu\rho\sigma}p_{\rho}\big(\partial_{x'\sigma}F_{\beta\nu}(x')\big)
	\partial_{p}^{\beta}f_{V}(p,x')|_{\text{c}}.
\end{align}

Given the electromagnetic fields expressed in terms of $n^{\mu}=(1,\bm 0)$,
\begin{eqnarray}
F_{\alpha\beta}=-\epsilon_{\mu\nu\alpha\beta}B^{\mu}n^{\nu}+n_{\beta}E_{\alpha}-n_{\alpha}E_{\beta},
\end{eqnarray}
it is found
\begin{eqnarray}\nonumber
\epsilon^{\mu\nu\rho\sigma}p_{\rho}\big(\partial_{\sigma}F_{\beta\nu}\big)&=&\delta^{\mu}_{\beta}(n\cdot\partial p\cdot B-n\cdot p\partial\cdot B)+(n\cdot p\partial_{\beta}-p_{\beta}n\cdot\partial)B^{\mu}
+n^{\mu}(p_{\beta}\partial\cdot B-\partial_{\beta}p\cdot B)
\\
&&+\epsilon^{\mu\nu\rho\sigma}p_{\rho}n_{[\nu}\partial_{\sigma}E_{\beta]}.
\end{eqnarray}
Assuming $f_{V}(p,x')=\tilde{f}_{V}(p_0,x'_0)$ with only energy and time dependence,  one finds
\begin{eqnarray}
	\tilde{a}^{i}(p,x)=\tilde{a}^{i}_{\rm eq}-\frac{\hbar e}{2p_0}\int^{\infty}_{-\infty} dx'_0\Theta(x_0-x'_0)\Theta(x_0')e^{-(x_0-x'_0)/\tau_{\rm R}}\epsilon^{ijk}p_{[0}\big(\partial_{x'k]}E_{j}(x')\big)
	\partial_{p0}\tilde{f}_{V}(p_0,x'_0)
	,
\end{eqnarray}
where we have used $\partial\cdot B=0$. It is found $\tilde{a}^{i}(p,x)$ can be induced by space-time variations of the electric field. When involving $x^{\prime i}_{\parallel}$ dependence, it is inevitable to have the momentum dependence for $f_{V}(p,x')$, which is neglected for simplification. For phenomenological applications, $\tilde{a}^{i}_{\rm eq}$ could be proportional to the kinetic vorticity in QGP albeit the negligence of spatial gradients on $f_{V}(p,x')$. Notably, the relaxation-time approximation also corresponds to the linearization of the collision term, for which the smallness of fluctuations from equilibrium distribution functions is required in the standard Boltzmann equation. Nevertheless, for AKE up to $\mathcal{O}(\hbar)$, the collision term is by default linear to $\tilde{a}^{\mu}$ usually accompanied by another term with the space-time gradients on $f_V$ stemming from spin-orbit interaction. In the case for gauge theories, the structure of the collision term could be more complicated, where the inverse relaxation times may have to be replaced by operators \cite{Li:2019qkf,Hattori:2019ahi,Fang:2022ttm}.  

In heavy ion collisions, there could locally exist strong background electromagnetic fields coming from colliding nuclei and dynamical ones generated in the QGP. When further considering spatial inhomogeneity of the electric fields, we may apply the Bianchi identity $\partial_{\mu}\tilde{F}^{\mu\nu}=0$, which leads to $\epsilon^{ijk}\partial_jE_k=\partial_0B^i$. 
One hence obtains
\begin{eqnarray}\label{eq:amu_EM_fields}
\delta\tilde{a}^{i}(p,x)=\frac{\hbar e}{2p_0}\int^{\infty}_{-\infty} dx'_0\Theta(x_0-x'_0)\Theta(x_0')e^{-(x_0-x'_0)/\tau_{\rm R}}\big(p_0\partial_0B^i(x')+\epsilon^{ijk}p_k\partial_0E_j(x')\big)
\partial_{p0}\tilde{f}_{V},
\end{eqnarray}
where $\delta \tilde{a}^{i}(p,x)=\tilde{a}^{i}(p,x)-\tilde{a}^{i}_{\rm eq}$.
In the collisionless limit such that $\tau_{\rm R}\rightarrow \infty$ and assuming the time variation of $\tilde{f}_{V}$ is sufficiently small compared to that of background fields (e.g. $|\partial_0B^i|/|B^i|\gg |\partial_0\tilde{f}_{V}|/\tilde{f}_{V}$)\footnote{Such a condition might be difficult to be justified in heavy ion collisions. However, provided the strong background fields decay rapidly before thermalization, at which $|\partial_0\tilde{f}_{V}|$ reaches the maximum, one may expect the contribution from e.g. $B^i\partial_0\tilde{f}_{V}$ in the integrand is relatively suppressed.} in early times, by using the integration by parts and dropping the vanishing surface terms, we arrive at
\begin{eqnarray}\label{eq:sol_ai}\nonumber
\delta\tilde{a}^{i}(p,x)&=&\frac{\hbar e}{2p_0}\Theta(x_0)\big[\big(p_0B^i(x_0)+\epsilon^{ijk}p_kE_j(x_0)\big)\partial_{p0}\tilde{f}_{V}(p_0,x_0)
\\
&&-
\big(p_0B^i(0)+\epsilon^{ijk}p_kE_j(0)\big)\partial_{p0}\tilde{f}_{V}(p_0,0)\big],
\end{eqnarray}
from which it is transparent to see that the spin polarization is induced by parallel magnetic fields and perpendicular electric fields as the spin Hall effect.
Here we implicitly hide the spatial dependence of electromagnetic fields for brevity. Nonetheless, one should recall here $B^i(0)\equiv B^i(0,x_j=x'_j)|_{\text{c},\,x'_0=0}$ and so does $E^i(0)$. In fact, we should set $\tilde{a}^{\mu}_{\rm eq}=0$ when collisions are suppressed. In contrast, when $\tau_{\rm R}\rightarrow 0$, one should find $\delta\tilde{a}^{\mu}=0$. To incorporate the approximate spin-relaxation effect, one may multiply the result in Eq.~(\ref{eq:sol_ai}) with $e^{-x_0/\tau_{\rm R}}$ albeit the over suppression for early-time contributions.  

Next, combining with the non-dynamical contribution, the full on-shell axial Wigner function becomes 
\begin{eqnarray}\nonumber\label{eq:Amu_onshell}
	\mathcal{A}^{\mu}({\bm p},x)&\equiv&\int\frac{dp_0}{2\pi}\Theta(p_0)\mathcal{A}^{\mu}(p,x)	
	\\
	&=&\frac{1}{2\epsilon_{\bm p}}\Big[\tilde{a}^{\mu}(p,x)
	-\frac{\hbar eB^{\mu}(x_0)}{2}\partial_{p0}\tilde{f}_V(p_0,x_0)
	\Big]_{p_0=\epsilon_{\bm p}\equiv \sqrt{|\bm p|^2+m^2}}.
\end{eqnarray} 
Then, $\mathcal{A}^{\mu}({\bm p},x)$ can be more explicitly written as 
\begin{eqnarray}
\mathcal{A}^{i}({\bm p},x)=\frac{\hbar e}{4\epsilon_{\bm p}}\bigg[-B^i(0)\partial_{p0}\tilde{f}_V(p_0,0)+\frac{\epsilon^{ijk}p_k}{\epsilon_{\bm p}}\big(E_j(x_0)\partial_{p0}\tilde{f}_V(p_0,x_0)-E_j(0)\partial_{p0}\tilde{f}_V(p_0,0)\big)\bigg]_{p_0=\epsilon_{\bm p}}
\end{eqnarray}
in the collisionless limit. 
In practice, it is expected that both electromagnetic fields are relatively suppressed at $x_0$ as the freeze-out time. Accordingly, one could approximate 
\begin{eqnarray}\label{eq:Ai_no_latetime_fields}
	\mathcal{A}^{i}({\bm p},x)\approx\frac{-\hbar e}{4\epsilon_{\bm p}^2}\big(\epsilon_{\bm p}B^i(0)+\epsilon^{ijk}p_kE_j(0)\big)\partial_{\epsilon_{\bm p}}f^{(0)}_V(\epsilon_{\bm p}),
\end{eqnarray}
where we introduced $\tilde{f}_V(\epsilon_{\bm p},0)=f^{(0)}_V(\epsilon_{\bm p})$ as the initial (quark) distribution function, which is dominated by the early-time contribution.
Furthermore, given $|B_z(0)|\sim 0$, $|E_z(0)|\sim 0$, and $p_z\sim 0$ at central rapidity with $z$ being the longitudinal (beam) direction, one could approximate
\begin{eqnarray}
\mathcal{A}^{x,y}({\bm p},x)\approx\frac{-\hbar e}{4\epsilon_{\bm p}}B^{x,y}(0)\partial_{\epsilon_{\bm p}}f^{(0)}_V(\epsilon_{\bm p})
\end{eqnarray}
and
\begin{eqnarray}
	\mathcal{A}^{z}({\bm p},x)\approx\frac{\hbar e}{4\epsilon_{\bm p}^2}p_{[x}E_{y]}(0)\partial_{\epsilon_{\bm p}}f^{(0)}_V(\epsilon_{\bm p})
	\approx\frac{\hbar e}{4\epsilon_{\bm p}^2}p_{x}E_{y}(0)\partial_{\epsilon_{\bm p}}f^{(0)}_V(\epsilon_{\bm p}),
\end{eqnarray}
where we further assume $|p_x|\gg |p_y|$ in peripheral collisions with transverse shear flow. In the collisionless scenario, the dynamical contribution from $\tilde{a}^{\mu}$ is frozen at the early time, while the vector-component of quark Wigner functions governed by $\tilde{f}_V(\epsilon_{\bm p},x_0)$ keeps evolving and reaches thermal equilibrium in the QGP phase. Considering the spin freeze-out at the QGP phase, as originally proposed in  Refs.~\cite{Becattini2013a,Fang:2016vpj}, the spin-polarization pseudo-vector of quarks from background electromagnetic fields is then given by
\begin{equation}
	\mathcal{P}^{x,y}({\bm p})=\frac{\int d\Sigma\cdot p\mathcal{J}_{5}^{x,y}(\bm p,X)}{2m\int d\Sigma\cdot\mathcal{N}(\bm p,X)}
	\approx\frac{\int d\Sigma\cdot peB^{x,y}(0)\partial_{\epsilon_{\bm p}}f^{(0)}_V(\epsilon_{\bm p})}{4m\int d\Sigma\cdot pf^{\rm th}_V(\epsilon_{\bm p})}
\end{equation}
and
\begin{equation}
	\mathcal{P}^{z}({\bm p})=\frac{\int d\Sigma\cdot p\mathcal{J}_{5}^{z}(\bm p,X)}{2m\int d\Sigma\cdot\mathcal{N}(\bm p,X)}
	\approx\frac{\int d\Sigma\cdot pp^{x}eE^{y}(0)\partial_{\epsilon_{\bm p}}f^{(0)}_V(\epsilon_{\bm p})}{4m\epsilon_{\bm p}\int d\Sigma\cdot pf^{\rm th}_V(\epsilon_{\bm p})},
\end{equation}
where $d\Sigma^{\mu}$ denotes the normal vector of a freeze-out hyper surface and we introduce
\begin{eqnarray}
\mathcal{J}_{5}^{\mu}(\bm p,X)\equiv 4\mathcal{A}^{\mu}({\bm p},X),\quad
\mathcal{N}^{\mu}(\bm p,X)\equiv 4\int\frac{dp_0}{2\pi}\Theta(p_0)\mathcal{V}^{\mu}(p,X)=\frac{4p^{\mu}f_{V}(p,X)|_{p_0=\epsilon_{\bm p}}}{2\epsilon_{\bm p}},
\end{eqnarray}
and $f^{\rm th}_V(\epsilon_{\bm p})=1/(e^{\beta\epsilon_{\bm p}}+1)$ represents the vector-charge distribution function in thermal equilibrium as the Fermi-Dirac distribution. For convenience of later computations, we alternatively use $X^{\mu}$ to represent the space-time coordinates.   

However, in heavy ion collision experiments, we have so far not found the evidence supporting global spin polarization induced by magnetic fields. Based on our findings with the inclusion of dynamical spin polarization dominated by the contributions from initial electromagnetic fields, the suppression of spin polarization from electromagnetic fields may not solely stem from the rapid decay of such fields. Alternatively, it may also be suppressed by the strong spin-relaxation rate from collisions, which efficiently washed out the early-time contributions. Although the early-time electromagnetic fields are stronger with higher collision energies, the lifetime of QGP is also longer, which accordingly enhances the spin-relaxation effects. In addition to the spin relaxation, the initial magnetic fields also drop more rapidly at high energies in the pre-equilibrium state and become saturated with finite electric conductivity. Since the dynamical spin polarization is induced by the time derivatives upon electromagnetic fields as shown in the integrand of Eq.~(\ref{eq:amu_EM_fields}), the spin polarization of quarks and anti-quarks produced later than the abrupt decay of magnetic fields may not be affected.      

\subsection{Background color fields}

In the case when color degrees of freedom are included, both the Wigner functions and QKT of quarks are more involved. Generically, we have to decompose an arbitrary color object into $O=O^sI+O^at^a$, where $O^s$ and $O^a$ denotes the color-singlet and color-octet components, respectively, and $t^a$ are the SU($N_c$) generators and $I$ is the identity matrix in color space. Before introducing the QKT, we should reanalyze how spin polarization and correlation are computed when considering color degrees of freedom.

Given the lowest-order contributions to singlet and octet vector-charge distribution functions are of $\mathcal{O}(g^0)$ and $\mathcal{O}(g)$ at weak coupling, respectively, the singlet and octet SKEs and AKEs are given by 
\begin{eqnarray}
	p^{\rho}\Big(\partial_{\rho}f^{\rm s}_V+\bar{C}_2gF^a_{\nu\rho}\partial_{p}^{\nu}f^{a}_V\Big)
	=\mathcal{C}_{\rm s},
\end{eqnarray}
\begin{eqnarray}
	p^{\rho}\Big(\partial_{\rho}f^{a}_V+gF^a_{\nu\rho}\partial_{p}^{\nu}f^{\rm s}_V\Big)
	=\mathcal{C}^{a}_{\rm o},
\end{eqnarray}
and
\begin{eqnarray}
	p^{\rho}\Big(\partial_{\rho}\tilde{a}^{\rm s \mu}+\bar{C}_2gF^a_{\nu\rho}\partial_{p}^{\nu}\tilde{a}^{a \mu}\Big)
	-\frac{\hbar\bar{C}_2}{2}\epsilon^{\mu\nu\rho\sigma}p_{\rho}
	(\partial_{\sigma}gF^a_{\beta\nu})\partial_{p}^{\beta}f_{V}^a=\mathcal{C}^{\mu}_{\rm s},
\end{eqnarray}
\begin{eqnarray}\label{eq:octet_AKE}
	p^{\rho}\Big(\partial_{\rho}\tilde{a}^{a \mu}+gF^a_{\nu\rho}\partial_{p}^{\nu}\tilde{a}^{\rm s \mu}\Big)
	-\frac{\hbar}{2}\epsilon^{\mu\nu\rho\sigma}p_{\rho}
	(\partial_{\sigma}gF^a_{\beta\nu})\partial_{p}^{\beta}f_{V}^{\rm s}=\mathcal{C}^{a\mu}_{\rm o},
\end{eqnarray}
where $\bar{C}_2=1/(2N_c)$ and we have dropped the higher-order terms in $g$ responsible for the gauge links between color fields for brevity. Here we introduce the collision terms characterized by $\mathcal{C}_{\rm s}$, $\mathcal{C}^{a}_{\rm o}$, $\mathcal{C}^{\mu}_{\rm s}$, and $\mathcal{C}^{a\mu}_{\rm o}$, which however depend on details of scattering processes. On the other hand, the color-singlet and color-octet axial Wigner functions are given by
\begin{eqnarray}\label{eq:axial_WF_s}
	\mathcal{A}^{{\rm s}\mu}&=&2\pi\Big[\delta(p^2-m^2)\tilde{a}^{{\rm s}\mu}
	+\hbar \bar{C}_2p_{\nu}\delta'(p^2-m^2)g\tilde{F}^{a\mu\nu}f^a_{V}\Big],
\end{eqnarray} 
\begin{eqnarray}\label{eq:axial_WF_o}
	\mathcal{A}^{a\mu}&=&2\pi\Big[\delta(p^2-m^2)\tilde{a}^{a\mu}
	+\hbar p_{\nu}\delta'(p^2-m^2)g\tilde{F}^{a\mu\nu}f^{\rm s}_{V}\Big],
\end{eqnarray} 
where we have also applied $f^{\rm s}_{V}\sim \mathcal{O}(g^0)$ and $f^a_{V}\sim \mathcal{O}(g)$ to drop the higher-order terms.

Since we are only interested in how the spin polarization is dynamically induced, the dynamics of $f_{V}$ is not our primary concern. Instead of constructing the proper collision terms for $\mathcal{C}_{\rm s}$ and $\mathcal{C}^{a}_{\rm o}$, we will simply introduce particular forms of $f^{\rm s}_{V}$ and $f^{a}_{V}$ as the solutions of SKEs. On the other hand, for AKEs, we may now postulate the relaxation-time forms,
\begin{eqnarray}
	\mathcal{C}^{\mu}_{\rm s}\approx -\frac{p_0(\tilde{a}^{{\rm s}\mu}-\tilde{a}^{{\rm s}\mu}_{\rm eq})}{\tau^{\rm s}_{\rm R}},\quad 
	\mathcal{C}^{a\mu}_{\rm o}\approx -\frac{p_0\tilde{a}^{a\mu}}{\tau^{\rm o}_{\rm R}},
\end{eqnarray} 
where we have assumed the absence of mixing terms and $\tilde{a}^{a\mu}_{\rm eq}=0$. A comprehensive analysis for the color-singlet AKE has been presented in Refs.~\cite{Yang:2021fea,Kumar:2022ylt} albeit with the omission of collisions. We will hence focus on the color-octet one. It is worthwhile to note the color-octet AKE in Eq.~(\ref{eq:octet_AKE}) with the suppressed diffusion term, $gF^a_{\nu\rho}\partial_{p}^{\nu}\tilde{a}^{\rm s \mu}$, at weak coupling (or equivalently with weak fields) reduces to the form same as Eq.~(\ref{eq:AKE}) by simply adding the color indices for $\tilde{a}^{\mu}$ and $F^{\mu\nu}$ and setting $\tilde{a}^{\mu}_{\rm eq}=0$. Therefore, the solution of the color-octet AKE gives rise to an analogous solution,
\begin{align}
	\tilde{a}^{a\mu}(p,x)=\frac{\hbar g}{2p_0}\int^{\infty}_{-\infty} dx'_0\Theta(x_0-x'_0)\Theta(x_0')e^{-(x_0-x'_0)/\tau^{\rm o}_{\rm R}}\epsilon^{\mu\nu\rho\sigma}p_{\rho}\big(\partial_{x'\sigma}F^a_{\beta\nu}(x')\big)
	\partial_{p}^{\beta}f^{\rm s}_{V}(p,x')|_{\text{c}}.
\end{align}
In addition, one also find the analogous form for Eq.~(\ref{eq:axial_WF_o}) and Eq.~(\ref{eq:Amu_WF_QED}). In the collisionless limit, similarly assuming $f^{\rm s}_{V}(p,x')=\tilde{f}_{V}(p_0,x'_0)$ and taking $\epsilon^{ijk}\partial_jE^a_k=\partial_0B^{ai}$ by ignoring the nonlinear terms as an Abelianized approximation for color fields, we can follow the same procedure as in the case with electromagnetic fields to obtain
\begin{eqnarray}\nonumber\label{eq:aioct_sol}
\tilde{a}^{ai}(p,x)&=&\frac{\hbar g}{2}\bigg[B^{ai}(x_0)\partial_{\epsilon_{\bm p}}\tilde{f}_V(\epsilon_{\bm p},x_0)-B^{ai}(0)\partial_{\epsilon_{\bm p}}\tilde{f}_V(\epsilon_{\bm p},0)
\\
&&\quad +\frac{\epsilon^{ijk}p_k}{\epsilon_{\bm p}}\Big(E^a_j(x_0)\partial_{\epsilon_{\bm p}}\tilde{f}_V(\epsilon_{\bm p},x_0)-E^a_j(0)\partial_{\epsilon_{\bm p}}\tilde{f}_V(\epsilon_{\bm p},0)\Big)\bigg],
\end{eqnarray}
which yields
\begin{eqnarray}\label{eq:J5_color_fields}
	\mathcal{J}_{5}^{ai}(\bm p,x)&=&\frac{2}{\epsilon_{\bm p}}\Big(\tilde{a}^{ai}(\bm p,x)-\frac{\hbar g B^{ai}(x_0)}{2}\partial_{\epsilon_{\bm p}}\tilde{f}_V(\epsilon_{\bm p},x_0)\Big)
	\\\nonumber
	&=&\frac{\hbar g}{\epsilon_{\bm p}}\bigg[-B^{ai}(0)\partial_{\epsilon_{\bm p}}\tilde{f}_V(\epsilon_{\bm p},0)
	 +\frac{\epsilon^{ijk}p_k}{\epsilon_{\bm p}}\Big(E^a_j(x_0)\partial_{\epsilon_{\bm p}}\tilde{f}_V(\epsilon_{\bm p},x_0)-E^a_j(0)\partial_{\epsilon_{\bm p}}\tilde{f}_V(\epsilon_{\bm p},0)\Big)\bigg]
\end{eqnarray}
by also incorporating the non-dynamical contribution.

In Refs.~\cite{Muller:2021hpe,Yang:2021fea}, in light of the original form for relativistic fermions \cite{Becattini2013a,Fang:2016vpj}, it is proposed that the spin polarization of a single quark (or an antiquark) takes the form,
\begin{eqnarray}\label{eq:original_P}
	\mathcal{P}^{\mu}({\bm p})=\frac{\int d\Sigma\cdot p{\rm Tr_c}\big(\mathcal{J}_{5}^{\mu}(\bm p,X)\big)}{2m\int d\Sigma_{\nu}{\rm Tr_c}\big(\mathcal{N}^{\nu}(\bm p,X)\big)}=\frac{\int d\Sigma\cdot p\mathcal{J}_{5}^{\rm s\mu}(\bm p,X)}{2m\int d\Sigma\cdot\mathcal{N}^{\rm s}(\bm p,X)},
\end{eqnarray}
where $\rm Tr_c$ denotes the trace over color space and
\begin{eqnarray}
	\mathcal{N}^{\rm s\mu}(\bm p,X)=\frac{4p^{\mu}f^{\rm s}_{V}(p,X)|_{p_0=\epsilon_{\bm p}}}{2\epsilon_{\bm p}}=\frac{2p^{\mu}f^{\rm s}_{V}({\bm p},X)}{\epsilon_{\bm p}}
\end{eqnarray}
with $f^{\rm s}_{V}({\bm p},X)\equiv f^{\rm s}_{V}(p,X)|_{p_0=\epsilon_{\bm p}}$. Here the color fields encoded in $\mathcal{J}_{5}^{\rm s\mu}(\bm p,X)$ should be regarded as the field operators and one has to further take an ensemble average or the quantum expectation value $\langle\,\rangle$ for the field operators therein to acquire the spin polarization pseudo-vector $\langle \mathcal{P}^{\mu}({\bm p})\rangle$. Considering the effect led by strong color fields in the glasma sate in early times, only the dynamical contribution from $\tilde{a}^{{\rm s}\mu}$ could possibly affect the spin polarization. The explicit form of $\tilde{a}^{{\rm s}\mu}$ induced by background color fields can be found in Refs.~\cite{Yang:2021fea,Kumar:2022ylt}. It is however also shown that the corresponding spin polarization actually vanishes and only the non-vanishing spin correlation, as will be discussed later, is present. On the other hand, $\tilde{a}^{a\mu}$ does not affect the spin polarization, whereas it could modify the spin correlation associated with spin alignment as will be discussed in the next section.

\section{Spin density matrix from quark coalescence in Wigner functions}
\label{sec:SPD}

\begin{figure}
	\begin{center}
		\includegraphics[width=0.9\hsize]{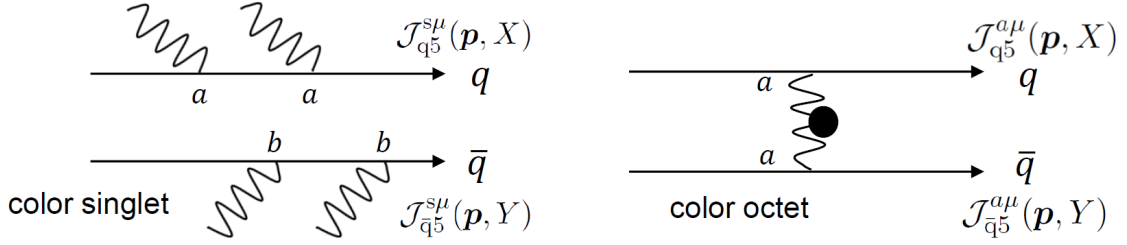}
	\end{center}
	\caption{Left: color-singlet contribution for the spin correlation affected by color fields represented by curvy lines. Right: color-octet contribution is shown on the right panel, where the blob represents possible corrections from the medium or from higher-order loops on correlated color fields.
	}
	\label{fig_in_ex_fields}
\end{figure}

For spin alignment, when spin quantization axis is set to be along the $y$ direction \footnote{In heavy ion collisions, the spin quantization axis is chosen to be perpendicular to the reaction plane of the collision. However, it can be also chosen along different directions depending on the experimental setup. The theoretical construction in this section is independent of the experimental choices.}, it is proposed that the normalized spin density matrix can be written as \cite{Liang:2004xn,Yang:2017sdk,Sheng:2022ffb,Kumar:2022ylt}
\begin{eqnarray}\label{eq:rho00_modified}
	\rho_{00}\approx \frac{1+\sum_{j=x,y,z}\langle\mathcal{P}^{j}_{q}\mathcal{P}^{j}_{\bar{q}}\rangle-2\langle\mathcal{P}^{y}_{q}\mathcal{P}^{y}_{\bar{q}}\rangle}{3+\sum_{j=x,y,z}\langle\mathcal{P}^{j}_{q}\mathcal{P}^{j}_{\bar{q}}\rangle},
\end{eqnarray}
where the subscripts $q$ and $\bar{q}$ correspond to the quark and antiquark, respectively.
When $|\langle\mathcal{P}^{j}_{q}\mathcal{P}^{j}_{\bar{q}}\rangle|\ll 1$, it further reduces to
\begin{eqnarray}\label{eq:rho00_smallP}
	\rho_{00}\approx \frac{1}{3}+\frac{2}{9}\big(\langle\mathcal{P}^{x}_{q}\mathcal{P}^{x}_{\bar{q}}\rangle+\langle\mathcal{P}^{z}_{q}\mathcal{P}^{z}_{\bar{q}}\rangle-2\langle\mathcal{P}^{y}_{q}\mathcal{P}^{y}_{\bar{q}}\rangle\big).
\end{eqnarray}
Here $\langle\mathcal{P}^{j}_{q}\mathcal{P}^{j}_{\bar{q}}\rangle$ represents the quantum expectation value of spin correlation, which is not necessary to be equal to the product of the expectation values of spin-polarization pseudovectors and we may elaborate on its explicit expressions in various forms later. Since the spin-polarization pseudovector of a single quark should be color singlet, when including the color degrees of freedom, the spin correlation associated with spin alignment is proposed to be \cite{Kumar:2022ylt}
\begin{equation}
	\langle\mathcal{P}_{q}^{\mu}({\bm p})\mathcal{P}_{\bar{q}}^{\mu}({\bm p})\rangle =\frac{\int d\Sigma_X\cdot p\int d\Sigma_Y\cdot p\langle\mathcal{J}_{{q}5}^{\rm s\mu}(\bm p,X)\mathcal{J}_{\bar{q}5}^{\rm s\mu}(\bm p,Y)\rangle}{4m^2\Big(\int d\Sigma_X\cdot\mathcal{N}^{\rm s}_{{q}}(\bm p,X)\int d\Sigma_Y\cdot\mathcal{N}^{\rm s}_{\bar{q}}(\bm p,Y)\Big)},
\end{equation}
based on a phenomenological construction in the quark model.
Note that here only the color-singlet components of $\mathcal{J}_{{q}5}^{\mu}$ and $\mathcal{J}_{\bar{q}5}^{\mu}$ contribute to both the spin polarization and correlation. By the symmetry of color-charge conjugation, we should have $\mathcal{J}_{\bar{q}5}^{{\rm s} \mu}({\bm p},X)=\mathcal{J}_{{q}5}^{{\rm s}\mu}({\bm p},X)$ and $\langle\mathcal{P}_{q}^{\mu}({\bm p})\mathcal{P}_{\bar{q}}^{\mu}({\bm p})\rangle$ is expected to be positive when having equal number of quarks and antiquarks (no corrections from the quark chemical potential). Furthermore, the color fields in $\mathcal{J}_{{q}5}^{{\rm s}\mu}({\bm p},X)$ and $\mathcal{J}_{\bar{q}5}^{{\rm s} \mu}({\bm p},Y)$ are not directly connected albeit the indirect correlation originates from the same color source such as the case for color fields coming from the same nucleus in glasma \cite{Kumar:2022ylt}. Such a scenario is schematically illustrated in the left panel of Fig.~\ref{fig_in_ex_fields}. Nonetheless, as will be more rigorously shown from the derivation of quark coalescence in the Wigner functions and kinetic theory of vector mesons, there exist extra contributions led by the color-octet contribution depicted in the right panel of Fig.~\ref{fig_in_ex_fields}, which turns out to play a central role in this paper.

\subsection{Spin density matrix from the vector-meson kinetic equation}

We will follow the approach in Ref.~\cite{Sheng:2022ffb} to derive the spin density matrix from the coalescence scenario in Wigner functions and kinetic theory of vector mesons.
We begin with the vector-meson field in mode expansions,
\begin{eqnarray}
	V^{\mu}(x)=\sum_{\lambda=\pm 1,0}\int \frac{d^3k}{(2\pi)^3\sqrt{2E_k}}\big[\epsilon^{\mu}(\lambda,\bm k)a(\lambda,\bm k)e^{-ik\cdot x}+\epsilon^{*\mu}(\lambda,\bm k)b^{\dagger}(\lambda,\bm k)e^{ik\cdot x}\big],
\end{eqnarray}
where $E_k=\sqrt{|\bm k|^2+M^2}$ with $M$ being the mass of vector mesons and \cite{Sheng:2022ffb}
\begin{eqnarray}
	\epsilon^{\mu}(\lambda,\bm k)=\Big(\frac{\bm k\cdot \bm\epsilon_{\lambda}}{M},\bm\epsilon_{\lambda}+\frac{\bm k\cdot\bm\epsilon_{\lambda}}{M(E_k+M)}\bm k\Big)
\end{eqnarray}
represents the polarization vector with $\bm\epsilon_{\lambda}$ being the spin-state vector determined by the spin quantization axis in experiments, which satisfies $\epsilon^{\mu}(\lambda,\bm k)\epsilon^{*}_{\mu}(\lambda',\bm k)=-\bm\epsilon_{\lambda}\cdot \bm\epsilon^{*}_{\lambda'}$ and $\epsilon^{\mu}(\lambda,\bm k)k_{\mu}=0$. We also impose $\bm\epsilon_{\lambda}\cdot \bm\epsilon^{*}_{\lambda'}=\delta_{\lambda\lambda'}$. In the rest frame of the vector meson, we have $\epsilon^{\mu}(\lambda,0)=(0,\bm\epsilon_{\lambda})$.
For $\phi$ mesons, we have $b(\lambda,\bm k)=a(\lambda,\bm k)$.
We may construct the Wigner function in real time formalism (see e.g. Ref.~\cite{Hattori:2020gqh}) via
\begin{eqnarray}\label{eq:Wigner_V}
	W^{<\mu\nu}(q,X)&=&\int d^4Ye^{iq\cdot Y}\langle V^{\dagger\nu}(X-Y/2)V^{\mu}(X+Y/2)\rangle
	\\\nonumber
	&=&\pi\sum_{\lambda,\lambda'=\pm 1,0}\int\frac{d^3k_{-}}{(2\pi)^3}\frac{e^{-ik_-\cdot X}}{\Big[\Big(|\bm q|^2+\frac{|\bm k_-|^2}{4}\Big)^2-(\bm q\cdot \bm k_{-})^2+2M^2\Big(|\bm q|^2+\frac{|\bm k_-|^2}{4}\Big)+M^4\Big]^{1/4}}
	\\\nonumber
	&&\times\bigg[\epsilon^{\mu}\Big(\lambda, q+\frac{k_-}{2}\Big)\epsilon^{*\nu}\Big(\lambda', q-\frac{k_-}{2}\Big)\Big\langle a^{\dagger}\Big(\lambda', \bm q-\frac{\bm k_-}{2}\Big)a\Big(\lambda, \bm q+\frac{\bm k_-}{2}\Big)\Big\rangle\delta(q^0-k_+^0)
	\\\nonumber
	&&+\epsilon^{\nu}\Big(\lambda', -q+\frac{k_-}{2}\Big)\epsilon^{*\mu}\Big(\lambda, -q-\frac{k_-}{2}\Big)\Big\langle b\Big(\lambda', -\bm q+\frac{\bm k_-}{2}\Big)b^{\dagger}\Big(\lambda, -\bm q-\frac{\bm k_-}{2}\Big)\Big\rangle\delta(q^0+k_+^0)\bigg]
	,
\end{eqnarray}
where
\begin{eqnarray}
	k_+^0=\frac{1}{2}\Big(E_{{q}+\frac{k_-}{2}}+E_{{q}-\frac{k_-}{2}}\Big)
	\, , \qquad 
	k_-^0=\Big(E_{{q}+\frac{k_-}{2}}-E_{{q}-\frac{k_-}{2}}\Big).
\end{eqnarray} 
For brevity, we set $\hbar=1$.
To have the quasi-particle in a definite spin state, we may assume the expectation values of the creation and annihilation operators have non-vanishing values only for particles or antiparticles when $\lambda=\lambda'$. For example, $\langle a^{\dagger}(\lambda,q)b(\lambda,p)\rangle=0$ and $\langle a^{\dagger}(\lambda',q)a(\lambda,p)\rangle\propto \delta_{\lambda'\lambda}$.
Moreover, in order to perform  the $k_-$ integral analytically,
we expand the integrand with respect to $ k_- $ and retain the terms up to $\mathcal{O}(k_-)$ such as
\begin{eqnarray}
	\epsilon_{\mu}\left(\lambda,q+\frac{k_-}{2}\right)\epsilon^{*}_{\nu}\left(\lambda,q-\frac{k_-}{2}\right) 
	= \Pi_{\mu\nu}^{(0)}(\lambda,q) + \frac{k_-^{\alpha}}{2} \Pi_{\mu\nu\alpha}^{(1)}(\lambda,q) + \mathcal{O}(k_-^2)
	\,,
\end{eqnarray}
where 
\begin{eqnarray}
	\label{Pi}
	\Pi_{\mu\nu}^{(0)}(\lambda,q)\equiv\epsilon_{\mu}(\lambda,q)\epsilon^{*}_{\nu}(\lambda,q)\,,
	\qquad \Pi_{\mu\nu\alpha}^{(1)}(\lambda,q)\equiv \big(\partial_{q^{\alpha}}\epsilon_{\mu}(\lambda,q)\big)\epsilon^{*}_{\nu}(\lambda,q)-\epsilon_{\mu}(\lambda,q)\big(\partial_{q^{\alpha}}\epsilon^{*}_{\nu}(\lambda,q)\big)\,.
\end{eqnarray}
In the end, the expansion with respect to $k_-$ provides us with the Wigner functions up to $\mathcal{O}(\hbar)$. 
Plugging those expressions into Eq.~(\ref{eq:Wigner_V}), we then find 
\begin{eqnarray}
	W^{<\mu\nu}(q,X)&=&\sum_{\lambda=\pm 1,0}\tilde{W}^{<\mu\nu}(\lambda, q,X),
	\\\nonumber
	\tilde{W}^{<}_{\mu\nu}(\lambda, q,X)&=&2\pi\delta(q^2-M^2)\Bigg[\Theta(q_0)\left(\Pi_{\mu\nu}^{(0)}(\lambda,q)+\frac{\rm i\hbar}{2}\Pi_{\mu\nu\alpha}^{(1)}(\lambda,q)\partial^{\alpha}\right)
	\\
	&&-\Theta(-q_0)\left(\Pi_{\nu\mu}^{(0)}(\lambda,-q)+\frac{\rm i\hbar}{2}\Pi_{\nu\mu\alpha}^{(1)}(\lambda,-q)\partial^{\alpha}\right)\Bigg]\check{f}_{\lambda}(q,X)\,,
\end{eqnarray}
where we dropped the $\mathcal{O}(|{\bm k}_-|^2)$ terms in the integrand except for those contributing to the distribution functions. Here we retrieve $\hbar$ for power counting. The distribution function $ \check{f}_{\lambda}(q,X) $ 
is defined as 
\begin{eqnarray}
	\check{f}_{\lambda}(q,X) =
	\begin{cases}
		f_{\lambda}(\bm q,X) & (q_0>0) \\
		-\left[1+f_{\lambda}(-\bm q,X) \right] & (q_0<0)\,,
	\end{cases}
\end{eqnarray}
where
\begin{eqnarray}
	f_{\lambda}(\bm q,X)=\int\frac{{\rm d}^3k_-}{(2\pi)^3}\Big\langle a^{\dagger}\Big(\lambda,\bm q-\frac{\bm k_-}{2}\Big)a\Big(\lambda,\bm q+\frac{\bm k_-}{2}\Big)\Big\rangle {\rm e}^{-{\rm i}k_-\cdot X}
\end{eqnarray} 
and
\begin{eqnarray}
	\int\frac{{\rm d}^3k_-}{(2\pi)^3}{\rm e}^{-{\rm i}k_-\cdot X}\Big\langle b\Big(\lambda,-\bm q+\frac{\bm k_-}{2}\Big)b^{\dagger}\Big(\lambda,-\bm q-\frac{\bm k_-}{2}\Big)\Big\rangle=1+f_{\lambda}(-\bm q,X)
	\, .
\end{eqnarray}
from the commutation relation for bosons.
Note that the $\Theta(-q_0)$ part in $W^{<\mu\nu}$ characterizes the outgoing vector mesons.

For our purpose, we will only consider the symmetric Wigner function with positive energy and up to $\mathcal{O}(\hbar^0)$,
\begin{eqnarray}
	\tilde{W}^{<(\mu\nu)}(\lambda, q,X)=\frac{1}{2}\tilde{W}^{<(\mu\nu)}(\lambda, q,X)=\pi\delta(q^2-M^2)\theta(q_0)\Pi^{(0)(\nu\mu)}(\lambda,q)f_{\lambda}(q,X),
\end{eqnarray} 
where $A^{(\mu\nu)}=A^{\mu\nu}+A^{\nu\mu}$.
Note that we shall have 
\begin{eqnarray}
	\frac{1}{2}\sum_{\lambda=\pm 1,0}\Pi^{(0)(\nu\mu)}(\lambda,q)=\frac{q^{\mu}q^{\nu}}{M^2}-\eta^{\mu\nu},
\end{eqnarray}
where we have neglected the $\hbar$ corrections and anti-symmetric components. The corresponding on-shell kinetic equation reads
\begin{eqnarray}\label{eq:KE_mesons}
	q\cdot\partial f_{\lambda}=\Sigma^{<\mu\rho}\hat{P}_{\rho\mu}(\lambda,q)(1+f_{\lambda})-\Sigma^{>\mu\rho}\hat{P}_{\rho\mu}(\lambda,q)f_{\lambda},
\end{eqnarray}
where $\hat{P}^{\mu\nu}(\lambda,q)=\epsilon^{(\mu}(\lambda,q)\epsilon^{*\nu)}(\lambda,q)/2$. Here $\Sigma^{\lessgtr\mu\rho}$ corresponds to the self-energies for the scattering processes led by the effective quark-meson interaction.
In this framework, we have $\rho_{\lambda\lambda}\propto f_{\lambda}$. As proposed in Ref.~\cite{Sheng:2022ffb}, when there are no pre-existing vector mesons such that $f_{\lambda}\ll 1$ and the coalescence time $\Delta t$ is sufficiently short,\footnote{The spatial dependence of $f_{\lambda}$ is also neglected.} Eg.~(\ref{eq:KE_mesons}) gives rise to 
\begin{eqnarray}\label{eq:flambda}
	f_{\lambda}\approx \frac{\Delta t}{E_q}\Sigma^{<\mu\rho}\hat{P}_{\rho\mu}(\lambda,q).
\end{eqnarray}

\subsection{Quark coalescence scenario}

Applying the meson-quark interaction characterized by an effective Lagrangian $\mathcal{L}_{\rm int}=g_{\phi}\Gamma\cdot V\bar{\psi}\psi$ \cite{Zhao:1998fn} with $\Gamma^{\mu}$ begin an effective form factor, the self-energy can be rearranged into the form
\begin{eqnarray}\label{eq:Sigma_in_quarks}
	\Sigma^{<\mu\rho}= \int \frac{d^3k}{(2\pi)^2}{\rm Tr}\Big[\Gamma^{\mu}S_{q}^<\Big(\frac{\bm q}{2}+{ \bm k}\Big)\Gamma^{\rho}S_{\bar{q}}^<\Big(\frac{\bm q}{2}-{ \bm k}\Big)\Big]\delta\big(q_0-\epsilon_{q}({\bm q/2+\bm k})-\epsilon_{\bar{q}}({\bm q/2-\bm k})\big),
\end{eqnarray}
where $\epsilon_{q/\bar{q}}({\bm p})\equiv\sqrt{|\bm p|^2+m^2_{q/\bar{q}}}$. Here $S_{q}^<$ and $S_{\bar{q}}^<$ denote the onshell lesser propagators of quarks and antiquarks. More precisely, we introduce
\begin{eqnarray}
	S_{q/\bar{q}}^<(\bm p)=\int\frac{dp_0}{2\pi}\grave{S}_{q/\bar{q}}^<(p)
\end{eqnarray}
by integrating the off-shell Wigner functions $\grave{S}_{q/\bar{q}}^<(p)$ over the zeroth component of its four momentum.
Note that ${\rm Tr}$ also includes the trace over color space. For simplicity, we may assume $\Gamma^{\mu}=\gamma^{\mu}$. Based on the decomposition of the quark Wigner functions \cite{Vasak:1987um},
\begin{eqnarray}
	\grave{S}^<=\mathcal{F}^<+ i\mathcal{P}^<\gamma^5+ \mathcal{V}^{<\mu}\gamma_\mu+\mathcal{A}^{<\mu}\gamma^5\gamma_{\mu}+ \frac{\mathcal{S}^{<\mu\nu}}{2}\sigma_{\mu\nu},
\end{eqnarray}
where $\sigma_{\mu\nu}=i[\gamma_{\mu},\gamma_{\nu}]/2$, it is found that
\begin{eqnarray}\nonumber
	{\rm Tr}\Big[\gamma^{\mu}\grave{S}_{q}^<\gamma^{\rho}\grave{S}_{\bar{q}}^<\Big]&=&4{\rm Tr_{c}}\Big[\eta^{\mu\rho}\Big(\mathcal{F}^<_{q} \mathcal{F}^<_{\bar{q}}+\mathcal{P}^<_{q}\cdot \mathcal{P}^<_{\bar{q}}-\mathcal{V}^{<}_{q}\cdot \mathcal{V}^{<}_{\bar{q}}-\mathcal{A}^{<}_{q}\cdot \mathcal{A}^{<}_{\bar{q}}
	+\frac{\mathcal{S}_{q\alpha\beta}\mathcal{S}_{\bar{q}}^{\alpha\beta}}{2}
	\Big)
	\\
	&&+\mathcal{V}^{<(\mu}_{q}\mathcal{V}^{<\rho)}_{\bar{q}}+\mathcal{A}^{<(\mu}_{q}\mathcal{A}^{<\rho)}_{\bar{q}}
	-\mathcal{S}^{<(\mu\nu}_{q}\mathcal{S}^{<\rho)}_{\bar{q}\quad\nu}
	\Big].
\end{eqnarray}
Here we hide the momentum dependence above for brevity.
Taking \cite{Hattori:2019ahi}
\begin{eqnarray}
	m\mathcal{F}(q)=q\cdot\mathcal{V}(q),
	\quad
	\mathcal{P}=0,\quad
	m\mathcal{S}_{\mu\nu}(q)=-\epsilon_{\mu\nu\rho\sigma} q^{\rho}\mathcal{A}^{\sigma}(q)
	,
\end{eqnarray}
for free fermions since we only consider the tree-level interaction for coalescence and hence
\begin{eqnarray}\nonumber
	m_{q}m_{\bar{q}}\grave{\mathcal{S}}^{<\mu\nu}_{q}(p)\grave{\mathcal{S}}^{<\rho}_{\bar{q}\quad\nu}(p')&=&\eta^{\mu\rho}\big(p'\cdot\mathcal{A}^<_{q}p\cdot\mathcal{A}^<_{\bar{q}}-p\cdot p'\mathcal{A}^<_{q}\cdot\mathcal{A}^<_{\bar{q}}\big)+\mathcal{A}_{q}^{<\mu}\mathcal{A}_{\bar{q}}^{<\rho}p\cdot p'+p'^{\mu}p^{\rho}\mathcal{A}^<_{q}\cdot\mathcal{A}^<_{\bar{q}}
	\\
	&&-p'^{\mu}\mathcal{A}_{q}^{<\rho}p\cdot\mathcal{A}^<_{\bar{q}}-p^{\rho}\mathcal{A}_{\bar{q}}^{<\mu}p'\cdot\mathcal{A}^<_{q},
\end{eqnarray}
and
\begin{eqnarray}
	m_{q}m_{\bar{q}}\grave{\mathcal{S}}^{<\mu\nu}_{q}(p)\grave{\mathcal{S}}^{<}_{\bar{q}\mu\nu}(p')=2\big(p'\cdot\mathcal{A}^<_{q}p\cdot\mathcal{A}^<_{\bar{q}}-p\cdot p'\mathcal{A}^<_{q}\cdot\mathcal{A}^<_{\bar{q}}\big),
\end{eqnarray}
where $\mathcal{A}^{<\mu}_{q}=\mathcal{A}^{<\mu}_{q}(p)$ and $\mathcal{A}^{<\mu}_{\bar{q}}=\mathcal{A}^{<\mu}_{\bar{q}}(p')$,
one obtains
\begin{eqnarray}\nonumber\label{eq:trace_selfE}
	&&{\rm Tr}\Big[\gamma^{\mu}\grave{S}_{q}^<(p)\gamma^{\rho}\grave{S}_{\bar{q}}^<(p')\Big]
	\\\nonumber
	&&=4{\rm Tr_{c}}\Bigg\{\eta^{\mu\rho}\Bigg[\frac{p\cdot\mathcal{V}_{q}^{<}(p)p'\cdot\mathcal{V}_{\bar{q}}^{<}(p')}{m_{q}m_{\bar{q}}}-\mathcal{V}^{<}_{q}(p)\cdot \mathcal{V}^{<}_{\bar{q}}(p')-\mathcal{A}^{<}_{q}(p)\cdot \mathcal{A}^{<}_{\bar{q}}(p')\bigg(1-\frac{p\cdot p'}{m_{q}m_{\bar{q}}}\bigg)
	\\\nonumber
	&&\quad-\frac{p'\cdot\mathcal{A}_{q}^{<}(p)p\cdot\mathcal{A}_{\bar{q}}^{<}(p')}{m_{q}m_{\bar{q}}}\Bigg]
	+\mathcal{V}^{<(\mu}_{q}(p)\mathcal{V}^{<\rho)}_{\bar{q}}(p')+\mathcal{A}^{<(\mu}_{q}(p)\mathcal{A}^{<\rho)}_{\bar{q}}(p')\bigg(1-\frac{p\cdot p'}{m_{q}m_{\bar{q}}}\bigg)
	\\
	&&\quad -\frac{p'^{(\mu}p^{\rho)}}{m_qm_{\bar{q}}}\mathcal{A}^<_{q}(p)\cdot\mathcal{A}^<_{\bar{q}}(p')
	+\frac{p'^{(\mu}\mathcal{A}_{q}^{<\rho)}(p)}{m_qm_{\bar{q}}}p\cdot\mathcal{A}^<_{\bar{q}}(p')+\frac{p^{(\mu}\mathcal{A}_{\bar{q}}^{<\rho)}(p')}{m_qm_{\bar{q}}}p'\cdot\mathcal{A}^<_{q}(p)
	\Bigg\}.
\end{eqnarray}
The expression above also works for the onshell Wigner functions $S_{q}^<(p)$ and $S_{\bar{q}}^<(p')$. Since the contributions from vector and axial-vector components of quark/antiquark Wigner functions are disentangled, we make the decomposition
\begin{eqnarray}\label{eq:trSqSqb}
	{\rm Tr}\Big[\gamma^{\mu}S_{q}^<(\bm p)\gamma^{\rho}S_{\bar{q}}^<(\bm p')\Big]=\int\frac{dp_0}{2\pi}\frac{dp'_0}{2\pi}{\rm Tr}\Big[\gamma^{\mu}\grave{S}_{q}^<(p)\gamma^{\rho}\grave{S}_{\bar{q}}^<( p')\Big]=\hat{\Sigma}_{V}^{<\mu\rho}+\hat{\Sigma}_{A}^{<\mu\rho},
\end{eqnarray}
where
\begin{eqnarray}\label{eq:SigmaV_in_V}
	\hat{\Sigma}_{V}^{<\mu\rho}=4{\rm Tr_{c}}\Bigg\{\eta^{\mu\rho}\Bigg[\frac{p\cdot\mathcal{V}_{q}^{<}(\bm p)p'\cdot\mathcal{V}_{\bar{q}}^{<}(\bm p')}{m_{q}m_{\bar{q}}}-\mathcal{V}^{<}_{q}(\bm p)\cdot \mathcal{V}^{<}_{\bar{q}}(\bm p')\Bigg]+\mathcal{V}^{<(\mu}_{q}(\bm p)\mathcal{V}^{<\rho)}_{\bar{q}}(\bm p')\Bigg\}
\end{eqnarray}
and
\begin{eqnarray}\label{eq:SigmaA_in_A}
	\hat{\Sigma}_{A}^{<\mu\rho}&=&4{\rm Tr_{c}}\Bigg\{\eta^{\mu\rho}\Bigg[\mathcal{A}^{<}_{q}(\bm p)\cdot \mathcal{A}^{<}_{\bar{q}}(\bm p')\bigg(\frac{p\cdot p'}{m_{q}m_{\bar{q}}}-1\bigg)-\frac{p'\cdot\mathcal{A}_{q}^{<}(\bm p)p\cdot\mathcal{A}_{\bar{q}}^{<}(\bm p')}{m_{q}m_{\bar{q}}}\Bigg]+\mathcal{A}^{<(\mu}_{q}(\bm p)\mathcal{A}^{<\rho)}_{\bar{q}}(\bm p')
	\\\nonumber
	&&\times \bigg(1-\frac{p\cdot p'}{m_{q}m_{\bar{q}}}\bigg)-\frac{p'^{(\mu}p^{\rho)}}{m_qm_{\bar{q}}}\mathcal{A}^<_{q}(\bm p)\cdot\mathcal{A}^<_{\bar{q}}(\bm p')
	+\frac{p'^{(\mu}\mathcal{A}_{q}^{<\rho)}(\bm p)}{m_qm_{\bar{q}}}p\cdot\mathcal{A}^<_{\bar{q}}(\bm p')
	+\frac{p^{(\mu}\mathcal{A}_{\bar{q}}^{<\rho)}(\bm p')}{m_qm_{\bar{q}}}p'\cdot\mathcal{A}^<_{q}(\bm p)
	\Bigg\},
\end{eqnarray}
with $p_0=\epsilon_{\bm p}$, $p'_0=\epsilon_{\bm p'}$,
and the onshell Wigner functions,\footnote{We have assumed both $\mathcal{V}_{q/\bar{q}}^<(\bm p)$ and $\mathcal{A}_{q/\bar{q}}^<(\bm p)$ can be rearranged into the functions proportional to $\delta(p^2-m_{q/\bar{q}}^2)$.}
\begin{eqnarray}
	\mathcal{V}_{q/\bar{q}}^<(\bm p)=\int\frac{dp_0}{2\pi}\Theta(p_0)\mathcal{V}_{q/\bar{q}}^<(p),\quad
	\mathcal{A}_{q/\bar{q}}^<(\bm p)=\int\frac{dp_0}{2\pi}\Theta(p_0)\mathcal{A}_{q/\bar{q}}^<(p).
\end{eqnarray}
Given the explicit form of $\mathcal{V}_{q/\bar{q}}^<(\bm p)$ and $\mathcal{A}_{q/\bar{q}}^<(\bm p)$, one may derive $f_{\lambda}$ from Eq.~(\ref{eq:flambda}) by calculating Eq.~(\ref{eq:trSqSqb}).  

We may now take the explicit form of the vector-component for Wigner functions of quarks and antiquarks up to $\mathcal{O}(\hbar^0)$,
\begin{eqnarray}\label{eq:WignerV_classical}
	\mathcal{V}^{<\mu}_{q/\bar{q}}(\bm p)=\frac{p^{\mu}}{2p_0}f_{{\rm V}q/\bar{q}}\Big|_{p_0=\epsilon_{q/\bar{q}}({\bm p})\equiv\sqrt{|\bm p|^2+m^2_{q/\bar{q}}}}.
\end{eqnarray}
Furthermore, given $p=q/2+k$ and $p'=q/2-k$ in light of Eq.~(\ref{eq:Sigma_in_quarks}) and the onshell conditions for quarks and antiquarks, we have 
\begin{eqnarray}\label{eq:kinetic_rel}
	k^2=\frac{m_{q}^2+m_{\bar{q}}^2}{2}-\frac{q^2}{4},\quad q\cdot k=\frac{m_{q}^2-m_{\bar{q}}^2}{2}.
\end{eqnarray} 
Using Eqs.~(\ref{eq:WignerV_classical}) and (\ref{eq:kinetic_rel}), one finds
\begin{eqnarray}
	\hat{\Sigma}_{V}^{<\mu\rho}&=&\frac{ {\rm Tr_{c}}(f_{{\rm V}q}f_{{\rm V}\bar{q}})}{\epsilon_{q}\left(\frac{\bm q}{2}+\bm p\right)\epsilon_{\bar{q}}\left(\frac{\bm q}{2}-\bm p\right)}\Big[\frac{\eta^{\mu\rho}}{2}\big((m_{q}+m_{\bar{q}})^2-q^2\big)
	+\frac{q^{\mu}q^{\rho}}{2}-2k^{\mu}k^{\rho}\Big]
\end{eqnarray}
and
\begin{eqnarray}\nonumber
	\hat{\Sigma}_{A}^{<\mu\rho}&=&4{\rm Tr_{c}}\Bigg\{\eta^{\mu\rho}\Bigg[\mathcal{A}^{<}_{q}(\bm p)\cdot \mathcal{A}^{<}_{\bar{q}}(\bm p')\frac{\big(q^2-(m_{q}+m_{\bar{q}})^2\big)}{2m_{q}m_{\bar{q}}}+\frac{4k\cdot\mathcal{A}_{q}^{<}(\bm p)k\cdot\mathcal{A}_{\bar{q}}^{<}(\bm p')}{m_{q}m_{\bar{q}}}\Bigg]
	\\\nonumber
	&&+ \frac{\big((m_{q}+m_{\bar{q}})^2-q^2\big)}{2m_{q}m_{\bar{q}}}\mathcal{A}^{<(\mu}_{q}(\bm p)\mathcal{A}^{<\rho)}_{\bar{q}}(\bm p')-\Big(\frac{q^{\mu}q^{\rho}}{2m_qm_{\bar{q}}}-\frac{2k^{\mu}k^{\rho}}{m_qm_{\bar{q}}}\Big)\mathcal{A}^<_{q}(\bm p)\cdot\mathcal{A}^<_{\bar{q}}(\bm p')
	\\
	&&
	+\frac{\big(q-2k\big)^{(\mu}\mathcal{A}_{q}^{<\rho)}(\bm p)}{m_qm_{\bar{q}}}k\cdot\mathcal{A}^<_{\bar{q}}(\bm p')
	-\frac{\big(q+2k\big)^{(\mu}\mathcal{A}_{\bar{q}}^{<\rho)}(\bm p')}{m_qm_{\bar{q}}}k\cdot\mathcal{A}^<_{q}(\bm p)
	\Bigg\}_{p=\frac{q}{2}+k,p'=\frac{q}{2}-k},
\end{eqnarray}
where we have also utilized $p\cdot\mathcal{A}^{<}_{q/\bar{q}}(p)=0$ for free fermions, which results in
\begin{eqnarray}\label{eq:SigmaV}
	\hat{\Sigma}_V^{<\mu\rho}\hat{P}_{\rho\mu}(\lambda,q)=
	\frac{ {\rm Tr_{c}}(f_{{\rm V}q}f_{{\rm V}\bar{q}})}{\epsilon_{q}\left(\frac{\bm q}{2}+\bm k\right)\epsilon_{\bar{q}}\left(\frac{\bm q}{2}-\bm k\right)}\Big[\frac{1}{2}\big(q^2-(m_{q}+m_{\bar{q}})^2\big)
	-2|k\cdot\epsilon(\lambda, {\bm q})|^2\Big] 
\end{eqnarray}
and
\begin{eqnarray}\nonumber\label{eq:SigmaA}
	&&\hat{\Sigma}_A^{<\mu\rho}\hat{P}_{\rho\mu}(\lambda,q)
	\\\nonumber
	&&=4{\rm Tr_{c}}\Bigg\{\Bigg[\mathcal{A}^{<}_{q}(\bm p)\cdot \mathcal{A}^{<}_{\bar{q}}(\bm p')\frac{\big((m_{q}+m_{\bar{q}})^2-q^2\big)}{2m_{q}m_{\bar{q}}}-\frac{4k\cdot\mathcal{A}_{q}^{<}(\bm p)k\cdot\mathcal{A}_{\bar{q}}^{<}(\bm p')}{m_{q}m_{\bar{q}}}\Bigg]
	+\frac{2|k\cdot\epsilon(\lambda, {\bm q})|^2}{m_qm_{\bar{q}}}\mathcal{A}^<_{q}(\bm p)\cdot\mathcal{A}^<_{\bar{q}}(\bm p')
	\\\nonumber
	&&\quad+ \frac{\big((m_{q}+m_{\bar{q}})^2-q^2\big)}{m_{q}m_{\bar{q}}}{\rm Re}\Big[\epsilon(\lambda,{\bm q})\cdot\mathcal{A}^{<}_{q}(\bm p)\epsilon^*(\lambda,{\bm q})\cdot\mathcal{A}^{<}_{\bar{q}}(\bm p')\Big]
	-\frac{4{\rm Re}\Big[k\cdot\epsilon(\lambda,{\bm q})\mathcal{A}_{q}^{<}(\bm p)\cdot \epsilon^*(\lambda,{\bm q})\Big]}{m_qm_{\bar{q}}}k\cdot\mathcal{A}^<_{\bar{q}}(\bm p')
	\\
	&&
	\quad
	-\frac{4{\rm Re}\Big[k\cdot\epsilon(\lambda,{\bm q})\mathcal{A}_{\bar{q}}^{<}(\bm p')\cdot \epsilon^*(\lambda,{\bm q})\Big]}{m_qm_{\bar{q}}}k\cdot\mathcal{A}^<_{q}(\bm p)
	\Bigg\}\Bigg|_{p,p'},
\end{eqnarray}
where $|_{p,p'}=\{p=\frac{q}{2}+k,p'=\frac{q}{2}-k\}$.
Overall, $\big(\hat{\Sigma}_V^{<\mu\rho}+\hat{\Sigma}_A^{<\mu\rho}\big)\hat{P}_{\rho\mu}(\lambda,q)$ can be rearranged as
\begin{eqnarray}\label{eq:SigmaVA}
	&&\big(\hat{\Sigma}_V^{<\mu\rho}+\hat{\Sigma}_A^{<\mu\rho}\big)\hat{P}_{\rho\mu}(\lambda,q)
	\\\nonumber
	&&=N_{m}
	{\rm Tr_{c}}\Bigg\{\Bigg[\frac{ f_{{\rm V}q}(\bm p)f_{{\rm V}\bar{q}}(\bm p')}{\epsilon_{q}(\bm p)\epsilon_{\bar{q}}(\bm p')}
	-\frac{4}{m_{q}m_{\bar{q}}}\mathcal{A}^{<}_{q}(\bm p)\cdot \mathcal{A}^{<}_{\bar{q}}(\bm p')\Bigg]\Bigg(1-\frac{2|k\cdot\epsilon(\lambda, {\bm q})|^2}{N_m}\Bigg)
	\\\nonumber
	&&\quad -\frac{4}{m_{q}m_{\bar{q}}}\Bigg[2{\rm Re}\Big(\epsilon(\lambda,{\bm q})\cdot\mathcal{A}^{<}_{q}(\bm p)\epsilon^*(\lambda,{\bm q})\cdot\mathcal{A}^{<}_{\bar{q}}(\bm p')\Big)
	+\frac{2}{N_m}
	\Big(2k\cdot\mathcal{A}_{q}^{<}(\bm p)k\cdot\mathcal{A}_{\bar{q}}^{<}(\bm p')
	\\\nonumber
	&&\quad
	+2{\rm Re}\big(k\cdot\epsilon(\lambda,{\bm q})\mathcal{A}_{q}^{<}(\bm p)\cdot \epsilon^*(\lambda,{\bm q})\big)k\cdot\mathcal{A}^<_{\bar{q}}(\bm p')
	+2{\rm Re}\big(k\cdot\epsilon(\lambda,{\bm q})\mathcal{A}_{\bar{q}}^{<}(\bm p')\cdot \epsilon^*(\lambda,{\bm q})\big)k\cdot\mathcal{A}^<_{q}(\bm p)
	\Big)\Bigg]
	\Bigg\}\Bigg|_{p,p'}, 
\end{eqnarray}
where
\begin{eqnarray}
	N_{m}=\frac{1}{2}\big(M^2-(m_{q}+m_{\bar{q}})^2\big).
\end{eqnarray}

For the application to spin alignment, it is generally believed that the $\phi$ meson as our focus is an s-wave particle. Consequently, we only consider the contact interaction and classical collision term for quark coalescence and ignore contributions from the orbital angular momentum of constituent quarks. Nevertheless, it is recently pointed out in Ref.~\cite{Kim:2022vtt} by the operator product expansion that $J/\psi$ could have a non-negligible contribution from the orbital angular momentum of quarks to its spin and a similar scenario might be applicable to $\phi$ mesons. In order to address the involvement of the orbital angular momentum in our approach, we may need to modify the contact interaction from the effective Lagrangian for quark-meson interaction or incorporate the $\hbar$ correction pertinent to spin-orbit interaction in the collision term for the kinetic equation of $\phi$ mesons (see e.g. the construction of the collision term of QKT for photons \cite{Hattori:2020gqh}). Alternatively, one may incorporate the contribution from e.g. p-wave wave functions for vector mesons in the recombination model \cite{Kanada-Enyo:2006dxd} with further inclusion of spin degrees of freedom. Such generalization is however beyond the scope of current work and may be pursued in the future. Furthermore, the additional effect from the orbital angular momentum upon spin alignment should be associated with a certain source from the QGP medium like vorticity, which is believed to be suppressed in high-energy collisions yet relevant in low-energy collisions.

\subsection{Non-relativistic approximation}

We shall now make further simplification. By working in the rest frame of vector mesons, the polarization vector is aligned with the spin quantization axis, $\epsilon^{\mu}(\lambda,0)=(0,\bm\epsilon_{\lambda})\equiv\epsilon_{\lambda}^{\mu}$. The kinematic conditions in Eq.~(\ref{eq:kinetic_rel}) then give rise to
\begin{eqnarray}
	k_0=\frac{m_q^2-m_{\bar{q}}^2}{2M},\quad |\bm k|^2=\frac{1}{4M^2}\big[(m_q^2-m_{\bar{q}}^2)^2+M^4-2M^2(m_q^2+m_{\bar{q}}^2)\big].
\end{eqnarray} 
We may focus on the case when $m_q=m_{\bar{q}}=m$, which yields $k_0=0$ and $N_m=2|\bm k|^2$.
Next, we consider the non-relativistic limit for quarks and anti-quarks such that $k^i\rightarrow 0$, which allows us to approximate $\bm p\approx\bm p'\approx\bm q/2\rightarrow 0$ for $f_{{\rm V}q/\bar{q}}$, $\mathcal{A}_{q/\bar{q}}^{<}$, and $\epsilon_{q/\bar{q}}$.\footnote{When ignoring the energy conservation such that $N_m=2|\bm k|^2$, one may naively drop the higher-order terms of $\mathcal{O}(|\bm k|^2)$ in Eq.~(\ref{eq:SigmaVA}) such as $2|k\cdot\epsilon(\lambda, {\bm q})|^2/N_m$. However, such terms should be maintain as the leading-order contribution.}  Apparently, this approximation is only valid when $M-(m_{q}+m_{\bar{q}})\ll M$. Note that $\mathcal{A}^{<0}_{q/\bar{q}}$ are suppressed compared with $\mathcal{A}^{<i}_{q/\bar{q}}$ in the non-relativistic limit.\footnote{By using $\mathcal{A}_{q/\bar{q}}^{<0}(p)=p^i\mathcal{A}_{q/\bar{q}}^{<i}(p)/p_0$, we now have $\mathcal{A}_{q/\bar{q}}^{<0}(q/2\pm k)\approx \pm k^i\mathcal{A}_{q/\bar{q}}^{<i}(q)/(M/2\pm k_0)\approx 0$ when $k^i\ll M$.} Considering $\epsilon^{\mu}_{\lambda}$ as a real vector, we could make a replacement for the $k$-related terms in the integrand by employing the relations, 
\begin{eqnarray}
k^ik^j\rightarrow |\bm k|^2\Big[z^2\epsilon^{i}_{\lambda}\epsilon^{j}_{\lambda}-\frac{(1-z^2)}{2}\hat{\Theta}^{ij}_{\lambda}\Big],\quad \hat{\Theta}^{ij}_{\lambda}=\eta^{ij}+\epsilon^{i}_{\lambda}\epsilon^{j}_{\lambda},\quad
z=\frac{-k\cdot\epsilon_{\lambda}}{|\bm k|},
\end{eqnarray}
which yield
\begin{eqnarray}
&&-\frac{2|k\cdot\epsilon(\lambda, {\bm q})|^2}{N_m}\rightarrow -\frac{2|\bm k|^2z^2}{N_m},
\\
&&2k\cdot\mathcal{A}_{q}^{<}(\bm p)k\cdot\mathcal{A}_{\bar{q}}^{<}(\bm p')
\rightarrow |\bm k|^2\big[(3z^2-1)\big(\epsilon_{\lambda}\cdot \mathcal{A}_{q}^{<}(\bm p)\epsilon_{\lambda}\cdot \mathcal{A}_{\bar{q}}^{<}(\bm p')\big)
-(1-z^2)\mathcal{A}_{q}^{<}(\bm p)\cdot \mathcal{A}_{\bar{q}}^{<}(\bm p')\big],
\\&& {\rm Re}\big(k\cdot\epsilon(\lambda,{\bm q})\mathcal{A}_{q}^{<}(\bm p)\cdot \epsilon^*(\lambda,{\bm q})\big)k\cdot\mathcal{A}^<_{\bar{q}}(\bm p')
\rightarrow -z^2|\bm k|^2\epsilon_{\lambda}\cdot\mathcal{A}_{q}^{<}(\bm p)\epsilon_{\lambda}\cdot\mathcal{A}_{\bar{q}}^{<}(\bm p'),
\\
&&{\rm Re}\big(k\cdot\epsilon(\lambda,{\bm q})\mathcal{A}_{\bar{q}}^{<}(\bm p')\cdot \epsilon^*(\lambda,{\bm q})\big)k\cdot\mathcal{A}^<_{q}(\bm p)
\rightarrow -z^2|\bm k|^2\epsilon_{\lambda}\cdot\mathcal{A}_{q}^{<}(\bm p)\epsilon_{\lambda}\cdot\mathcal{A}_{\bar{q}}^{<}(\bm p'),
\end{eqnarray}
for $\bm p=\bm p'\approx\bm q/2$.
It turns out that
\begin{eqnarray}\nonumber
	\big(\hat{\Sigma}_V^{<\mu\rho}+\hat{\Sigma}_A^{<\mu\rho}\big)\hat{P}_{\rho\mu}(\lambda,q)
	=\frac{N_m(1-z^2)}{m^2}
	{\rm Tr_{c}}\big[f_{{\rm V}q}(\bm q/2)f_{{\rm V}\bar{q}}(\bm q/2)
	-4\big(\epsilon_{\lambda}\cdot \mathcal{A}_{q}^{<}(\bm q/2)\epsilon_{\lambda}\cdot \mathcal{A}_{\bar{q}}^{<}(\bm q/2)\big)\big].
	\\
\end{eqnarray}
For a complex $\epsilon^{\mu}_{\lambda}$, it is expected that one could simply replace $\big(\epsilon_{\lambda}\cdot \mathcal{A}_{q}^{<}(\bm q/2)\epsilon_{\lambda}\cdot \mathcal{A}_{\bar{q}}^{<}(\bm q/2)\big)$ by ${\rm Re}\big(\epsilon_{\lambda}\cdot \mathcal{A}_{q}^{<}(\bm q/2)\epsilon^*_{\lambda}\cdot \mathcal{A}_{\bar{q}}^{<}(\bm q/2)\big)$ in the final result.

Setting 
\begin{eqnarray}
	\bm\epsilon_0=(0,1,0),\quad \bm\epsilon_{+1}=-\frac{1}{\sqrt{2}}(i,0,1),\quad \bm\epsilon_{-1}=\frac{1}{\sqrt{2}}(-i,0,1),
\end{eqnarray}
we derive
\begin{eqnarray}
	f_{0}(q)\approx \frac{\tilde{N}\Delta t}{E_q}{\rm Tr_{c}}(f_{{\rm V}q}f_{{\rm V}\bar{q}})
	\Bigg[1-\frac{4{\rm Tr_{c}}\big(\mathcal{ A}^{<y}_{q}(\bm q/2)\mathcal{A}^{<y}_{\bar{q}}(\bm q/2)\big)}{{\rm Tr_{c}}(f_{{\rm V}q}f_{{\rm V}\bar{q}})}\Bigg]_{\bm q=0}
\end{eqnarray}
and
\begin{eqnarray}
	f_{\pm 1}(q)\approx \frac{\tilde{N}\Delta t}{E_q}{\rm Tr_{c}}(f_{{\rm V}q}f_{{\rm V}\bar{q}})
	\Bigg[1-\frac{2{\rm Tr_{c}}\big(\mathcal{ A}^{<x}_{q}(\bm q/2)\mathcal{A}^{<x}_{\bar{q}}(\bm q/2)+\mathcal{ A}^{<z}_{q}(\bm q/2)\mathcal{A}^{<z}_{\bar{q}}(\bm q/2)\big)}{{\rm Tr_{c}}(f_{{\rm V}q}f_{{\rm V}\bar{q}})}\Bigg]_{\bm q=0},
\end{eqnarray}
where 
\begin{eqnarray}
\tilde{N}=\int^{\infty}_0 \frac{d|\bm k|}{(2\pi)}\int^{1}_{-1}dz \frac{N_m|\bm k|^2(1-z^2)}{m^2}
\delta(M-2\sqrt{|\bm k|^2+m^2})
=\frac{M(M^2-4m^2)^{3/2}}{24\pi m^2}
\end{eqnarray}
is an overall constant, while its explicit form is unimportant for the normalized spin-density matrix.

Eventually, in the non-relativistic limit, it is found that
\begin{eqnarray}\label{eq:rho00_QKT_local}
	\rho_{00}(q,X)=\frac{f_{0}(q,X)}{f_{0}(q,X)+f_{+1}(q,X)+f_{-1}(q,X)}
	=\frac{1-\frac{4{\rm Tr_{c}}\big(\mathcal{ A}^{<y}_{q}(\bm q/2)\mathcal{A}^{<y}_{\bar{q}}(\bm q/2)\big)}{{\rm Tr_{c}}(f_{{\rm V}q}f_{{\rm V}\bar{q}})}}{3-\frac{4\sum_{j=x,y,z}{\rm Tr_{c}}\big(\mathcal{ A}^{<j}_{q}(\bm q/2)\mathcal{A}^{<j}_{\bar{q}}(\bm q/2)\big)}{{\rm Tr_{c}}(f_{{\rm V}q}f_{{\rm V}\bar{q}})}},
\end{eqnarray}
for $\bm q=0$.
When considering the global spin alignment, Eq.~(\ref{eq:rho00_QKT_local}) could be further revised as
\begin{eqnarray}\label{eq:rho00_QKT_global}\nonumber
	\rho_{00}(q)&=&\frac{\int d\Sigma_{X}\cdot qf_{0}(q,X)}{\int d\Sigma_{X}\cdot q\big(f_{0}(q,X)+f_{+1}(q,X)+f_{-1}(q,X)\big)}
	\\
	&=&\frac{1-{\rm Tr_c}\langle\hat{\mathcal{P}}_{q}^{y}({\bm q/2})\hat{\mathcal{P}}_{\bar{q}}^{y}({\bm q/2})\rangle_{\bm q=0}}{3-\sum_{i=x,y,z}{\rm Tr_c}\langle\hat{\mathcal{P}}_{q}^{i}({\bm q/2})\hat{\mathcal{P}}_{\bar{q}}^{i}({\bm q/2})\rangle_{\bm q=0}},
  \end{eqnarray}
where
\begin{eqnarray}
	{\rm Tr_c}\langle\hat{\mathcal{P}}_{q}^{i}({\bm p})\hat{\mathcal{P}}_{\bar{q}}^{i}({\bm p})\rangle
	&=&\frac{4\int d\Sigma_{X}\cdot p{\rm Tr_{c}}[\langle\mathcal{A}^{<i}_{q}(\bm p,X) \mathcal{A}^{<i}_{\bar{q}}(\bm p,X)\rangle]}{\int d\Sigma_{X}\cdot p{\rm Tr_{c}}[f_{{\rm V}q}(\bm p,X)f_{{\rm V}\bar{q}}(\bm p,X)]},
\end{eqnarray} 
which is equivalent to
\begin{eqnarray}
{\rm Tr_c}\langle\hat{\mathcal{P}}_{q}^{i}({\bm p})\hat{\mathcal{P}}_{\bar{q}}^{i}({\bm p})\rangle=\frac{\int d\Sigma_X\cdot p{\rm Tr_{c}}\big[\langle\mathcal{J}_{{q}5}^{i}(\bm p,X)\mathcal{J}_{\bar{q}5}^{i}(\bm p,X)\rangle\big]}{\int d\Sigma_X\cdot p{\rm Tr_{c}}\big[\mathcal{N}^0_{{q}}(\bm p,X)\mathcal{N}^{0}_{\bar{q}}(\bm p,X)\big]},
\end{eqnarray}
in the non-relativistic limit.
Equation (\ref{eq:rho00_QKT_global}) is found to be structurally similar to Eq.~(\ref{eq:rho00_modified}) yet with some subtle differences. For isotropic spin correlations, $\rho_{00}(q)=1/3$ for both Eq.~(\ref{eq:rho00_QKT_global}) and Eq.~(\ref{eq:rho00_modified}). With weak spin correlations, Eq.~(\ref{eq:rho00_QKT_global}) reduces to
\begin{eqnarray}\label{eq:rho00_weak_spin}
	\rho_{00}\approx \frac{1}{3}+\frac{1}{9}{\rm Tr_c}\big(\langle\hat{\mathcal{P}}^{x}_{q}\hat{\mathcal{P}}^{x}_{\bar{q}}\rangle+\langle\hat{\mathcal{P}}^{z}_{q}\hat{\mathcal{P}}^{z}_{\bar{q}}\rangle-2\langle\hat{\mathcal{P}}^{y}_{q}\hat{\mathcal{P}}^{y}_{\bar{q}}\rangle\big),
\end{eqnarray}
which is analogous to the form of Eq.~(\ref{eq:rho00_smallP}) despite an overall factor of $2$ difference for the spin-correlation corrections. Comparing ${\rm Tr_c}\langle\hat{\mathcal{P}}_{q}^{i}\hat{\mathcal{P}}_{\bar{q}}^{i}\rangle$ with $\langle\mathcal{P}^{i}_{q}\mathcal{P}^{i}_{\bar{q}}\rangle$, in addition to how we trace over the color degrees of freedom, as will be further expatiated below, one immediately notices a factor of $4$ difference and the integration of local and non-local correlations, for which the latter difference does not occur when the involved spin correlations are constant in position space. Nevertheless, due to non-locality of $\langle\mathcal{P}^{i}_{q}\mathcal{P}^{i}_{\bar{q}}\rangle$ as opposed to ${\rm Tr_c}\langle\hat{\mathcal{P}}_{q}^{i}\hat{\mathcal{P}}_{\bar{q}}^{i}\rangle$, they are physically distinct quantities. As previously proposed in e.g. Refs.~\cite{Pang:2016igs,Kumar:2022ylt}, $\langle\mathcal{P}^{i}_{q}\mathcal{P}^{i}_{\bar{q}}\rangle$ could be responsible for probing the correlation of spin polarization of a $\Lambda$ hyperon and of an $\bar{\Lambda}$, whereas it does not directly contribute to spin alignment of vector mesons. This newly derived $\rho_{00}(q)$ is also different from the one in Ref.~\cite{Sheng:2022wsy}, for which the spin correction on quark Wigner functions is presumably governed by the local spin-polarization pseudo-vector as a consistent treatment with the quark model, whereas we directly derive the spin-dependent corrections from the Wigner functions of quarks in AKT.

Tracing over color space, the relevant spin correlation for spin alignment reads
\begin{eqnarray}\label{eq:spin_corr_1}\nonumber
	{\rm Tr_c}\langle\hat{\mathcal{P}}_{q}^{i}({\bm q/2})\hat{\mathcal{P}}_{\bar{q}}^{i}({\bm q/2})\rangle=
	\frac{4\int d\Sigma_{X}\cdot q\big(2N_c^2\langle\mathcal{A}^{{\rm s}i}_{q}(\bm q/2,X) \mathcal{A}^{{\rm s}i}_{\bar{q}}(\bm q/2,X)\rangle
		+\langle\mathcal{A}^{ai}_{q}(\bm q/2,X) \mathcal{A}^{ai}_{\bar{q}}(\bm q/2,X)\rangle\big)}{\int d\Sigma_{X}\cdot q\big(2N_c^2f^{\rm s}_{Vq}(\bm q/2,X)f^{\rm s}_{V\bar{q}}(\bm q/2,X)+f^a_{Vq}(\bm q/2,X)f^a_{V\bar{q}}(\bm q/2,X)\big)},
	\\
\end{eqnarray}  
from which it is found that not only the color-singlet components but also the color-octet components of Wigner functions are involved. In high-energy nuclear collisions, the quark coalescence occurs at the late time when the vector component of Wigner functions reaches thermal equilibrium, for which $f^a_{Vq/\bar{q}}$ are suppressed.
On the contrary, non-equilibrium effects upon the axial-vector component should play an important role for spin polarization or correlation. In such a case, both $\langle\mathcal{A}^{{\rm s}i}_{q} \mathcal{A}^{{\rm s}i}_{\bar{q}}\rangle$ and $\langle\mathcal{A}^{ai}_{q} \mathcal{A}^{ai}_{\bar{q}}\rangle$ need to be considered for spin alignment. The scenarios for $\langle\mathcal{A}^{{\rm s}i}_{q} \mathcal{A}^{{\rm s}i}_{\bar{q}}\rangle$ and $\langle\mathcal{A}^{ai}_{q} \mathcal{A}^{ai}_{\bar{q}}\rangle$ triggered by color fields are schematically illustrated in Fig.~\ref{fig_in_ex_fields}. Moreover, unlike $\langle\mathcal{A}^{{\rm s}i}_{q} \mathcal{A}^{{\rm s}i}_{\bar{q}}\rangle$ expected to be positive, $\langle\mathcal{A}^{ai}_{q} \mathcal{A}^{ai}_{\bar{q}}\rangle$ should be negative based on the charge-conjugation symmetry implying $\mathcal{A}^{ai}_{\bar{q}}=-\mathcal{A}^{ai}_{q}$.

\section{Spin alignment from the glasma}
\label{sec:spin_alignment_glasma}

We now evaluate $\rho_{00}$ from Eq.~(\ref{eq:rho00_QKT_global}) in the glasma state, for which we shall compute the spin correlation, 
\begin{eqnarray}\nonumber\label{eq:spin_corr_aa}
	{\rm Tr_c}\langle\hat{\mathcal{P}}_{q}^{i}({\bm q/2})\hat{\mathcal{P}}_{\bar{q}}^{i}({\bm q/2})\rangle
	\approx\frac{\int d\Sigma_{X}\cdot q\big(\langle\tilde{a}^{{\rm s}i}_{q}(\bm q/2,X) \tilde{a}^{{\rm s}i}_{\bar{q}}(\bm q/2,X)\rangle
		+\langle\tilde{a}^{ai}_{q}(\bm q/2,X) \tilde{a}^{ai}_{\bar{q}}(\bm q/2,X)\rangle/(2N_c^2)\big)}{m_qm_{\bar{q}}\int d\Sigma_{X}\cdot q\big(f^{\rm s}_{{\rm V}q}(\bm q/2,X)f^{\rm s}_{{\rm V}\bar{q}}(\bm q/2,X)\big)},
	\\
\end{eqnarray}
where we have dropped the non-dynamical contribution in late times and the $f^{a}_{{\rm V}q/\bar{q}}$ in equilibrium and taken $\epsilon_{q/\bar{q}}(\bm q/2)\approx m_{q/\bar{q}}$ for the non-relativistic limit.

From Ref.~\cite{Guerrero-Rodriguez:2021ask} by solving the linearized Yang-Mills equation, with small rapidity, the non-vanishing color-field correlators can be written as
\begin{eqnarray}
	\langle E_{\perp}^{ai}(X')E_{\perp}^{a'j}(X'')\rangle&=&-\frac{1}{2}g^2N_c\delta^{aa'}\epsilon^{in}\epsilon^{jm}\int^{X'}_{\perp;q,u}\int^{X''}_{\perp;l,v}\Omega_{-}(u_{\perp},v_{\perp})
	\frac{q^n l^m}{ql}
	\times\!J_1(qX'_0)J_1(lX''_0),    \label{eq:eiejcorr_sim}\\
	\langle B_{\perp}^{ai}(X')B_{\perp}^{a'j}(X'')\rangle&=&-\frac{1}{2}g^2N_c\delta^{aa'}\epsilon^{in}\epsilon^{jm}\int^{X'}_{\perp;q,u}\int^{X''}_{\perp;l,v}
	\Omega_{+}(u_{\perp},v_{\perp})
	\frac{q^n l^m}{ql}
	\times\!J_1(qX'_0)J_1(lX''_0),    \label{eq:bibjcorr_sim}\\
	\langle E_{\perp}^{ai}(X')B^{a'z}(X'')\rangle&=&-i\frac{1}{2}g^2N_c\delta^{aa'}\epsilon^{in}\int^{X'}_{\perp;q,u}\int^{X''}_{\perp;l,v}
	\Omega_{-}(u_{\perp},v_{\perp})
	\frac{q^n}{q}
	\times\!J_1(qX'_0)J_0(lX''_0),      
	\label{eq:eib3corr_sim}\\
	\langle B_{\perp}^{ai}(X')E^{a'z}(X'')\rangle&=&-i\frac{1}{2}g^2N_c\delta^{aa'}\epsilon^{in}\int^{X'}_{\perp;q,u}\int^{X''}_{\perp;l,v}
	\Omega_{+}(u_{\perp},v_{\perp})
	\frac{q^n}{q}
	\times\!J_1(qX'_0)J_0(lX''_0),
	\label{eq:bie3corr_sim}\\
	\langle E^{az}(X')E^{a'z}(X'')\rangle&=&\frac{1}{2}g^2N_c\delta^{aa'}\int^{X'}_{\perp;q,u}\int^{X''}_{\perp;l,v}
	\Omega_{+}(u_{\perp},v_{\perp})
	\times\!J_0(qX'_0)J_0(lX''_0),
	\label{eq:e3e3corr_sim}\\
	\langle B^{az}(X')B^{a'z}(X'')\rangle&=&\frac{1}{2}g^2N_c\delta^{aa'}\int^{X'}_{\perp;q,u}\int^{X''}_{\perp;l,v}
	\Omega_{-}(u_{\perp},v_{\perp})
	\times\!J_0(qX'_0)J_0(lX''_0),   \label{eq:b3b3corr}
\end{eqnarray}
where $\Omega_{\mp}(u_{\perp},v_{\perp})$
\begin{eqnarray}
	\Omega_{\mp}(u_{\perp},v_{\perp})=\left[G_1(u_{\perp},v_{\perp})G_2(u_{\perp},v_{\perp})\mp h_1(u_{\perp},v_{\perp})h_2(u_{\perp},v_{\perp})\right]
\end{eqnarray}
with $G_{1,2}$ and $h_{1,2}$ corresponding to the unpolarized and linearly polarized gluon distribution functions of nuclei $1$ and $2$, respectively,
and
\begin{eqnarray}\nonumber
	\int^{X'}_{\perp;q,u}\equiv\int \!\frac{d^2q_{\perp}}{(2\pi)^2}\int \!d^2u_{\perp}e^{iq_{\perp}(X'-u)_{\perp}}.
\end{eqnarray}
Here $V^i_{\perp}$ represents the transverse component of an arbitrary spatial vector $V^i$ with respect to $z$ axis as the beam direction, where $A_{\perp}B_{\perp}=\sum_{i=x,y}A^iB^i$.
Consequently, for $X_0=Y_0=0$, only the correlations between longitudinal color fields exist, which take the form
\begin{eqnarray}
	\langle E^{az}(0,X_{\perp})E^{az}(0,Y_{\perp})\rangle&\approx&\frac{1}{2}g^2N_c(N_c^2-1)
	\Omega_{+}(X_{\perp},Y_{\perp}),
	\\
	\label{eq:Bz_corr}
	\langle B^{az}(0,X_{\perp})B^{az}(0,Y_{\perp})\rangle&\approx&\frac{1}{2}g^2N_c(N_c^2-1)
	\Omega_{-}(X_{\perp},Y_{\perp}).
\end{eqnarray}
We may further adopt the Golec-Biernat W\"usthoff (GBW) dipole distribution such that \cite{Golec-Biernat:1998zce,Guerrero-Rodriguez:2021ask}
\begin{align}\label{eq:GBW}
	\Omega_{\pm}(u_{\perp},v_{\perp})=\Omega(u_{\perp},v_{\perp})=\frac{Q_s^4}{g^4N_c^2}\left(\frac{1-e^{-Q_s^2|u_{\perp}-v_{\perp}|^2/4}}{Q_s^2|u_{\perp}-v_{\perp}|^2/4}\right)^2,
\end{align}
where $Q_s$ denotes the saturation momentum. 

\begin{figure}[t]
	\begin{minipage}{7cm}
		\begin{center}
			{\includegraphics[width=1.1\hsize,height=6cm,clip]{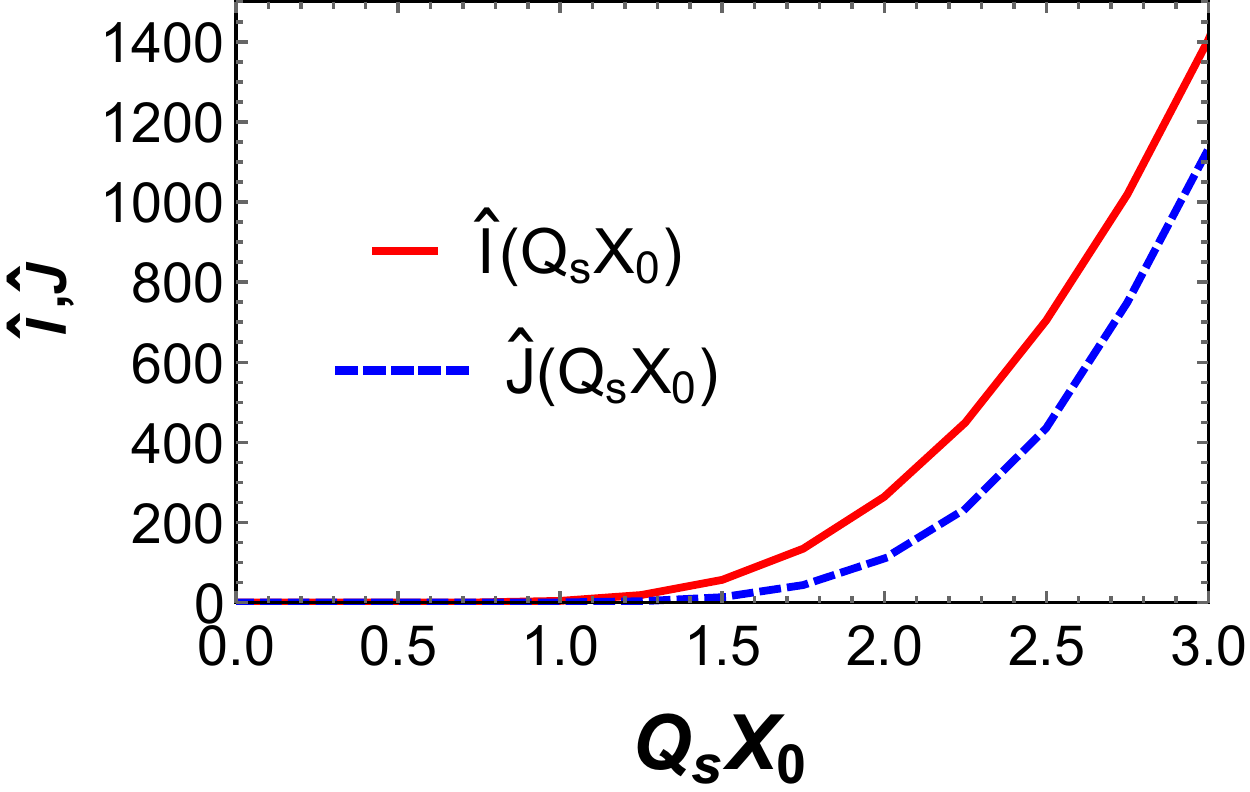}}
			
			\caption{Numerical results for $\hat{\mathcal{I}}(Q_sX_0)$ and $\hat{\mathcal{J}}(Q_sX_0)$ at small $Q_sX_0$.
			}
			\label{fig_IJhat_smallX0}
		\end{center}
	\end{minipage}
	\hspace {1cm}
	\begin{minipage}{7cm}
		\begin{center}
			{\includegraphics[width=1.1\hsize,height=6cm,clip]{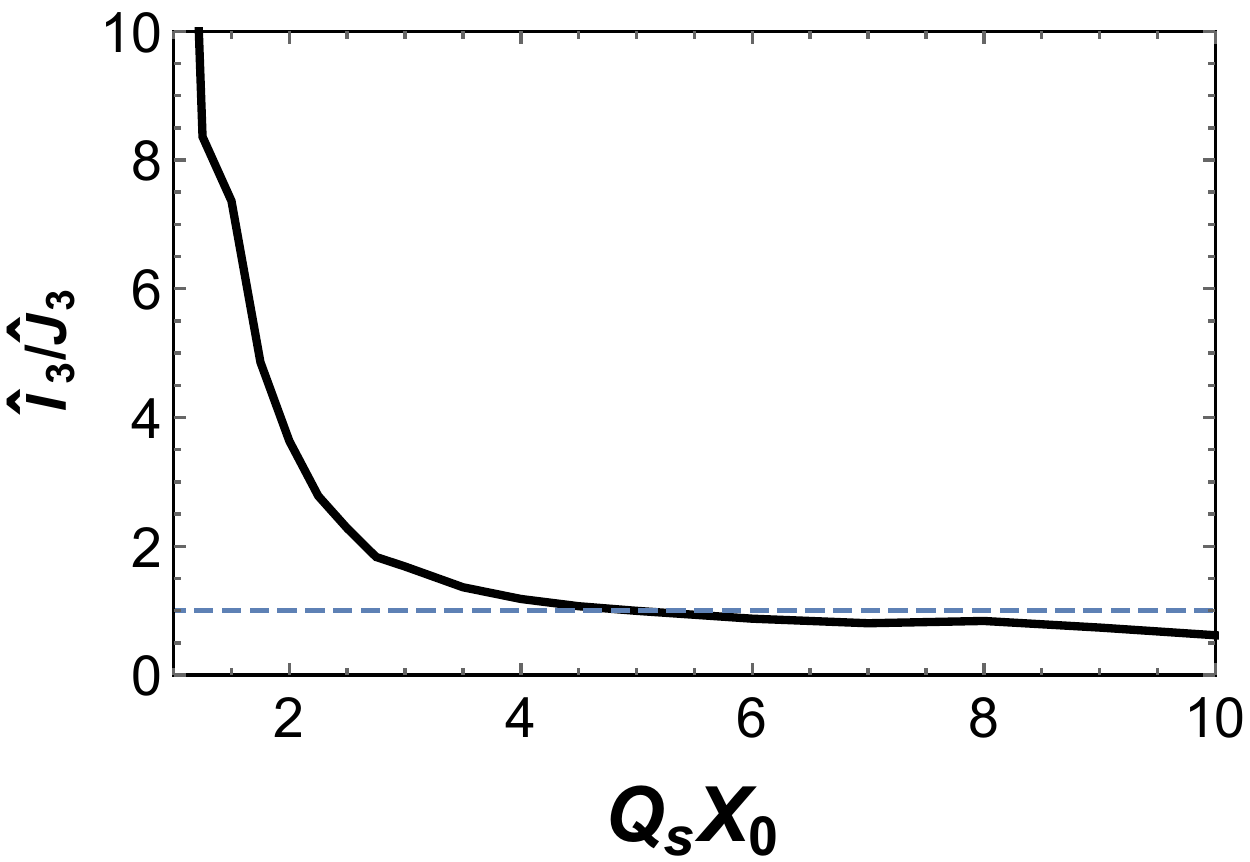}}
			
			\caption{The ratio of $\hat{\mathcal{I}}(Q_sX_0)$ and $\hat{\mathcal{J}}(Q_sX_0)$.}
			\label{fig_IJhat_ratio}
		\end{center}
	\end{minipage}
\end{figure}

\begin{figure}[t]
	\begin{minipage}{7cm}
		\begin{center}
			{\includegraphics[width=1.1\hsize,height=6cm,clip]{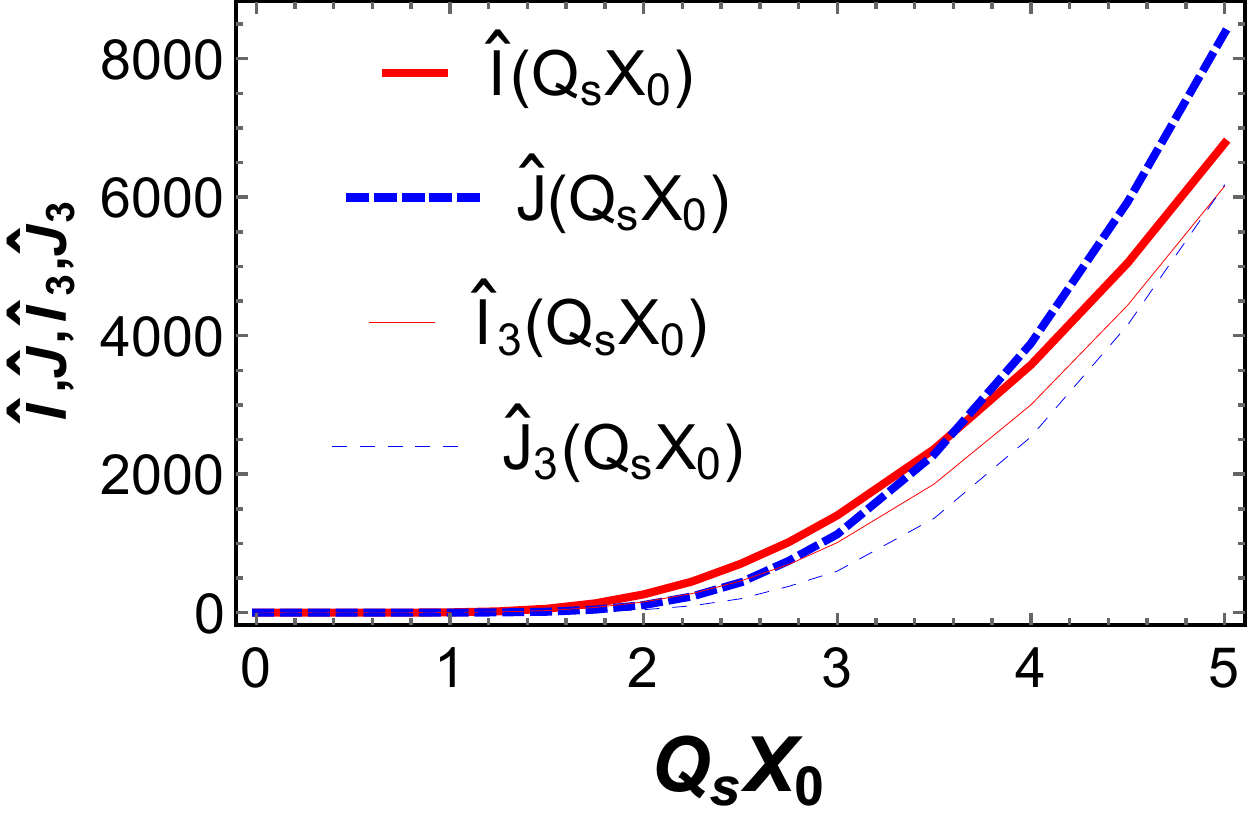}}
			
			\caption{Numerical results for $\hat{\mathcal{I}}(Q_sX_0)$, $\hat{\mathcal{J}}(Q_sX_0)$, $\hat{\mathcal{I}}_3(Q_sX_0)$, and $\hat{\mathcal{J}}_3(Q_sX_0)$ up to $Q_sX_0=5$.
			}
			\label{fig_IJhat_largerX0}
		\end{center}
	\end{minipage}
	\hspace {1cm}
	\begin{minipage}{7cm}
		\begin{center}
			{\includegraphics[width=1.1\hsize,height=6cm,clip]{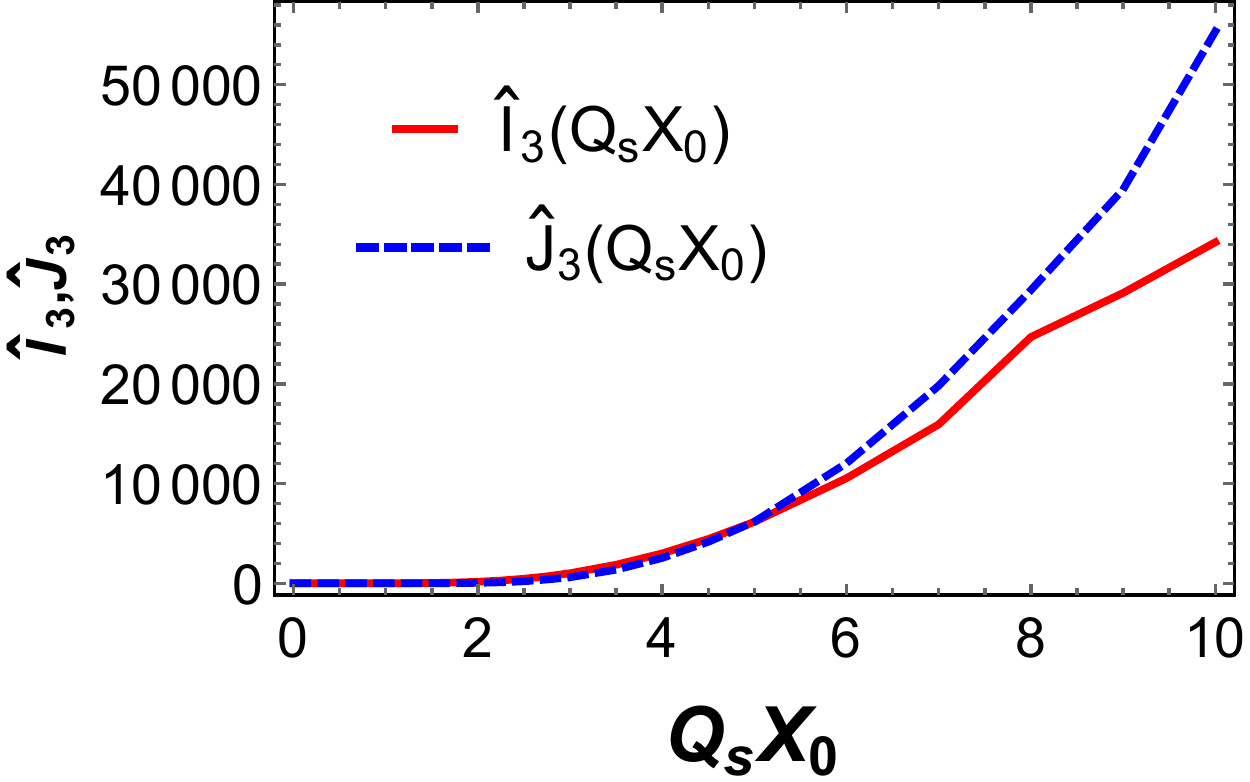}}
			
			\caption{Numerical results for $\hat{\mathcal{I}}_3(Q_sX_0)$ and $\hat{\mathcal{J}}_3(Q_sX_0)$ at up to $Q_sX_0=10$. }
			\label{fig_IJhat3}
		\end{center}
	\end{minipage}
\end{figure}

Since the color fields from the glamsa decay in time, we only need to consider the dynamical contribution on spin correlations led by strong color fields in early times. 
From Eqs.~(\ref{eq:J5_color_fields}) and (\ref{eq:Bz_corr}) and the GBW distribution giving rise to $\Omega(X,X)=Q_s^4/(g^4N_c^2)$, in the non-relativistic limit, it is found that
\footnote{In fact, when including finite $\bm k$ beyond the non-relativistic approximation, the color-field correlators involved could be non-local. E.g., one should consider the integration of $\langle B^{ai}(0,X_{\perp}-k_{\perp}X_0/M)B^{ai}(0,X_{\perp}+k_{\perp}X_0/M)\rangle$ over $\bm k$ for $\bm q=0$. With energy conservation and the GBW distribution, $\langle B^{ai}(0,X_{\perp})B^{ai}(0,X_{\perp})\rangle$ in Eq.~(\ref{eq:spin_corr_int}) should be replaced by $C_{B}(Q_sX_0)\langle B^{ai}(0,X_{\perp})B^{ai}(0,X_{\perp})\rangle$, where $C_B(Q_sX_0)=(1-{\rm{exp}}[-Q_s^2X_0^2(M^2-4m^2)/M^2])/(Q_s^2X_0^2(M^2-4m^2)/M^2)$ and finally one should take $X_0$ as the freeze-out time. Nonetheless, one also needs to include the contributions from chromo-electric fields at finite $\bm k$.}
\begin{eqnarray}\nonumber\label{eq:spin_corr_int}
	\frac{1}{2N_c^2}\langle\tilde{a}^{ai}_{q}(\bm q/2,X) \tilde{a}^{ai}_{\bar{q}}(\bm q/2,X)\rangle
	&\approx& -\frac{\hbar^2g^2}{8N_c^2}\big(\partial_{\epsilon_{\bm q/2}}f^{(0)}_V(\epsilon_{\bm q/2})\big)^2\langle B^{ai}(X)B^{ai}(X)\rangle_{X_0=0}
	\\
	&\approx& -\frac{\hbar^2Q_s^4(N_c^2-1)}{16N_c^3}\delta^{iz}\big(\partial_{\epsilon_{\bm q/2}}f^{(0)}_V(\epsilon_{\bm q/2})\big)^2,
\end{eqnarray}
where we introduce a shorthand notation $f^{(0)}_V(\epsilon_{\bm p})=\tilde{f}^{\rm s}_V(\epsilon_{\bm p},0)$.
On the other hand, from the color-singlet contribution led by local four-field correlations (see Ref.~\cite{Kumar:2022ylt} and similar calculation of the longitudinal correlation in appendix.~\ref{app:azaz_calc}), we obtain
\begin{eqnarray}\label{eq:spin_corr_ext_yy}
	\langle\tilde{a}^{{\rm s}y}_{q}(\bm q/2,X) \tilde{a}^{{\rm s}y}_{\bar{q}}(\bm q/2,X)\rangle\approx \frac{\hbar^2(N_c^2-1)Q_s^6}{64(2\pi)^4N_c^4m^2}\big(\partial_{\epsilon_{\bm q/2}}f^{(0)}_V(\epsilon_{\bm q/2})\big)^2\hat{\mathcal{I}}(Q_sX^{\rm th}_0),
\end{eqnarray}
where $\hat{\mathcal{I}}(Q_sX^{\rm th}_0)$ corresponds to a dimensionless factor depending upon $Q_sX^{\rm th}_0$ with $X^{\rm th}_0$ denoting a thermalization time as the ending time of the glasma phases\footnote{In principle, Eq.~(\ref{eq:spin_corr_int}) should also depend on $X^{\rm th}_0$, where $\langle B^{ai}(X)B^{ai}(X)\rangle_{X_0=0}$ should be more precisely replaced by $\langle (B^{ai}(0,\bm X)-B^{ai}(X^{\rm th}_0,\bm X))(B^{ai}(0,\bm X)-B^{ai}(X^{\rm th}_0,\bm X))\rangle$. We consider the case for $|B^{ai}(0,\bm X)|\gg |B^{ai}(X^{\rm th}_0,\bm X)|$.}. For simplicity, we neglect the transition period between the glasma and QGP. The exact value of $\hat{\mathcal{I}}(Q_sX^{\rm th}_0)$ has to be numerically computed from the multi-dimensional integral as shown in Ref.~\cite{Kumar:2022ylt}. From the symmetry of color fields in the glasma, we expect that $\langle\tilde{a}^{{\rm s}x}_{q}(\bm q/2,X) \tilde{a}^{{\rm s}x}_{\bar{q}}(\bm q/2,X)\rangle$ is equal to $\langle\tilde{a}^{{\rm s}y}_{q}(\bm q/2,X) \tilde{a}^{{\rm s}y}_{\bar{q}}(\bm q/2,X)\rangle$ in Eq.~(\ref{eq:spin_corr_ext_yy}). On the other hand,  $\langle\tilde{a}^{{\rm s}z}_{q}(\bm q/2,X) \tilde{a}^{{\rm s}z}_{\bar{q}}(\bm q/2,X)\rangle$ corresponds to $\langle\tilde{a}^{{\rm s}y}_{q}(\bm q/2,X) \tilde{a}^{{\rm s}y}_{\bar{q}}(\bm q/2,X)\rangle$ in Eq.~(\ref{eq:spin_corr_ext_yy}) by replacing $\hat{\mathcal{I}}(Q_sX^{\rm th}_0)$ therein with $\hat{\mathcal{J}}(Q_sX^{\rm th}_0)$ calculated in appendix.~\ref{app:azaz_calc}. The numerical results of $\hat{\mathcal{I}}(Q_sX_0)$ and $\hat{\mathcal{J}}(Q_sX_0)$ and their ratio are shown in Fig.~\ref{fig_IJhat_smallX0} and Fig.~\ref{fig_IJhat_ratio}, respectively. Also, as shown in Fig.~\ref{fig_IJhat_largerX0}, $\hat{\mathcal{I}}_3(Q_sX_0)$ and $\hat{\mathcal{J}}_3(Q_sX_0)$ correspond to the dominant terms contributing to $\hat{\mathcal{I}}(Q_sX_0)$ and $\hat{\mathcal{J}}(Q_sX_0)$ at large $Q_sX_0$. See Ref.~\cite{Kumar:2022ylt} and appendix.~\ref{app:azaz_calc} for the explicit definitions of $\hat{\mathcal{I}}_3$ and $\hat{\mathcal{J}}_3$. Consequently, for $Q_sX_0\gtrsim 5$, we could approximate $\hat{\mathcal{I}}(Q_sX_0)\approx\hat{\mathcal{I}}_3(Q_sX_0)$ and $\hat{\mathcal{J}}(Q_sX_0)\approx\hat{\mathcal{J}}_3(Q_sX_0)$ with the numerical results illustrated in Fig.~\ref{fig_IJhat3}. Except for the contributions from color fields, we also have
\begin{eqnarray}
	\langle\tilde{a}^{{\rm s}i}_{q}(\bm q/2,X) \tilde{a}^{{\rm s}i}_{\bar{q}}(\bm q/2,X)\rangle_{\rm EM}\approx -\frac{\hbar^2}{4}e^2B^2_i(0)\big(\partial_{\epsilon_{\bm q/2}}f^{(0)}_V(\epsilon_{\bm q/2})\big)^2
\end{eqnarray}
from $U(1)$ magnetic fields generated by colliding nuclei. Accordingly, in light of Eq.~(\ref{eq:spin_corr_aa}), we make a decomposition for the spin correlatiors contributing to spin alignment (in the non-relativistic limit) induced by color fields from the glasma and electromagnetic fields,
\begin{eqnarray}
{\rm Tr_c}\langle\hat{\mathcal{P}}_{q}^{i}({\bm q/2})\hat{\mathcal{P}}_{\bar{q}}^{i}({\bm q/2})\rangle=\Pi^{ii}_{\rm oct}+\Pi^{ii}_{\rm sin}+\Pi^{ii}_{\rm EM},
\end{eqnarray}
with
\begin{eqnarray}
\Pi^{ii}_{\rm oct}=-\frac{\hbar^2Q_s^4(N_c^2-1)\delta^{iz}\big(\partial_{\epsilon_{\bm q/2}}f^{(0)}_V(\epsilon_{\bm q/2})\big)^2}{16N_c^3m^2f^{\rm th}_{Vq}(\epsilon_{\bm q/2})f^{\rm th}_{V\bar{q}}(\epsilon_{\bm q/2})},
\end{eqnarray}
\begin{eqnarray}
	\Pi^{yy}_{\rm sin}=\Pi^{xx}_{\rm sin}=\Pi^{zz}_{\rm sin}\frac{\hat{\mathcal{I}}(Q_sX^{\rm th}_0)}{\hat{\mathcal{J}}(Q_sX^{\rm th}_0)}=\frac{\hbar^2(N_c^2-1)Q_s^6\big(\partial_{\epsilon_{\bm q/2}}f^{(0)}_V(\epsilon_{\bm q/2})\big)^2\hat{\mathcal{I}}(Q_sX^{\rm th}_0)}{64(2\pi)^4N_c^4m^4f^{\rm th}_{Vq}(\epsilon_{\bm q/2})f^{\rm th}_{V\bar{q}}(\epsilon_{\bm q/2})},
\end{eqnarray}
and
\begin{eqnarray}
\Pi^{ii}_{\rm EM}=-\frac{\hbar^2e^2B^2_i(0)\big(\partial_{\epsilon_{\bm q/2}}f^{(0)}_V(\epsilon_{\bm q/2})\big)^2}{4m^2f^{\rm th}_{Vq}(\epsilon_{\bm q/2}2)f^{\rm th}_{V\bar{q}}(\epsilon_{\bm q/2})},
\end{eqnarray}
where $f^{\rm th}_{Vq/\bar{q}}(\epsilon_{\bm q/2})\approx 1/(e^{m/T}+1)$ as the thermal distribution for quarks and antiquarks at zero chemical potentials and non-relativistic limit with $T$ being the freeze-out temperature on a coalescence hyper-surface. 

We then estimate the order of magnitude for the spin alignment of $\phi$ mesons as an example in RHIC and LHC at sufficiently high collision energies that the glasma phase could exist. Consequently, we will consider three sets of saturation momenta, $Q_s=1$, $2$, and $3$ GeV. 
For other approximations, we adopt the same setup in Ref.~\cite{Kumar:2022ylt}. We take $\hbar=1$ and postulate $f^{(0)}_{V}=1/(e^{\epsilon_{\bm q/2}/\Lambda}+1)$ as an early-time distribution function of quarks and antiquarks with $\Lambda\sim Q_s\gg \epsilon_{\bm q/2}$ such that $\partial_{\epsilon_{\bm q/2}}f^{(0)}_{V}\approx -1/(4Q_s)$. For other numerical parameters, we take $m\approx 500$ MeV as the constituent quark mass for strange quarks, $T\approx 150$ MeV as the freeze-out temperature at chemical equilibrium, and $X^{\rm th}_0=0.5$ fm as the thermalization time at the end of the glasma phase. For the maximum collision energy at RHIC, we anticipate $Q_s=1$ GeV, which yields $\hat{\mathcal{I}}\approx 1.6\hat{\mathcal{J}}\approx 700$ for $Q_sX^{\rm th}_0\approx 2.5$, and approximate\footnote{In fact, the event-by-event fluctuating electromagnetic fields can engender sizable contributions for $\Pi^{xx}_{\rm EM}\sim B_{x}^{2}$ \cite{Deng:2012pc}, but the magnetic-field contribution is relatively suppressed by those from color fields and thus not our primary interest in the present work.} $B^{i}\approx B(0)\delta^{iy}$ with $|eB(0)|\approx \, m_{\pi}^2$ and $m_{\pi}\approx 140$ MeV \cite{Skokov:2009qp}, which results in
\begin{eqnarray}
	\Pi^{ii}_{\rm oct}\approx -3.9\delta^{iz},\quad \Pi^{yy}_{\rm sin}=\Pi^{xx}_{\rm sin}\approx 1.6\Pi^{zz}_{\rm sin}\approx 0.58,\quad \Pi^{ii}_{\rm EM}\approx-0.02\delta^{iy}.
\end{eqnarray}
For LHC energies, we could have $Q_s\approx 2\sim 3$ GeV. Considering $Q_s=2$ GeV, which yields $\hat{\mathcal{I}}\approx 0.8\hat{\mathcal{J}}\approx 6800$ for $Q_sX^{\rm th}_0\approx 5$, and $|eB(0)|\approx \, 10 m_{\pi}^2$, it is found that
\begin{eqnarray}
\Pi^{ii}_{\rm oct}\approx -15.6\delta^{iz},\quad \Pi^{yy}_{\rm sin}=\Pi^{xx}_{\rm sin}\approx 0.8\Pi^{zz}_{\rm sin}\approx 90.8,\quad \Pi^{ii}_{\rm EM}\approx-0.5\delta^{iy}.
\end{eqnarray}
For $Q_s=3$ GeV with the same setup of the case for $Q_s=2$ GeV, which yields $\hat{\mathcal{I}}\approx 0.8\hat{\mathcal{J}}\approx 19700$ for $Q_sX^{\rm th}_0\approx 7.5$, one obtains 
\begin{eqnarray}
 	\Pi^{ii}_{\rm oct}\approx -35.1\delta^{iz},\quad \Pi^{yy}_{\rm sin}=\Pi^{xx}_{\rm sin}\approx 0.8\Pi^{zz}_{\rm sin}\approx 1331,\quad \Pi^{ii}_{\rm EM}\approx-0.22\delta^{iy},
\end{eqnarray}
where the change of $\Pi^{ii}_{\rm EM}$ with the same magnitude of $|eB(0)|$ stems from the $Q_s^{-2}$ suppression due to the approximation, $\partial_{\epsilon_{\bm q/2}}f^{(0)}_{V}\approx -1/(4Q_s)$. As opposed to $\Pi^{ii}_{\rm oct}$, here $\Pi^{ii}_{\rm sin}$ are rather sensitive to the value of $X^{\rm th}_0$. When choosing $X^{\rm th}_0=0.2$ fm, we then have 
$Q_sX^{\rm th}_0\approx 2$ with $\hat{\mathcal{I}}\approx 2.4\hat{\mathcal{J}}\approx 260$ for $Q_s=2$ GeV and 
$Q_sX^{\rm th}_0\approx 3$ with $\hat{\mathcal{I}}\approx 1.2\hat{\mathcal{J}}\approx 1400$ for $Q_s=3$ GeV.
We accordingly acquire
\begin{eqnarray}
\Pi^{yy}_{\rm sin}=\Pi^{xx}_{\rm sin}\approx 2.4\Pi^{zz}_{\rm sin}\approx 3.5
\end{eqnarray}
for $Q_s=2$ GeV and 
\begin{eqnarray}
	\Pi^{yy}_{\rm sin}=\Pi^{xx}_{\rm sin}\approx 1.2\Pi^{zz}_{\rm sin}\approx 94
\end{eqnarray}
for $Q_s=3$ GeV. Superficially, the results seem to be unrealistically large even when focusing on just the contributions from $\Pi^{ii}_{\rm oct}$, while all these values should be further suppressed by the spin-relaxation effects in the QGP phase. Note that $\Pi^{yy}_{\rm sin}$ here is much larger than $\langle\mathcal{P}^{y}_{q}\mathcal{P}^{y}_{\bar{q}}\rangle$ obtained in Ref.~\cite{Kumar:2022ylt} due to the absence of strong suppression coming from  $1/(Q_s^2A_{\rm T})$ with $A_{\rm T}$ the transverse area of the QGP led by non-locality. 

On the other hand, $\Pi^{ii}_{\rm oct}$ led by two-field correlations should be in principle more dominant than $\Pi^{ii}_{\rm sin}$ induced by four-field correlations according to the weak-field expansion of the QKT. Nevertheless, due to non-perturbtive properties of the glasma, the hierarchy is not guaranteed and the four-field correlations could surpass two-field correlations at larger $Q_s$. From the numerical results, around $Q_s=2$ GeV, depending on the choice of $X^{\rm th}_0$, one could have either $|\Pi^{ii}_{\rm sin}|>|\Pi^{ii}_{\rm oct}|$ or $|\Pi^{ii}_{\rm sin}|<|\Pi^{ii}_{\rm oct}|$, where the former implies the breakdown of our perturbtive approach. We may still estimate $\rho_{00}$ based on the primary contribution from anisotropic $\Pi^{ii}_{\rm oct}$ at $Q_s=2$ GeV with a certain assumption of $X^{\rm th}_0$. For $Q_s=3$ GeV, $\Pi^{ii}_{\rm sin}$ overwhelms $\Pi^{ii}_{\rm oct}$ and thus our estimate becomes invalid. A non-perturbative treatment is presumably needed at higher collision energies. 

As discussed in the previous section, the quarks and antiquarks may emerge at the time later than the initial time with strongest color fields and electromagnetic fields. In practice, we shall consider $\tilde{f}^{\rm s}_V(\epsilon_{\bm p},X_0)\approx f^{(0)}_V(\epsilon_{\bm p})\Theta(X_0-X^{\rm q}_0)$ with $X^{\rm q}_0$ being the time for emergence of quarks or antiquarks in the glamsa state and thus we have to evaluate
\begin{eqnarray}
	\frac{1}{2N_c^2}\langle\tilde{a}^{ai}_{q}(\bm q/2,X) \tilde{a}^{ai}_{\bar{q}}(\bm q/2,X)\rangle
	&\approx& -\frac{\hbar^2 g^2}{8N_c^2}\big(\partial_{\epsilon_{\bm q/2}}f^{(0)}_V(\epsilon_{\bm q/2})\big)^2\langle B^{ai}(X)B^{ai}(X)\rangle_{X_0=X_0^{\rm q}}.
\end{eqnarray}  
In such a case, we have not only non-vanishing $\langle B^{az}(X)B^{az}(X)\rangle$ but also $\langle B^{ax,y}(X)B^{ax,y}(X)\rangle$. 

To analyze the possible correction from a nonzero $X^{\rm q}_0$, we first calculate $\langle B^{az}(X)B^{az}(X)\rangle$ with $X_0\neq 0$. By using $\int d^2q_{\perp}=\int dqqd\theta_{q}$, $\int d^2l_{\perp}=\int dlld\theta_{l}$, and
\begin{eqnarray}
	\int d\theta_q\int d\theta_{l}e^{iq_{\perp}(X'-u)_{\perp}}e^{il_{\perp}(Y'-v)_{\perp}}=(2\pi)^2J_0(q|X_{\perp}-u_{\perp}|)J_0(l|Y_{\perp}-v_{\perp}|),
\end{eqnarray}
we find
\begin{eqnarray}\nonumber
	&&\int^{X}_{\perp;q,u}\int^{X}_{\perp;l,v}
	\Omega_{-}(u_{\perp},v_{\perp})
	\times\!J_0(qX_0)J_0(lX_0)
	\\\nonumber
	&&=\int\frac{dqdl}{(2\pi)^2}ql\int \!d^2u_{\perp}\int \!d^2v_{\perp}\Omega_{-}(u_{\perp},v_{\perp})
	\!J_0(qX_0)J_0(lX_0)J_0(q|X_{\perp}-u_{\perp}|)J_0(l|X_{\perp}-v_{\perp}|)
	\\
	&&=\int\frac{d^2\bar{u}_{\perp}d^2\bar{v}_{\perp}}{(2\pi)^2}\Omega_{-}(u_{\perp},v_{\perp})\frac{\delta(X_0-|\bar{u}_{\perp}|)\delta(X_0-|\bar{v}_{\perp}|)}{|\bar{u}_{\perp}||\bar{v}_{\perp}|},
\end{eqnarray}
where we have applied the orthogonal condition for Bessel
functions,
\begin{eqnarray}\label{eq:Bessel_rel}
	\int^{\infty}_{0}dr rJ_{\nu}(kr)J_{\nu}(sr)=\frac{\delta(k-s)}{s},
\end{eqnarray}
and the change of variables, $\bar{u}_{\perp}=X_{\perp}-u_{\perp}$ and $\bar{v}_{\perp}=X_{\perp}-v_{\perp}$, to reach the second equality. When adopting the GBW distribution, we have $\Omega_{-}(u_{\perp},v_{\perp})=\Omega(u_{\perp},v_{\perp})=\Omega(\bar{u}_{\perp},\bar{v}_{\perp})$. It is more convenient to work in polar coordinates,
\begin{eqnarray}
	\int d^2\bar{u}_{\perp}\int d^2\bar{v}_{\perp}=\int^{\infty}_0 d|\bar{u}_{\perp}||\bar{u}_{\perp}|\int^{\infty}_0 d|\bar{v}_{\perp}||\bar{v}_{\perp}|\int^{2\pi}_0 d\theta_{\bar{u}}\int^{2\pi}_0 d\theta_{\bar{v}}.
\end{eqnarray} 
For an integrand depending on only $\Theta_{\bar{u},\bar{v}}=\theta_{\bar{v}}-\theta_{\bar{u}}$, we could make the change of variables such that (see appendix.~\ref{app:int_derivation} for the derivation)
\begin{eqnarray}\nonumber
	\int^{2\pi}_0 d\theta_{\bar{u}}\int^{2\pi}_0 d\theta_{\bar{v}}\mathcal{F}(\theta_{\bar{u}}-\theta_{\bar{v}})
	=\int^{2\pi}_{-2\pi}d\Theta_{\bar{u},\bar{v}}2\pi\mathcal{F}(\Theta_{\bar{u},\bar{v}})-
	\int^{2\pi}_{0}d\Theta_{\bar{u},\bar{v}}\Theta_{\bar{u},\bar{v}}\big[\mathcal{F}(\Theta_{\bar{u},\bar{v}})+\mathcal{F}(-\Theta_{\bar{u},\bar{v}})\big].
	\\
\end{eqnarray} 
Given that $\Omega(u_{\perp},v_{\perp})$ only depends on
\begin{eqnarray}
	|\bar{u}_{\perp}-\bar{v}_{\perp}|^2=|\bar{u}_{\perp}|^2+|\bar{v}_{\perp}|^2-2|\bar{u}_{\perp}||\bar{v}_{\perp}|\cos\Theta_{\bar{u},\bar{v}},
\end{eqnarray}
it is found that
\begin{eqnarray}
	g^2\langle B^{az}(X)B^{az}(X)\rangle
	=\frac{Q_s^4(N_c^2-1)}{4\pi^2 N_c}\mathcal{I}_{Bz}(Q_sX_0)
\end{eqnarray}
and accordingly
\begin{eqnarray}
	\frac{1}{2N_c^2}\langle\tilde{a}^{az}_{q}(\bm q/2,X) \tilde{a}^{az}_{\bar{q}}(\bm q/2,X)\rangle
	&\approx& -\frac{\hbar^2Q_s^4(N_c^2-1)}{16N_c^3}\Big(\frac{\mathcal{I}_{Bz}(Q_sX_0)}{2\pi^2}\Big)\big(\partial_{\epsilon_{\bm q/2}}f^{(0)}_V(\epsilon_{\bm q/2})\big)^2,
\end{eqnarray}
where 
\begin{eqnarray}
	\mathcal{I}_{Bz}(Q_sX_0)=\int^{2\pi}_{0}d\Theta_{\bar{u},\bar{v}}(2\pi-\Theta_{\bar{u},\bar{v}})\left(\frac{1-e^{-Q_s^2X_0^2(1-\cos\Theta_{\bar{u},\bar{v}})/2}}{Q_s^2X_0^2(1-\cos\Theta_{\bar{u},\bar{v}})/2}\right)^2
\end{eqnarray}
can be evaluated numerically. Note that $\mathcal{I}_{Bz}(Q_sX_0)=2\pi^2$ when $Q_sX_0\rightarrow 0$.

Next, we shall consider the contribution from dynamically generated transverse chromo-magnetic fields,
\begin{eqnarray}
	\langle B_{\perp}^{ay}(X)B_{\perp}^{ay}(X)\rangle=-\frac{1}{2}g^2N_c(N_c^2-1)\int^{X}_{\perp;q,u}\int^{X}_{\perp;l,v}
\Omega_{+}(u_{\perp},v_{\perp})
\frac{q^x l^x}{ql}
\times\!J_1(qX_0)J_1(lX_0),
\end{eqnarray}
for which
\begin{eqnarray}\nonumber
&&\int^{X}_{\perp;q,u}\int^{X}_{\perp;l,v}
\Omega_{+}(u_{\perp},v_{\perp})
\frac{q^x l^x}{ql}
\times\!J_1(qX_0)J_1(lX_0)
\\\nonumber
&&=-\int\frac{dqdl}{(2\pi)^2}ql\int \!d^2u_{\perp}\int \!d^2v_{\perp}\Omega_{+}(u_{\perp},v_{\perp})\frac{(X-u)^x_{\perp}(X-v)^x_{\perp}}{|X_{\perp}-u_{\perp}||X_{\perp}-v_{\perp}|}
\\\nonumber
&&\quad \times\!J_1(qX_0)J_1(lX_0)J_1(q|X_{\perp}-u_{\perp}|)J_1(l|X_{\perp}-v_{\perp}|)
\\
&&=-\int\frac{d^2\bar{u}_{\perp}d^2\bar{v}_{\perp}}{(2\pi)^2}\Omega_{-}(u_{\perp},v_{\perp})\frac{\bar{u}_{\perp}^x\bar{v}_{\perp}^x\delta(X_0-|\bar{u}_{\perp}|)\delta(X_0-|\bar{v}_{\perp}|)}{|\bar{u}_{\perp}|^2|\bar{v}_{\perp}|^2},
\end{eqnarray}
by using
\begin{eqnarray}\nonumber
&&\int d\theta_q\int d\theta_{l}\frac{q^jl^i }{ql}e^{iq_{\perp}(X-u)_{\perp}}e^{il_{\perp}(Y-v)_{\perp}}
\\
&&=-(2\pi)^2\frac{(X-u)_{\perp}^i(Y-v)_{\perp}^j}{|X_{\perp}-u_{\perp}||Y_{\perp}-v_{\perp}|}J_1(q|X_{\perp}-u_{\perp}|)J_1(l|Y_{\perp}-v_{\perp}|).
\end{eqnarray}
Given
\begin{eqnarray}
\bar{u}_{\perp}^x\bar{v}_{\perp}^x=\frac{|\bar{u}_{\perp}||\bar{v}_{\perp}|}{2}\big(\cos\Theta_{\bar{u},\bar{v}}+\cos\theta_{\bar{u},\bar{v}} \big),
\end{eqnarray}
where $\theta_{\bar{u},\bar{v}}=\theta_{\bar{v}}+\theta_{\bar{u}}$, and taking the GBW distribution, it is found that
\begin{eqnarray}\nonumber
&&-\int\frac{d^2\bar{u}_{\perp}d^2\bar{v}_{\perp}}{(2\pi)^2}\Omega_{-}(u_{\perp},v_{\perp})\frac{\bar{u}_{\perp}^x\bar{v}_{\perp}^x\delta(X_0-|\bar{u}_{\perp}|)\delta(X_0-|\bar{v}_{\perp}|)}{|\bar{u}_{\perp}|^2|\bar{v}_{\perp}|^2}
\\
&&=\frac{1}{4\pi^2}\frac{Q_s^4}{g^4N_c^2}
\int^{2\pi}_0d\Theta_{\bar{u},\bar{v}}\Big[\sin\Theta_{\bar{u},\bar{v}}-(2\pi-\Theta_{\bar{u},\bar{v}})\cos\Theta_{\bar{u},\bar{v}}\Big]\left(\frac{1-e^{-Q_s^2X_0^2(1-\cos\Theta_{\bar{u},\bar{v}})/2}}{Q_s^2X_0^2(1-\cos\Theta_{\bar{u},\bar{v}})/2}\right)^2,
\end{eqnarray}
by using the relations in appendix.~\ref{app:int_derivation}.
Consequently, one finds
\begin{eqnarray}
g^2\langle B_T^{ay}(X)B_T^{ay}(X)\rangle=\frac{Q_s^4(N_c^2-1)}{4\pi^2 N_c}\mathcal{I}_{By}(Q_sX_0),
\end{eqnarray}
which yields
\begin{eqnarray}
	\frac{1}{2N_c^2}\langle\tilde{a}^{ay}_{q}(\bm q/2,X) \tilde{a}^{ay}_{\bar{q}}(\bm q/2,X)\rangle
	&\approx& -\frac{\hbar^2Q_s^4(N_c^2-1)}{16N_c^3}\Big(\frac{\mathcal{I}_{By}(Q_sX_0)}{2\pi^2}\Big)\big(\partial_{\epsilon_{\bm q/2}}f^{(0)}_V(\epsilon_{\bm q/2})\big)^2,
\end{eqnarray}
where 
\begin{eqnarray}
\mathcal{I}_{By}(Q_sX_0)=\int^{2\pi}_0\frac{d\Theta_{\bar{u},\bar{v}}}{2}\Big[(2\pi-\Theta_{\bar{u},\bar{v}})\cos\Theta_{\bar{u},\bar{v}}-\sin\Theta_{\bar{u},\bar{v}}\Big]\left(\frac{1-e^{-Q_s^2X_0^2(1-\cos\Theta_{\bar{u},\bar{v}})/2}}{Q_s^2X_0^2(1-\cos\Theta_{\bar{u},\bar{v}})/2}\right)^2.
\end{eqnarray}
One can consistently check $\mathcal{I}_{By}(Q_sX_0)=0$ when $Q_sX_0\rightarrow 0$. 

\begin{figure}
	\begin{center}
		\includegraphics[width=0.5\hsize]{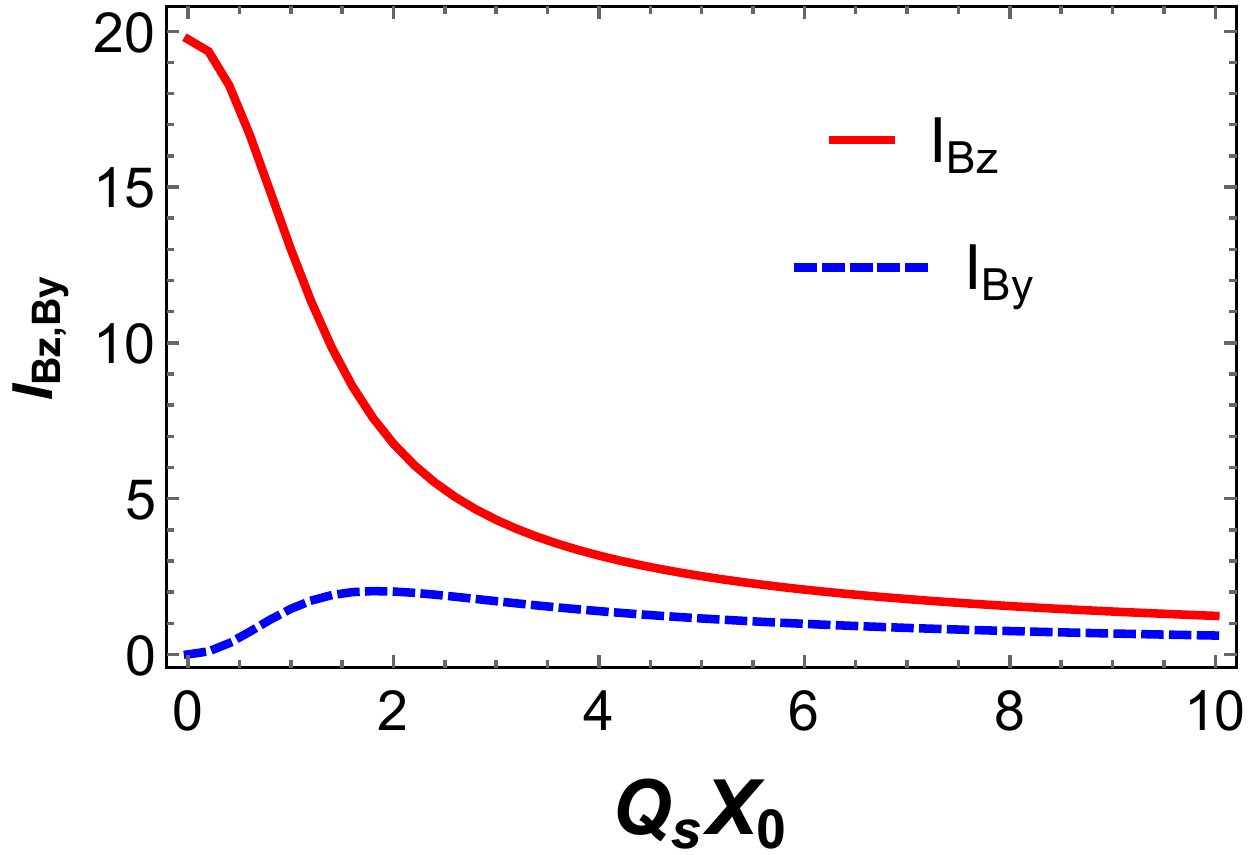}
	\end{center}
	\caption{Numerical results for $\mathcal{I}_{Bz}(Q_sX_0)$ and $\mathcal{I}_{By}(Q_sX_0)$.
	}
	\label{fig_integral}
\end{figure}

By symmetry, it is expected that $\langle\tilde{a}^{ay}_{q}(\bm q/2,X) \tilde{a}^{ay}_{\bar{q}}(\bm q/2,X)\rangle=\langle\tilde{a}^{ax}_{q}(\bm q/2,X) \tilde{a}^{ax}_{\bar{q}}(\bm q/2,X)\rangle$. The numerical results of $\mathcal{I}_{Bz}(Q_sX_0)$ and $\mathcal{I}_{By}(Q_sX_0)$ are shown in Fig.~\ref{fig_integral}. Notably, even at late time up to $Q_sX_0\approx 10$ such that $I_{Bz}$ is about $10$ times smaller, we still have $\Pi^{zz}_{\rm oct}\sim 1$. When assuming $\tilde{f}^{\rm s}_V(\epsilon_{\bm p},X_0)$ is created after $X^{\rm q}_0=0.1$ fm, one finds $\Pi^{zz}_{\rm oct}\approx -10.3$ for $Q_s=2$ GeV and $\Pi^{ii}_{\rm sin}$ are nearly unchanged because of the dominant contribution in late times within the glasma \cite{Kumar:2022ylt}, while the magnitude of $U(1)$ magnetic fields rapidly drops to $|eB^i(X^{\rm q}_0)|\approx 0.01m_{\pi}^2$ \cite{Deng:2012pc} in LHC energies, from which $\Pi^{ii}_{\rm EM}$ becomes negligible. The same scenario is applicable to the high-energy collisions at RHIC. In principle, a more rigorous estimation of $\Pi^{zz}_{\rm oct}$ is proportional to $\mathcal{I}_{B_z}(Q_sX_0^{\rm q})-\mathcal{I}_{B_z}(Q_sX_0^{\rm th})$, while this should not give significant suppression by order of magnitude provided $X_0^{\rm q}$ is not too close to $X_0^{\rm th}$. Finally, we may roughly conclude the spin alignment of $\phi$ mesons from the glasma for $Q_s\approx 1\sim 2$ GeV in an approximate equation,
\begin{eqnarray}\label{eq:rho00_approx}
\rho_{00}\sim\frac{1}{3+10e^{-2X^{\rm eq}_0/\tau^{\rm o}_{\rm R}}},
\end{eqnarray}  
where $X^{\rm eq}_0$ represents the freeze-out time at chemical equilibrium of the QGP and recall $\tau^{\rm o}_{\rm R}$ is an unknown parameter characterizing the effect of spin relaxation. Assuming the spin relaxation results in about 10 times suppression of the dynamical spin correlation, the contribution from color fields for spin alignment will be around the same order as the experimental measurement.

In practice, the first-principle study of the spin relaxation potentially applicable to heavy ion collisions has been so far conducted in weakly-coupled QGP up to the leading logarithmic order in coupling \cite{Li:2019qkf,Hattori:2019ahi,Hongo:2022izs}, where the corresponding collision term for dynamical spin relaxation in AKE is far from a relaxation-time form. In the heavy-quark limit $m\gg T$, we may naively adopt the relaxation rate derived in Ref.~\cite{Hongo:2022izs} and approximate \footnote{As also found in Refs.~\cite{Hattori:2019ahi, Hongo:2022izs}, the $\mathcal{O}(T/m)$ term can be written as a momentum diffusion term and neglected here.}
	\begin{eqnarray}
		(\tau^{\rm o}_{\rm R})^{-1}\approx \frac{g^2C_2(F)m_D^2T}{6\pi m^2}\ln g,
	\end{eqnarray}
	where $C_2(F)=(N_c^2-1)/(2N_c)$ and $m_D^2=g^2T^2(2N_c+N_f)/6$. Taking $N_c=N_f=3$, $\alpha_{s}=g^2/(4\pi)\approx 1/3$, and $T=200$ MeV as the average temperature of QGP, one obtains $(\tau^{\rm o}_{\rm R})^{-1}\approx 0.04$ GeV. For $X^{\rm eq}_0\approx 5$ fm, it is found $e^{-2X^{\rm eq}_0/\tau^{\rm o}_{\rm R}}\approx 0.11$ and $\rho_{00}\approx 0.24$ from Eq.~(\ref{eq:rho00_approx}). Albeit the rough agreement with the experimental measurement for $\rho_{00}$ of $\phi$ mesons at small transverse momenta in LHC \cite{ALICE:2019aid}, we emphasize that the estimation is subject to several approximations and phenomenological postulations and a more sophisticated analysis is required for quantitative comparisons. There have been great efforts devoted to modeling the dynamical spin polarization and relaxation from collisional effects in QGP and our result of spin polarization (correlation) from glasma effects can be used as an initial conditions for future simulations in the QGP phase. 

\begin{figure}[t]
	\begin{minipage}{7cm}
		\begin{center}
			{\includegraphics[width=1.1\hsize,height=6cm,clip]{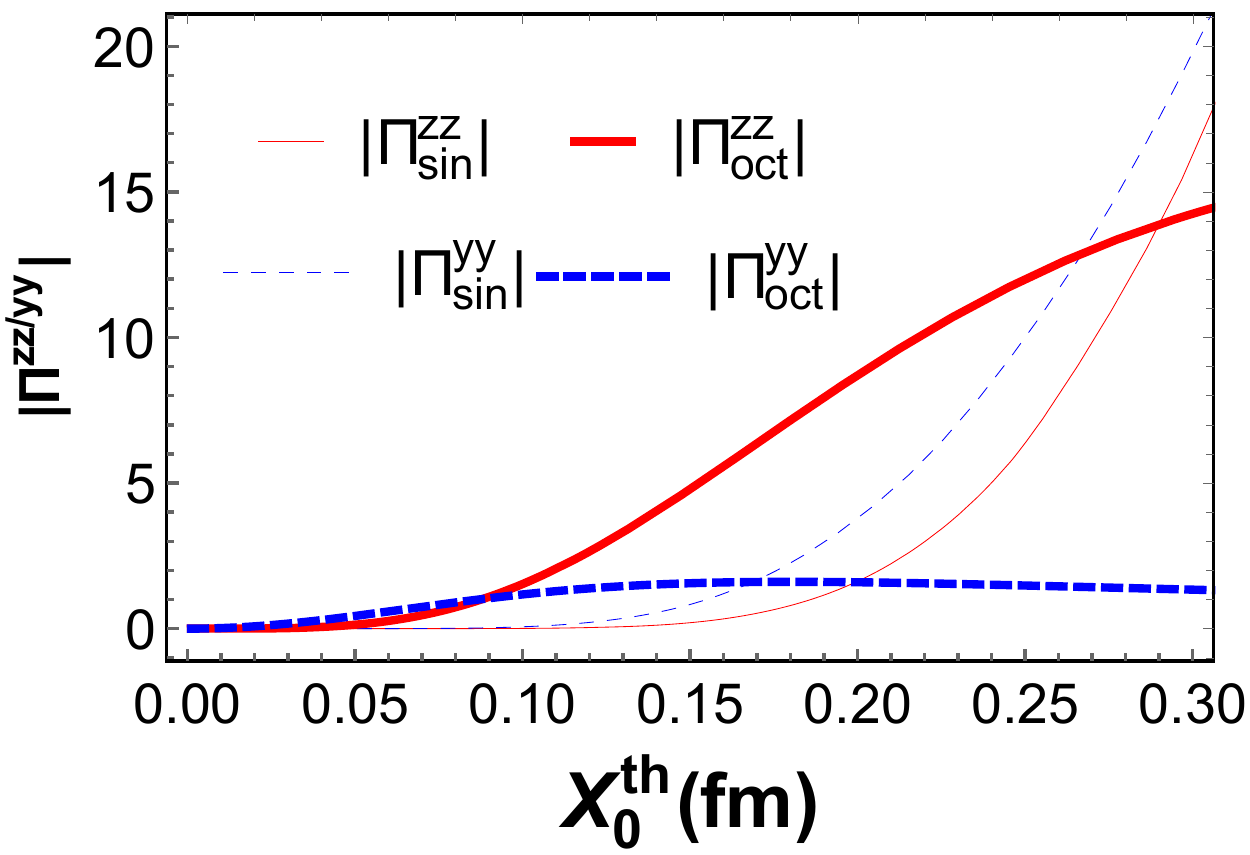}}
			
			\caption{Magnitudes of spin correlations from the color-octet and color-singlet contributions for $Q_s=2$ GeV.
			}
			\label{fig_Pi_fixedQs}
		\end{center}
	\end{minipage}
	\hspace {1cm}
	\begin{minipage}{7cm}
		\begin{center}
			{\includegraphics[width=1.1\hsize,height=6cm,clip]{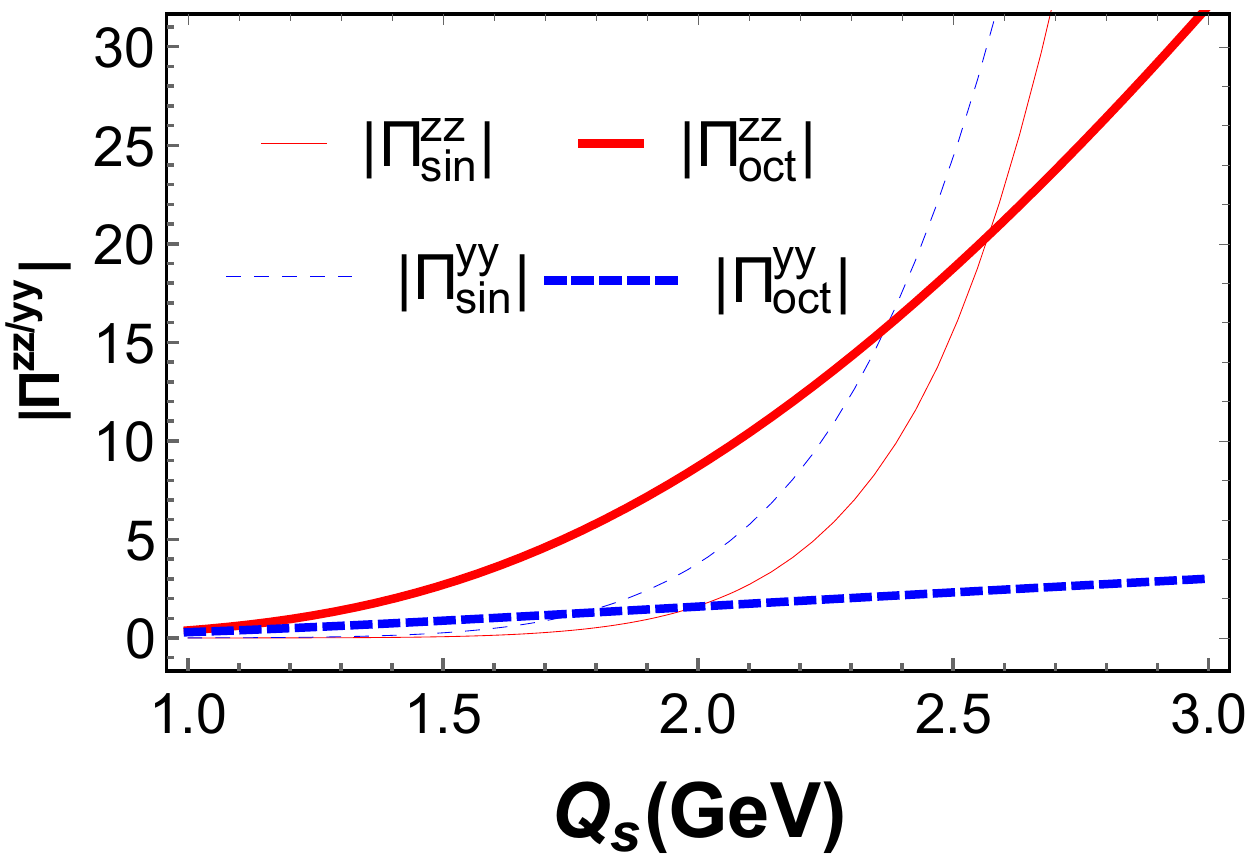}}
			
			\caption{Magnitudes of spin correlations from the color-octet and color-singlet contributions for $X^{\rm th}_0=0.2$ fm.}
			\label{fig_Pi_fixedX0th}
		\end{center}
	\end{minipage}
\end{figure}

Before ending this section, we further elaborate how our result is contingent on the choices of $Q_s$ and $X^{\rm th}_0$. In general, at a fixed $Q_s$, choosing a smaller $X^{\rm th}_0$ seems to result in the dominance of $|\Pi^{ii}_{\rm oct}|$ over $|\Pi^{ii}_{\rm sin}|$. Nonetheless, when $X^{\rm th}_0$ is too close to the initial time $X_0=0$, $|\Pi^{ii}_{\rm oct}|$ also drops since the initial color field encoded in $|\Pi^{ii}_{\rm oct}|$ is actually the difference between the initial color field at $X_0=0$ and the one at $X_0=X^{\rm th}_0$ as manifested by Eq.~(\ref{eq:aioct_sol}). Consequently, in our approximation, we also have to choose a sufficiently large $X^{\rm th}_0$ such that $\mathcal{O}(|B^{ai}(0,\bm X)|-|B^{ai}(X^{\rm th}_0,\bm X)|)\sim \mathcal{O}(|B^{ai}(0,\bm X)|)$ as the validity for neglecting late-time fields at the end of glasma phase when evaluating $\Pi^{ii}_{\rm oct}$. Moreover, as mentioned previously, we ignore the transition between the glasma and QGP phases, whereas the adopted $X^{\rm th}_0=0.2$ fm is same as the proper time for matching the glasma phase and pre-equilibrium state described by effective kinetic theory at the LHC energy in Ref.~\cite{Kurkela:2018wud}. To clarify the valid region for $|\Pi^{zz}_{\rm oct}|>|\Pi^{ii}_{\rm sin}|,|\Pi^{yy}_{\rm oct}|$ in our estimation, we plot the spin correlations from the color-octet and color-singlet contributions with $X^{\rm th}_0$ dependence at fixed $Q_s=2$ GeV and with $Q_s$ dependence for fixed $X^{\rm th}_0=0.2$ fm in Fig.~\ref{fig_Pi_fixedQs} and Fig.~\ref{fig_Pi_fixedX0th}, respectively. Here $\Pi^{ii}_{\rm oct}$ are calculated by including the field difference between $X_0=0$ and $X_0=X^{\rm th}_0$, which yield \footnote{Here $\Pi^{zz}_{\rm oct}(X^{\rm th}_0)$ is proportional to $\langle (B^{az}(0)-B^{az}(X^{\rm th}_0))(B^{az}(0)-B^{az}(X^{\rm th}_0))\rangle$, where we omit the spatial dependence. Accordingly, the terms associated with $\Omega(X^{\rm th}_{0})$ and $\mathcal{I}_{Bz}(Q_sX^{\rm th}_0)$ therein are led by the contributions from $\langle B^{az}(0)B^{az}(X^{\rm th}_0)\rangle$ and $\langle B^{az}(X^{\rm th}_0)B^{az}(X^{\rm th}_0)\rangle$, respectively.}
\begin{eqnarray}
\Pi^{zz}_{\rm oct}(X^{\rm th}_0)=\Pi^{zz}_{\rm oct}(0)\left(1-2\Omega(X^{\rm th}_0)+\frac{\mathcal{I}_{Bz}(Q_sX^{\rm th}_0)}{(2\pi^2)}\right),\quad 
\Pi^{yy}_{\rm oct}(X^{\rm th}_0)=\Pi^{zz}_{\rm oct}(0)\frac{\mathcal{I}_{By}(Q_sX^{\rm th}_0)}{(2\pi^2)},
\end{eqnarray}
and thus manifest the $X^{\rm th}_0$ dependence. It is found that our estimation is approximately valid when $0.1\lesssim X^{\rm th}_0\lesssim 0.25$ fm at $Q_s=2$ GeV or $1\lesssim Q_s\lesssim 2.3$ GeV for $X^{\rm th}_0=0.2$ fm.

Finally, we further comment on higher-order corrections in $\hbar$ expansion upon spin alignment. In principle, one could further incorporate the $\hbar^2$ corrections for AKE and $\mathcal{A}^{<}_{q/\bar{q}}$ albeit unavailable in literature at this moment, which may give rise to $\mathcal{O}(\hbar^3)$ corrections on $\rho_{00}$ from Eq.~(\ref{eq:SigmaA_in_A}) in our model of coalescence. Such corrections should be suppressed provided the $\hbar$ expansion holds. On the other hands, there could be possible $\mathcal{O}(\hbar^2)$ corrections for $\rho_{00}$ from $\mathcal{V}^{<}_{q/\bar{q}}$ led by Eq.~(\ref{eq:SigmaV_in_V}). Nevertheless, the existence of such corrections implies that $\mathcal{V}^{<}_{q/\bar{q}}$ are out of equilibrium (for the vector-charge degrees of freedom) and the effects could be probed by spin-independent observables. As a result, at least in high-energy nuclear collisions, it is unlikely that such corrections could be prominent for light quarks including strange quarks at small transverse momenta. We hence assumed $\mathcal{V}^{<}_{q/\bar{q}}$ in thermal equilibrium without quantum corrections in our setup.

\section{Spin alignment for vector mesons with finite momenta : qualitative analyses}
\label{sec:SAV:QA}

Previously, we focused on the spin alignment of vector mesons in the rest frame. \footnote{In Sec.~\ref{sec:spin_alignment_glasma}, we calculate the spin correlations from color fields of the glasma in the lab frame along with vector mesons in the rest frame. Accordingly, the estimation of spin alignment therein is actually for vector mesons with nearly zero momenta in the lab frame.} We may now investigate its momentum dependence in the lab frame.   
As stated in the previous section, the contribution from color-singlet correlations should be theoretically regarded as a higher-order correction compared with the color-octet ones. Consequently, we focus on the color-octet contribution, 
\begin{eqnarray}\nonumber
	{\rm Tr_c}\langle\hat{\mathcal{P}}_{q}^{i}({\bm q/2})\hat{\mathcal{P}}_{\bar{q}}^{i}({\bm q/2})\rangle
	&\approx&\frac{\int d\Sigma_{X}\cdot q\langle\tilde{a}^{ai}_{q}(\bm q/2,X) \tilde{a}^{ai}_{\bar{q}}(\bm q/2,X)\rangle}{2N_c^2m^2\int d\Sigma_{X}\cdot qf^{\rm s}_{{\rm V}q}(\bm q/2,X)f^{\rm s}_{{\rm V}\bar{q}}(\bm q/2,X)}
	\\
	&\approx&\frac{-\hbar^2g^2\int d\Sigma_{X}\cdot q\langle B_{\rm r}^{ai}(0,\bm X)B_{\rm r}^{ai}(0,\bm X)\rangle (\partial_{\epsilon_{\bm q/2}}\tilde{f}_V(\epsilon_{\bm q/2},0))^2}{8N_c^2m^2\int d\Sigma_{X}\cdot qf^{\rm s}_{{\rm V}q}(\bm q/2,X)f^{\rm s}_{{\rm V}\bar{q}}(\bm q/2,X)}
	,
\end{eqnarray}
for $\bm q=0$ in the non-relativistic limit for quarks and antiquarks, where $B^{ai}_{\rm r}$ denotes the chromo-magnetic field in the rest frame of vector mesons. Utilizing Eq.~(\ref{eq:rho00_weak_spin}) in the weak-correlation limit by augmentation with the spin-relaxation correction from a color-octet relaxation time, we have
\begin{eqnarray}\label{eq:deltarho00_Br}
	\rho_{00}- \frac{1}{3}\approx \frac{-\hbar^2g^2e^{-2X^{\rm eq}_0/\tau^{\rm o}_{\rm R}}\int d\Sigma_{X}\cdot q\Pi_{B}({\bm X}) (\partial_{\epsilon_{\bm q/2}}\tilde{f}_V(\epsilon_{\bm q/2},0))^2}{72N_c^2m^2\int d\Sigma_{X}\cdot qf^{\rm th}_{{\rm V}q}(\epsilon_{\bm q/2})f^{\rm th}_{{\rm V}\bar{q}}(\epsilon_{\bm q/2})},
\end{eqnarray}
where $\Pi_{B}({\bm X})=\langle B^{ax}_{\rm r}(0,\bm X)B^{ax}_{\rm r}(0,\bm X)\rangle +\langle B^{az}_{\rm r}(0,\bm X)B^{az}_{\rm r}(0,\bm X)\rangle -2\langle B^{ay}_{\rm r}(0,\bm X)B^{ay}_{\rm r}(0,\bm X)\rangle $.
Despite the negligible size of momentum corrections from quarks and antiquarks, we may simply conduct the Lorentz boost on the color fields to approximate the spin alignment of vector mesons with finite momenta in the lab frame, which will be helpful to qualitatively understand transverse-momentum and centrality dependence of spin alignment.

We could rewrite $B^{ai}_{\rm r}$ in terms of the color fields in the lab frame through 
\begin{eqnarray}
	B^{ai}_{\rm r}=\gamma (B^{ai}+\epsilon^{ijk}v_jE^a_{k})-(\gamma-1){\bm B}^a\cdot\hat{{\bm v}}\hat{v}^i,
\end{eqnarray} 
where $\gamma=1/\sqrt{1-|{\bm v}|^2}$ with $v^i=q^i/\sqrt{|\bm q|^2+M^2}$ and $\hat{v}^i=v^i/|\bm v|$. In principle, $|\bm q|$ here cannot be too large; otherwise the relativistic corrections upon quarks and antiquarks should be considered. By dropping the correlations between a chromo-magnetic field and an electric one and those between color fields along different directions, we accordingly find
\begin{eqnarray}
	\langle B^{ai}_{\rm r}(X)B^{ai}_{\rm r}(X)\rangle&=&\gamma^2\big(\langle B^{ai}(X)B^{ai}(X)\rangle +\epsilon^{ijk}v_j\epsilon^{ij'k'}v_{j'}\langle E^{a}_k(X)E^{a}_{k'}(X)\rangle\big)
	\\\nonumber
	&&-2\gamma(\gamma-1)\langle B^{ai}(X)B^{ai}(X)\rangle\hat{v}_i^2+(\gamma-1)^2\hat{v}_i^2\sum_{j=x,y,z}\langle B^{aj}(X)B^{aj}(X)\rangle
\end{eqnarray}	
and thus
\begin{eqnarray}
	\langle B^{ai}_{\rm r}(X)B^{ai}_{\rm r}(X)\rangle\approx\tilde{\gamma}_i^2\langle B^{ai}(X)B^{ai}(X)\rangle +\epsilon^{ijk}v_j\epsilon^{ij'k'}v_{j'}\langle E^{a}_k(X)E^{a}_{k'}(X)\rangle+\mathcal{O}(|\bm v|^4),
\end{eqnarray}
where $\tilde{\gamma}_i^2=1+|\bm v|^2-v_i^2\approx \gamma^2-v_i^2+\mathcal{O}(|\bm v|^4)$,
which yields
\begin{eqnarray}
	\langle B^{ax}_{\rm r}(X)B^{ax}_{\rm r}(X)\rangle&\approx&\tilde{\gamma}^2_{x}\langle B^{ax}(X)B^{ax}(X)\rangle +v_y^2\langle E^{az}(X)E^{az}(X)\rangle,
	\\
	\langle B^{ay}_{\rm r}(X)B^{ay}_{\rm r}(X)\rangle&\approx&\tilde{\gamma}^2_{y}\langle B^{ay}(X)B^{ay}(X)\rangle +v_x^2\langle E^{az}(X)E^{az}(X)\rangle,
	\\
	\langle B^{az}_{\rm r}(X)B^{az}_{\rm r}(X)\rangle&\approx&\gamma^2\langle B^{az}(X)B^{az}(X)\rangle
	+v_y^2\langle E^{ax}(X)E^{ax}(X)\rangle+v_x^2\langle E^{ay}(X)E^{ay}(X)\rangle,
\end{eqnarray}
up to $\mathcal{O}(|\bm v|^2)$ for $|v_{x,y}|\gg |v_z|$ at central rapidity. 

For color fields from the glasma, as shown in the previous section, the correlators of longitudinal color fields dominate over those of transverse ones and $\langle E^{az}(X)E^{az}(X)\rangle=\langle B^{az}(X)B^{az}(X)\rangle$ with the GBW distribution. Then Eq.~(\ref{eq:deltarho00_Br}) becomes
\begin{eqnarray}
	\rho_{00}- \frac{1}{3}\approx \frac{\hbar^2g^2(v_x^2-2v_y^2-1)e^{-2X^{\rm eq}_0/\tau^{\rm o}_{\rm R}}\int d\Sigma_{X}\cdot q\langle B^{az}(0,\bm X)B^{az}(0,\bm X)\rangle (\partial_{\epsilon_{\bm q/2}}\tilde{f}_V(\epsilon_{\bm q/2},0))^2}{72N_c^2m^2\int d\Sigma_{X}\cdot qf^{\rm th}_{{\rm V}q}(\epsilon_{\bm q/2})f^{\rm th}_{{\rm V}\bar{q}}(\epsilon_{\bm q/2})}.
\end{eqnarray}
In practice, $v_{x,y}$ could depend on spatial coordinates, but here we may consider just the average velocities. Recall that $x$ and $y$ correspond to the directions parallel and perpendicular to the reaction plane, respectively. It is hence anticipated that $v_x^2\geq v_y^2$ in most cases. Therefore, in the high-energy nuclear collisions with the presence of glasma, we expect $\rho_{00}<1/3$ and the deviation decreases with larger transverse momenta (but not too large) and less central collisions, for which $v_x^2-2v_y^2$ increases. 

Generically, we may consider two potential sources of color fields. One stems from the color fields generated by the glasma state, while the other comes from only the internal color fields characterizing an effective potential that binds the pair of a quark and an antiquark. The spin alignment induced by the glasma only exists in relatively high-energy nuclear collisions. On the other hand, the effective potential could play a more dominant role in low-energy collisions, where the contribution from external color fields vanishes as well. However, the magnitude of such a potential term entails non-perturbative calculations such as the lattice simulations, which is beyond the scope of the present work. For simplicity, we also assume the screening effect in the QGP phase such that the non-dynamical contribution of internal color fields at late time can be neglected.  Unlike the color fields from the glasma, the effective potential should be approximately isotropic. We hence postulate $\langle E^{ai}(X)E^{ai}(X)\rangle=\langle B^{ai}(X)B^{ai}(X)\rangle=\langle B^{a}(X)B^{a}(X)\rangle$. In the weak-correlation limit, Eq.~(\ref{eq:rho00_weak_spin}) accordingly yields
\begin{eqnarray}
	\rho_{00}- \frac{1}{3}\approx \frac{\hbar^2g^2(v_x^2-2v_y^2)e^{-2X^{\rm eq}_0/\tau^{\rm o}_{\rm R}}\int d\Sigma_{X}\cdot q\langle B^{a}(0,\bm X)B^{a}(0,\bm X)\rangle (\partial_{\epsilon_{\bm q/2}}\tilde{f}_V(\epsilon_{\bm q/2},0))^2}{36N_c^2m^2\int d\Sigma_{X}\cdot qf^{\rm th}_{{\rm V}q}(\epsilon_{\bm q/2})f^{\rm th}_{{\rm V}\bar{q}}(\epsilon_{\bm q/2})}.
\end{eqnarray}
As opposed to the case in high-energy collisions, we could possibly have $\rho_{00}>1/3$ given $v_x^2>2v_y^2$ from the effective potential at low-energy non-central collisions and the deviation may increase with larger transverse momenta ${\rm P}_{\rm T}$ (but not too large) and more peripheral collisions. Nonetheless, the effective potential also gives rise to $\rho_{00}<1/3$ in central collisions. In practice, the effect from the glasma and from the effective potential possibly co-exist for $\phi$ mesons and $J/\psi$, which should compete with each other in sufficiently high-energy collisions, while the latter effect is unlikely present for $K^{*0}$ although the glasma effect on $K^{*0}$ needs to be further investigated. As a result, the spin alignment for $K^{*0}$ may only occur at high-energy collisions with the glasma effect that yields $\rho_{00}<1/3$. Nonetheless, these two effects could be more prominent in distinct kinematic regions or centrality conditions. In table~\ref{table1}, we roughly summarize the qualitative behaviors of $\rho_{00}$ led by individual effects, where we also expect that the spin alignment of $J/\psi$ at high collision energies follows similar behaviors as those of $\phi$ mesons although it is unlikely that charm and anti-charm quarks will reach thermal equilibrium.   

\begin{table}
\centering
	\setlength{\tabcolsep}{6pt}
	\renewcommand{\arraystretch}{1.5}
	\begin{tabular}{ | l | l | l |l|l|}
		\hline
		& small-${\rm P}_{\rm T}$ & large-${\rm P}_{\rm T}$ & central & non-central \\ \hline
		glasma & $\rho^{\phi,J/\psi}_{00}<1/3$ & $\rho^{\phi,J/\psi}_{00}\lesssim 1/3$ & $\rho^{\phi,J/\psi}_{00}<1/3$ &  $\rho^{\phi,J/\psi}_{00}\lesssim 1/3$ \\ \hline
		effective potential & $|\rho^{\phi,J/\psi}_{00}-1/3|\gtrsim 0$ & $|\rho^{\phi,J/\psi}_{00}-1/3|>0$  & $\rho^{\phi,J/\psi}_{00}< 1/3$ &  $\rho^{\phi,J/\psi}_{00}>1/3$
		\\
		\hline
	\end{tabular}
 \caption{Competing effects for spin alignment from color fields}\label{table1}
\end{table}

\section{Conclusions and outlook}
\label{sec:conc}

In this paper, we estimate the spin alignment of vector mesons induced by color fields in the glasma phase via the newly derived equation with local spin correlation in the quark coalescence scenario. We find that both the color-singlet and color-octet components of the axial-charge current densities for quarks and antiquarks contribute to the associated spin correlators, which are dynamically generated through the background color fields. Based on our estimates the resulting spin alignment could be significant. We identify and discuss the limitations of our perturbative approach contingent upon the saturation momentum and lifetime of the glasma. We also qualitatively analyze the spin alignment of vector mesons with nonzero momentum in a self-consistent framework with color fields originating from both the glasma and effective potential characterized by isotropic internal color fields, which may result in opposite signs for $\rho_{00}-1/3$ for different transverse meson momenta and collisions of different centrality. The differences for spin alignment between these two scenarios stem from the intrinsic spatial anisotropy of the color fields and the momentum anisotropy of vector mesons, respectively.         
As briefly discussed in Sec.~\ref{sec:spin_alignment_glasma}, our estimates for spin correlations are subject to several approximations. Here we reiterate some potential issues and propose future research directions. Most importantly, the validity of our estimate is sensitive to the value of the thermalization time $X^{\rm th}_0$ where the glasma phase ends. Our numerical study indicates that our estimate breaks down around $Q_sX_0^{\rm th}\approx 2\sim 5$. However, the order-of-magnitude estimates for spin correlations still reveal non-negligible contributions to spin alignment from the glasma effect. In addition to the need for developing a more rigorous approach to treat the non-perturbative dynamics of color fields for spin transport of quarks, it is also crucial to have more reliable estimates for the spin relaxation after the end of the glasma phase. As shown in weakly coupled gauge theories, the collision terms responsible for spin relaxation are far more complicated than the relaxation-time form \cite{Li:2019qkf,Hattori:2019ahi,Fang:2022ttm,Wang:2022yli,Hongo:2022izs}. 

The color-field induced diffusion terms that are neglected in the weak-field limit may further cause the suppression of spin correlations. Furthermore, the sudden truncation of the glasma phase is unrealistic, which further raises the issue of the connection between spin transport of quarks in the glasma phase and in the QGP in the framework of QKT. On the other hand, the non-relativistic approximation for constituent quarks and antiquarks is adopted here, which reduces the non-local correlator of color fields to the local one. For a quantitative estimation of the spin correlation via non-perturbative approaches like lattice simulations, the spatial separation between color fields should be taken into account. Overall, a more precise estimation for color-field effects beyond the non-relativistic approximation will be required for a reliable comparison with the growing data for relativistic heavy ion collisions. Our formalism provides for a framework in which this is, in principle, possible.

\acknowledgments
D.-L. Y. would like to thank S. H. Lee for helpful discussions.
B. M. was supported by the U. S. Department of Energy under Grant No. DE-FG02-05ER41367.
D.-L. Y. was supported by National Science and Technology Council (Taiwan) under Grant No. MOST 110-2112-M-001-070-MY3.

\appendix

\section{Analytic solution from AKE}

Considering
\begin{eqnarray}
p\cdot\partial \tilde{a}^{\mu}(p,x)=G^{\mu}(p,x)-\frac{p_0\tilde{a}^{\mu}(p,x)}{\tau},
\end{eqnarray}
one finds
\begin{eqnarray}
\tilde{a}^{\mu}(p,X)=i\int d^4k\int\frac{d^4 X'}{(2\pi)^4}\frac{e^{-ik\cdot(X-X')}G^{\mu}(X,X')}{k\cdot p+ip_0\tau^{-1}+i\epsilon},
\end{eqnarray}
which can be decomposed into
\begin{eqnarray}
	\tilde{a}^{\mu}(p,X)=\tilde{a}^{\mu}_{(1)}(p,X)+\tilde{a}^{\mu}_{(2)}(p,X),
\end{eqnarray}
where 
\begin{eqnarray}
	\tilde{a}^{\mu}_{(1)}(p,X)=i\int d^4k\int\frac{d^4 X'}{(2\pi)^4}\frac{e^{-ik\cdot(X-X')}(k\cdot p+ip_0\tau^{-1})G^{\mu}(X,X')}{(k\cdot p+ip_0\tau^{-1})^2+\epsilon^2}\Big|_{\epsilon\rightarrow 0},
\end{eqnarray}
and
\begin{eqnarray}
	\tilde{a}^{\mu}_{(2)}(p,X)=\int d^4k\int\frac{d^4 \delta X}{(2\pi)^4}\pi \delta(k\cdot p+ip_0\tau^{-1})e^{-ik\cdot\delta X}G^{\mu}(X,X'),
\end{eqnarray}
where $\delta X\equiv X-X'$. Note that the convention for Fourier transformation here is 
\begin{eqnarray}
f(p,k)=\int \frac{d^4X}{(2\pi)^4}e^{ik\cdot X'}f(p,X').
\end{eqnarray}
Assigning $p_{\mu}=(p_0,0,0,p_z)$ hence we obtain
\begin{eqnarray}
	\tilde{a}^{\mu}_{(1)}(p,X)=i\int dk_0dk_z\int\frac{d \delta X_0d \delta X_z}{(2\pi)^2}\frac{e^{-ik_0\delta X_0+ik_z\delta X_z}}{k_0p_0-k_zp_z+ip_0\tau^{-1}}G^{\mu}(X,X')|_{\delta X_{x,y}=0},
\end{eqnarray}
which yields
\begin{eqnarray}\nonumber
	\tilde{a}^{\mu}_{(1)}(p,X)&=&\pi\int dk_z\int\frac{d\delta X_0d\delta X_z}{(2\pi)^2}\frac{{\rm sgn}(\delta X_0)}{p_0}e^{ik_z(\delta X_z-\delta X_0p_z/p_0)-\delta X_0/\tau}G^{\mu}(X,\delta X)|_{\delta X_{x,y}=0}
	\\\nonumber
	&=&\pi\int\frac{d\delta X_0d\delta X_z}{(2\pi)}\frac{{\rm sgn}(\delta X_0)}{p_0}\delta\big(\delta X_z-\delta X_0p_z/p_0\big)e^{-\delta X_0/\tau}G^{\mu}(X,\delta X)|_{\delta X_{x,y}=0}
	\\
	&=&\frac{1}{2p_0}\int d\delta X_0\Big[{\rm sgn}(\delta X_0)G^{\mu}(X,\delta X)e^{-\delta X_0/\tau}\Big]_{\delta X_{x,y}=0,\delta X_z=p_z\delta X_z/p_0}
\end{eqnarray}
by using
\begin{eqnarray}
	\int_{-\infty}^{\infty} dke^{-ikx}/(k+a)=-i\pi{\rm sgn}(x)e^{ixa}.
\end{eqnarray}
Similarly, it is found that
\begin{eqnarray}\nonumber
	\tilde{a}^{\mu}_{(2)}(p,X)&=&\int dk_0dk_z\int\frac{d \delta X_0d \delta X_z}{(2\pi)^2} \frac{\pi\delta(k_0-k_zp_z/p_0+i\tau^{-1})}{p_0}e^{-ik_0\delta X_0+ik_z\delta X_z}G^{\mu}(X,X')
	\\
	&=&\int\frac{d \delta X_0}{2p_0} e^{-\delta X_0/\tau}G^{\mu}(X,X')|_{\delta X_{x,y}=0,\delta X_z=p_z\delta X_z/p_0}
	,
\end{eqnarray}
and thus
\begin{eqnarray}
\tilde{a}^{\mu}(p,X)=\int\frac{d \delta X_0}{p_0}\Theta(\delta X_0) e^{-\delta X_0/\tau}G^{\mu}(X,X')|_{\delta X_{x,y}=0,\delta X_z=p_z\delta X_z/p_0},
\end{eqnarray}
where we have used $1+{\rm sgn}(x)=2\Theta(x)$.

\section{Derivation of the integral}
\label{app:int_derivation}

Considering the integral
\begin{eqnarray}
I=\int^{a}_0 dx\int^{a}_0 dy\mathcal{F}(x,y)=\int^{a/2}_{-a/2} d\bar{x}\int^{a/2}_{-a/2} d\bar{y}\mathcal{F}(\bar{x}+a/2,\bar{y}+a/2),
\end{eqnarray}
we can introduce $V=\bar{y}-\bar{x}$ and $U=\bar{y}+\bar{x}$ and rewrite the integral as
\begin{eqnarray}\nonumber\label{eq:general_I}
I&=&\frac{1}{2}\Big(\int^a_0dV\int^{a-V}_0dU+\int^a_0dV\int^{0}_{V-a}dU+\int^0_{-a}dV\int^{0}_{-V-a}dU+\int^0_{-a}dV\int^{a+V}_0dU\Big)
\\
&&\mathcal{F}\big((U-V+a)/2,(U+V+a)/2\big),
\end{eqnarray}
where the overall $1/2$ factor comes from the  Jacobian determinant.  

When $\mathcal{F}(x,y)=\mathcal{F}(y-x)=\mathcal{F}(V)$, the integral in Eq.~(\ref{eq:general_I}) reduces to
\begin{eqnarray}\nonumber
I&=&\Big(\int^a_0dV(a-V)+\int^0_{-a}dV(a+V)\Big)\mathcal{F}(V)
\\
&=&\int^a_{-a}dVa\mathcal{F}(V)-\int^a_0dVV\big(\mathcal{F}(V)+\mathcal{F}(-V)\big).
\end{eqnarray}
It is found that $I=0$ when $\mathcal{F}(V)$ is an odd function with $V$.

On the other hand, when $\mathcal{F}(x,y)=\mathcal{F}(V)\cos U$, the integral in Eq.~(\ref{eq:general_I}) becomes
\begin{eqnarray}\nonumber
	I&=&\Big(\int^a_0dV\sin(a-V)+\int^0_{-a}dV\sin(a+V)\Big)\mathcal{F}(V)
	\\
	&=&\int^a_{-a}dV\sin a\cos V\mathcal{F}(V)-\int^a_0dV\sin V\cos a\big(\mathcal{F}(V)+\mathcal{F}(-V)\big).
\end{eqnarray}
For $a=2\pi$, the integral further reduces to
\begin{eqnarray}
I=-\int^{2\pi}_0dV\sin V\big(\mathcal{F}(V)+\mathcal{F}(-V)\big).
\end{eqnarray}

\section{Calculation of the longitudinal spin correlation}
\label{app:azaz_calc}

At mid-rapidity $\eta\rightarrow0$, in the small-momentum limit such that $\hat{p}^{\mu}_{\perp}\equiv p^{\mu}_{\perp}/p_0\ll 1$ for $p_0=\epsilon_{\bm p}\equiv\sqrt{\bm p^2+m^2}$ being onshell, the longitudinal component of color singlet spin four-vector is obtained in Ref.~\cite{Kumar:2022ylt}

\begin{eqnarray}
	\tilde{a}^{sz}(p,X)
	&=&-\Bigg(\frac{g^2\bar{C}_2}{2}	p_0\big(\partial_{p0}f_V(p_0)\big)\Bigg)\int^{p,X}_{k,X'}\int^{p,X'}_{k',X''}\Big[\partial_{X''0}\left(E_{[2}^a(X') E_{1]}^a (X'')\right)
	\nonumber\\
	&& -\partial_{X'1}\left(B^{a}_{[3}(X')E^{a}_{1]}(X'')\right) -\partial_{X'2}\left(B^{a}_{[3}(X')E^{a}_{2]}(X'')\right)\nonumber\\&&+(X_0''-X_0')\Big(\partial^2_{X''1}\left( E_1^a(X') E_2^a (X'')\right)
	-\partial_{X''1}\partial_{X''2}\left( E_1^a(X') E_1^a (X'')\right)\nonumber\\&&
	+\partial_{X''2}\partial_{X''1}\left( E_2^a(X') E_2^a (X'')\right)
	-\partial^2_{X''2}\left( E_2^a(X') E_1^a (X'')\right)\Big)\Big].\label{eq:azaz}
\end{eqnarray}
Equation (\ref{eq:azaz}) can be used to obtain $\langle\tilde{a}^{sz}(p,X)\tilde{a}^{sz}(p,Y)\rangle$. Since, in the end we will integrate over spatial $X$ and $Y$ on the freeze-out hyper-surface with $X_0=Y_0$, a further simplification can be made by symmetry, $\langle\tilde{a}^{sz}(p,X)\tilde{a}^{sz}(p,Y)\rangle=\langle\tilde{a}^{sz}(p,Y)\tilde{a}^{sz}(p,X)\rangle$ which suggest that the integrals involving variables $X', Y',X'',$ and $Y''$ should remain invariant under $(X'\leftrightarrow Y',X''\leftrightarrow Y'')$. It turns out that we can write 
\begin{eqnarray}
	\langle\tilde{a}^{sz}(p,X)\tilde{a}^{sz}(p,Y)\rangle
	&=&\langle\tilde{a}^{sz}(p,X)\tilde{a}^{sz}(p,Y)\rangle_{I}+\langle\tilde{a}^{sz}(p,X)\tilde{a}^{sz}(p,Y)\rangle_{II}+\langle\tilde{a}^{sz}(p,X)\tilde{a}^{sz}(p,Y)\rangle_{III},\nonumber\\
	\label{eq:azaz_correlation}
\end{eqnarray}
where
\begin{eqnarray}\nonumber
	\langle\tilde{a}^{sz}(p,X)\tilde{a}^{sz}(p,Y)\rangle_{I} 
	&=&\Bigg(\frac{g^2\bar{C}_2}{2}	p_0\big(\partial_{p0}f_V(p_0)\big)\Bigg)^2\int^{p,X}_{k,X'}\int^{p,X'}_{k',X''}\int^{p,Y}_{\bar{k},Y'}\int^{p,Y'}_{\bar{k}',Y''}\nonumber\\
	&&\Big[\partial_{X''0}\partial_{Y''0}\left\langle E_{[2}^a(X') E_{1]}^a (X'')E_{[2}^b(Y') E_{1]}^b (Y'')\right\rangle 
	\nonumber\\
	&& -2\partial_{X''0}\partial_{Y'1}\left\langle E_{[2}^a(X') E_{1]}^a (X'')B_{[3}^{b}(Y') E_{1]}^{b} (Y'')\right\rangle
	\nonumber\\
	&&-2\partial_{X''0}\partial_{Y'2}\left\langle  E_{[2}^a(X') E_{1]}^a (X'')B_{[3}^{b}(Y') E_{2]}^{b} (Y'')\right\rangle
	\nonumber\\
	&&
	+\partial_{X'1}\partial_{Y'1}\left\langle B_{[3}^a(X')E_{1]}^a(X'')B_{[3}^b(Y')E_{1]}^b(Y'')\right\rangle
	\nonumber\\&&
	+2\partial_{X'1}\partial_{Y'2}\left\langle B_{[3}^a(X')E_{1]}^a(X'')B_{[3}^b(Y')E_{2]}^b(Y'')\right\rangle
	\nonumber\\
	&&
	+\partial_{X'2}\partial_{Y'2}\left\langle B_{[3}^a(X')E_{2]}^a(X'')B_{[3}^b(Y')E_{2]}^b(Y'')\right\rangle
	\Big],\label{eq:azazI}
\end{eqnarray}
\begin{eqnarray}\nonumber
	\langle\tilde{a}^{sz}(p,X)\tilde{a}^{sz}(p,Y)\rangle_{II} 
	&=&-\Bigg(\frac{g^2\bar{C}_2}{2}	p_0\big(\partial_{p0}f_V(p_0)\big)\Bigg)^2\int^{p,X}_{k,X'}\int^{p,X'}_{k',X''}\int^{p,Y}_{\bar{k},Y'}\int^{p,Y'}_{\bar{k}',Y''}\nonumber\\
	&&\Big[-(Y_0''-Y_0')\partial_{X''0}\partial_{Y''1}\partial_{Y''1}\left\langle E_{[2}^a(X') E_{1]}^a (X'')E_1^b(Y') E_2^b (Y'')\right\rangle
	\nonumber\\&&
	+(Y_0''-Y_0')\partial_{X''0}\partial_{Y''1}\partial_{Y''2}\left\langle E_{[2}^a(X') E_{1]}^a (X'')E_1^b(Y') E_1^b (Y'')\right\rangle
	\nonumber\\&&
	-(Y_0''-Y_0')\partial_{X''0}\partial_{Y''2}\partial_{Y''1}\left\langle E_{[2}^a(X') E_{1]}^a (X'')E_2^b(Y') E_2^b (Y'')\right\rangle
	\nonumber\\&&
	+(Y_0''-Y_0')\partial_{X''0}\partial_{Y''2}\partial_{Y''2}\left\langle E_{[2}^a(X') E_{1]}^a (X'')E_2^b(Y') E_1^b (Y'')\right\rangle
	\nonumber\\&&
	+2(Y_0''-Y_0')\partial_{X'1}\partial_{Y''1}\partial_{Y''1}\left\langle B_{[3}^a(X')E_{1]}^a(X'')E_1^b(Y')E_2^b(Y'')\right\rangle
	\nonumber\\&&
	-2(Y_0''-Y_0')\partial_{X'1}\partial_{Y''1}\partial_{Y''2}\left\langle B_{[3}^a(X')E_{1]}^a(X'')E_1^b(Y')E_1^b(Y'')\right\rangle
	\nonumber\\&&
	+2(Y_0''-Y_0')\partial_{X'1}\partial_{Y''2}\partial_{Y''1}\left\langle B_{[3}^a(X')E_{1]}^a(X'')E_2^b(Y')E_2^b(Y'')\right\rangle
	\nonumber\\&&
	-2 (Y_0''-Y_0')\partial_{X'1}\partial_{Y''2}\partial_{Y''2}\left\langle B_{[3}^a(X')E_{1]}^a(X'')E_2^b(Y')E_1^b(Y'')\right\rangle
	\nonumber\\&&
	+2 (Y_0''-Y_0')\partial_{X'2}\partial_{Y''1}\partial_{Y''1}\left\langle B_{[3}^a(X')E_{2]}^a(X'')E_1^b(Y')E_2^b(Y'')\right\rangle
	\nonumber\\&&
	-2 (Y_0''-Y_0')\partial_{X'2}\partial_{Y''1}\partial_{Y''2}\left\langle B_{[3}^a(X')E_{2]}^a(X'')E_1^b(Y')E_1^b(Y'')\right\rangle
	\nonumber\\&&
	+2 (Y_0''-Y_0')\partial_{X'2}\partial_{Y''2}\partial_{Y''1}\left\langle B_{[3}^a(X')E_{2]}^a(X'')E_2^b(Y')E_2^b(Y'')\right\rangle
	\nonumber\\&&
	-2 (Y_0''-Y_0')\partial_{X'2}\partial_{Y''2}\partial_{Y''2}\left\langle B_{[3}^a(X')E_{2]}^a(X'')E_2^b(Y')E_1^b(Y'')\right\rangle
	\Big],\label{eq:azazII}
\end{eqnarray}
and 
\begin{eqnarray}
	\langle\tilde{a}^{sz}(p,X)\tilde{a}^{sz}(p,Y)\rangle_{III}
	&=&\Bigg(\frac{g^2\bar{C}_2}{2}	p_0\big(\partial_{p0}f_V(p_0)\big)\Bigg)^2\int^{p,X}_{k,X'}\int^{p,X'}_{k',X''}\int^{p,Y}_{\bar{k},Y'}\int^{p,Y'}_{\bar{k}',Y''}
	\nonumber\\&&
	\Big[+(X_0''-X_0')(Y_0''-Y_0')\partial_{X''1}\partial_{X''1}\partial_{Y''1}\partial_{Y''1}\left\langle E^a_{1}(X') E^a_{2} (X'')E_1^b(Y')E_2^b(Y'')\right\rangle
	\nonumber\\&&
	+(X_0''-X_0')(Y_0''-Y_0')\partial_{X''1}\partial_{X''2}\partial_{Y''1}\partial_{Y''2}\left\langle E^a_{1}(X') E^a_{1} (X'')E_1^b(Y')E_1^b(Y'')\right\rangle
	\nonumber\\&&
	+(X_0''-X_0')(Y_0''-Y_0')\partial_{X''2}\partial_{X''1}\partial_{Y''2}\partial_{Y''1}\left\langle E^a_{2}(X') E^a_{2} (X'')E_2^b(Y')E_2^b(Y'')\right\rangle
	\nonumber\\&&
	+(X_0''-X_0')(Y_0''-Y_0')\partial_{X''2}\partial_{X''2}\partial_{Y''2}\partial_{Y''2}\left\langle E^a_{2}(X') E^a_{1} (X'')E_2^b(Y')E_1^b(Y'')\right\rangle
	\nonumber\\&&
	-2 (X_0''-X_0')(Y_0''-Y_0')\partial_{X''1}\partial_{X''1}\partial_{Y''1}\partial_{Y''2}\left\langle E^a_{1}(X') E^a_{2} (X'')E_1^b(Y')E_1^b(Y'')\right\rangle
	\nonumber\\&&
	+2 (X_0''-X_0')(Y_0''-Y_0')\partial_{X''1}\partial_{X''1}\partial_{Y''2}\partial_{Y''1}\left\langle E^a_{1}(X') E^a_{2} (X'')E_2^b(Y')E_2^b(Y'')\right\rangle
	\nonumber\\&&
	-2 (X_0''-X_0')(Y_0''-Y_0')\partial_{X''1}\partial_{X''1}\partial_{Y''2}\partial_{Y''2}\left\langle E^a_{1}(X') E^a_{2} (X'')E_2^b(Y')E_1^b(Y'')\right\rangle
	\nonumber\\&&
	-2 (X_0''-X_0')(Y_0''-Y_0')\partial_{X''1}\partial_{X''2}\partial_{Y''2}\partial_{Y''1}\left\langle E^a_{1}(X') E^a_{1} (X'')E_2^b(Y')E_2^b(Y'')\right\rangle
	\nonumber\\&&
	+2 (X_0''-X_0')(Y_0''-Y_0')\partial_{X''1}\partial_{X''2}\partial_{Y''2}\partial_{Y''2}\left\langle E^a_{1}(X') E^a_{1} (X'')E_2^b(Y')E_1^b(Y'')\right\rangle
	\nonumber\\&&
	-2(X_0''-X_0')(Y_0''-Y_0')\partial_{X''2}\partial_{X''1}\partial_{Y''2}\partial_{Y''2}\left\langle E^a_{2}(X') E^a_{2} (X'')E_2^b(Y')E_1^b(Y'')\right\rangle
	\Big].\nonumber\\\label{eq:azazIII}
\end{eqnarray}
In the above Eqs.~(\ref{eq:azazI}), (\ref{eq:azazII}) and (\ref{eq:azazIII}) the four-field correlators can be written in terms of two-field correlators as follows
\begin{eqnarray}
	\left\langle \alpha_1^a(X') \alpha_2^a (X'')\alpha_3^b(Y') \alpha_4^b (Y'')\right\rangle&=&\left\langle \alpha_1^a(X') \alpha_2^a (X'')\right\rangle\left\langle \alpha_3^b(Y') \alpha_4^b (Y'')\right\rangle+\left\langle \alpha_1^a(X') \alpha_3^b(Y')\right\rangle \left\langle\alpha_2^a (X'')\alpha_4^b (Y'')\right\rangle\nonumber\\&&+\left\langle \alpha_1^a(X') \alpha_4^b (Y'')\right\rangle \left\langle\alpha_2^a (X'')\alpha_3^b(Y')\right\rangle. \label{eq:wicks_decompo}
\end{eqnarray}
Now keeping in mind $\langle\tilde{a}^{sz}(p,X)\rangle$ vanishes (see Ref.~\cite{Kumar:2022ylt}), the terms with color structure $\langle \alpha_1^a(X') \alpha_2^a (X'')\rangle\langle \alpha_3^b(Y') \alpha_4^b(Y'')\rangle$ will not contribute in Eqs.~(\ref{eq:azazI}), (\ref{eq:azazII}), and (\ref{eq:azazIII}). Moreover, all the terms associated with $\partial_{X0''} E_{[2}^a(X') E_{1]}^a (X'')$ in  Eqs.~(\ref{eq:azazI}) and (\ref{eq:azazII}) will vanish since such terms involve the integral
\begin{eqnarray}\nonumber
	\int^{X'_0}_{X''_0}\partial_{X''0}J_1(qX'_0)J_1(lX''_0)\Theta(X'_0)\Theta(X''_0)-\int^{X'_0}_{X''_0}\partial_{X''0}J_1(qX''_0)J_1(lX'_0)\Theta(X'_0)\Theta(X''_0)
	=0.
\end{eqnarray}
In the end, summing over all the non-vanishing contribution from Eqs. (\ref{eq:azazI}), (\ref{eq:azazII}), and (\ref{eq:azazIII}) one can obtain
\begin{eqnarray}
	\langle\tilde{a}^{sz}(p,X)\tilde{a}^{sz}(p,Y)\rangle 
	&\approx&\Bigg(\frac{g^2\bar{C}_2}{2}	p_0\big(\partial_{p0}f_V(p_0)\big)\Bigg)^2\Big[\mathcal{J}_{1}+\mathcal{J}_{2}+\mathcal{J}_{3}\Big],\label{eq:azaz2}
\end{eqnarray}
where the terms $\mathcal{J}_{1}=\mathcal{J}_{A}+\mathcal{J}_{B}$, $\mathcal{J}_{2}=\mathcal{J}_{C}+\mathcal{J}_{D}$ and $\mathcal{J}_{3}=\mathcal{J}_{E}+\mathcal{J}_{F}$ are the nonvanishing contribution from Eqs. (\ref{eq:azazI}),  (\ref{eq:azazII}), and  (\ref{eq:azazIII}) respectively. The expressions for $\mathcal{J}_{A}$, $\mathcal{J}_{B}$, $\mathcal{J}_{C}$, $\mathcal{J}_{D}$, $\mathcal{J}_{E}$, $\mathcal{J}_{F}$ in terms with two field correlators are as follows:
\begin{eqnarray}\nonumber
	\mathcal{J}_{A}&=&\int^{p,X}_{k,X'}\int^{p,X'}_{k',X''}\int^{p,Y}_{\bar{k},Y'}\int^{p,Y'}_{\bar{k}',Y''}\nonumber\\
	&&\Big[\partial_{X'1}\partial_{Y'1}\left(\left\langle B^{a3}(X')B^{b3}(Y')\right\rangle\left\langle E^{a1}(X'')E^{b1}(Y'')\right\rangle+\left\langle B^{a1}(X')B^{b1}(Y')\right\rangle\left\langle E^{a3}(X'')E^{b3}(Y'')\right\rangle \right)
	\nonumber\\&&
	+2\partial_{X'1}\partial_{Y'2}\left(\left\langle B^{a3}(X')B^{b3}(Y')\right\rangle\left\langle E^{a1}(X'')E^{b2}(Y'')\right\rangle
	+\left\langle B^{a1}(X')B^{b2}(Y')\right\rangle\left\langle E^{a3}(X'')E^{b3}(Y'')\right\rangle\right)
	\nonumber\\&&
	+\partial_{X'2}\partial_{Y'2}\left(\left\langle B^{a3}(X')B^{b3}(Y')\right\rangle\left\langle E^{a2}(X'')E^{b2}(Y'')\right\rangle
	+\left\langle B^{a2}(X')B^{b2}(Y')\right\rangle\left\langle E^{a3}(X'')E^{b3}(Y'')\right\rangle\right)\Big],\nonumber\\\label{eq:JA}
\end{eqnarray}
\begin{eqnarray}\nonumber
	\mathcal{J}_{B}&=&\int^{p,X}_{k,X'}\int^{p,X'}_{k',X''}\int^{p,Y}_{\bar{k},Y'}\int^{p,Y'}_{\bar{k}',Y''}
	\nonumber\\&&
	\Big[\partial_{X'1}\partial_{Y'1}\left(\left\langle B^{a3}(X')E^{b1}(Y'')\right\rangle\left\langle E^{a1}(X'')B^{b3}(Y')\right\rangle
	+\left\langle B^{a1}(X')E^{b3}(Y'')\right\rangle\left\langle E^{a3}(X'')B^{b1}(Y')\right\rangle\right)
	\nonumber\\&&
	+2\partial_{X'1}\partial_{Y'2}\left(\left\langle B^{a3}(X')E^{b2}(Y'')\right\rangle\left\langle E^{a1}(X'')B^{b3}(Y')\right\rangle
	+\left\langle B^{a1}(X')E^{b3}(Y'')\right\rangle\left\langle E^{a3}(X'')B^{b2}(Y')\right\rangle\right)
	\nonumber\\&&
	+\partial_{X'2}\partial_{Y'2}\left(\left\langle B^{a3}(X')E^{b2}(Y'')\right\rangle\left\langle E^{a2}(X'')B^{b3}(Y')\right\rangle
	+\left\langle B^{a2}(X')E^{b3}(Y'')\right\rangle\left\langle E^{a3}(X'')B^{b2}(Y')\right\rangle\right)\Big],\nonumber\\\label{eq:JB}
\end{eqnarray}
\begin{eqnarray}\nonumber
	\mathcal{J}_{C} &=&-\int^{p,X}_{k,X'}\int^{p,X'}_{k',X''}\int^{p,Y}_{\bar{k},Y'}\int^{p,Y'}_{\bar{k}',Y''}\nonumber\\
	&&\Big[
	+2(Y_0''-Y_0')\partial_{X'1}\partial_{Y''1}\partial_{Y''1}\left\langle B_{3}^a(X')E_1^b(Y')\right\rangle\left\langle E_{1}^a(X'')E_2^b(Y'')\right\rangle
	\nonumber\\&&
	-2(Y_0''-Y_0')\partial_{X'1}\partial_{Y''1}\partial_{Y''2}\left\langle B_{3}^a(X')E_{1}^a(Y')\right\rangle\left\langle E_1^b(X'')E_1^b(Y'')\right\rangle
	\nonumber\\&&
	+2(Y_0''-Y_0')\partial_{X'1}\partial_{Y''2}\partial_{Y''1}\left\langle B_{3}^a(X')E_2^b(Y')\right\rangle\left\langle E_{1}^a(X'') E_2^b(Y'')\right\rangle
	\nonumber\\&&
	-2 (Y_0''-Y_0')\partial_{X'1}\partial_{Y''2}\partial_{Y''2}\left\langle B_{3}^a(X')E_2^b(Y')\right\rangle\left\langle E_{1}^a(X'')E_1^b(Y'')\right\rangle
	\nonumber\\&&
	+2 (Y_0''-Y_0')\partial_{X'2}\partial_{Y''1}\partial_{Y''1}\left\langle B_{3}^a(X')E_1^b(Y')\right\rangle\left\langle E_{2}^a(X'')E_2^b(Y'')\right\rangle
	\nonumber\\&&
	-2 (Y_0''-Y_0')\partial_{X'2}\partial_{Y''1}\partial_{Y''2}\left\langle B_{3}^a(X')E_1^b(Y')\right\rangle\left\langle E_{2}^a(X'')E_1^b(Y'')\right\rangle
	\nonumber\\&&
	+2 (Y_0''-Y_0')\partial_{X'2}\partial_{Y''2}\partial_{Y''1}\left\langle B_{3}^a(X')E_2^b(Y')\right\rangle\left\langle E_{2}^a(X'')E_2^b(Y'')\right\rangle
	\nonumber\\&&
	-2 (Y_0''-Y_0')\partial_{X'2}\partial_{Y''2}\partial_{Y''2}\left\langle B_{3}^a(X')E_2^b(Y')\right\rangle\left\langle E_{2}^a(X'')E_1^b(Y'')\right\rangle
	\Big],\label{eq:JC}
\end{eqnarray}
\begin{eqnarray}\nonumber
	\mathcal{J}_{D} &=&-\int^{p,X}_{k,X'}\int^{p,X'}_{k',X''}\int^{p,Y}_{\bar{k},Y'}\int^{p,Y'}_{\bar{k}',Y''}\nonumber\\
	&&\Big[
	+2(Y_0''-Y_0')\partial_{X'1}\partial_{Y''1}\partial_{Y''1}\left\langle B_{3}^a(X')E_2^b(Y'')\right\rangle\left\langle E_{1}^a(X'')E_1^b(Y')\right\rangle
	\nonumber\\&&
	-2(Y_0''-Y_0')\partial_{X'1}\partial_{Y''1}\partial_{Y''2}\left\langle B_{3}^a(X')E_1^b(Y'')
	\right\rangle\left\langle E_{1}^a(X'')E_1^b(Y')\right\rangle
	\nonumber\\&&
	+2(Y_0''-Y_0')\partial_{X'1}\partial_{Y''2}\partial_{Y''1}\left\langle B_{3}^a(X')E_2^b(Y'')\right\rangle\left\langle E_{1}^a(X'')E_2^b(Y')\right\rangle
	\nonumber\\&&
	-2 (Y_0''-Y_0')\partial_{X'1}\partial_{Y''2}\partial_{Y''2}\left\langle B_{3}^a(X')E_1^b(Y'')\right\rangle\left\langle E_{1}^a(X'') E_2^b(Y')\right\rangle
	\nonumber\\&&
	+2 (Y_0''-Y_0')\partial_{X'2}\partial_{Y''1}\partial_{Y''1}\left\langle B_{3}^a(X')E_2^b(Y'')\right\rangle\left\langle E_{2}^a(X'') E_1^b(Y')\right\rangle
	\nonumber\\&&
	-2 (Y_0''-Y_0')\partial_{X'2}\partial_{Y''1}\partial_{Y''2}\left\langle B_{3}^a(X')E_1^b(Y'')\right\rangle\left\langle E_{2}^a(X'') E_1^b(Y')\right\rangle
	\nonumber\\&&
	+2 (Y_0''-Y_0')\partial_{X'2}\partial_{Y''2}\partial_{Y''1}\left\langle B_{3}^a(X')E_2^b(Y'')\right\rangle\left\langle E_{2}^a(X'') E_2^b(Y')\right\rangle
	\nonumber\\&&
	-2 (Y_0''-Y_0')\partial_{X'2}\partial_{Y''2}\partial_{Y''2}\left\langle B_{3}^a(X')E_1^b(Y'')\right\rangle\left\langle E_{2}^a(X'') E_2^b(Y')\right\rangle
	\Big],\label{eq:JD}
\end{eqnarray}
\begin{eqnarray}\nonumber
	\mathcal{J}_E&=& \int^{p,X}_{k,X'}\int^{p,X'}_{k',X''}\int^{p,Y}_{\bar{k},Y'}\int^{p,Y'}_{\bar{k}',Y''}(X_0''-X_0')(Y_0''-Y_0')\nonumber\\
	&&\times\Big[\partial_{X''1}\partial_{X''1}\partial_{Y''1}\partial_{Y''1}\Big(\left\langle E_1^a(X') E_1^b(Y')\right\rangle \left\langle E_2^a(X'') E_2^b(Y'')\right\rangle\Big)\nonumber\\
	&&+\partial_{X''1}\partial_{X''2}\partial_{Y''1}\partial_{Y''2}\Big(\left\langle E_1^a(X') E_1^b(Y')\right\rangle \left\langle E_1^a(X'') E_1^b(Y'')\right\rangle\Big)\nonumber\\
	&&+\partial_{X''1}\partial_{X''2}\partial_{Y''1}\partial_{Y''2}\Big(\left\langle E_2^a(X') E_2^b(Y')\right\rangle \left\langle E_2^a(X'') E_2^b(Y'')\right\rangle\Big)\nonumber\\
	&&+\partial_{X''2}\partial_{X''2}\partial_{Y''2}\partial_{Y''2}\Big(\left\langle E_2^a(X') E_2^b(Y')\right\rangle \left\langle E_1^a(X'') E_1^b(Y'')\right\rangle\Big)\nonumber\\
	&&-2\partial_{X''1}\partial_{X''1}\partial_{Y''1}\partial_{Y''2}\Big(\left\langle E_1^a(X') E_1^b(Y')\right\rangle \left\langle E_2^a(X'') E_1^b(Y'')\right\rangle\Big)\nonumber\\
	&&+2\partial_{X''1}\partial_{X''1}\partial_{Y''1}\partial_{Y''2}\Big(\left\langle E_1^a(X') E_2^b(Y')\right\rangle \left\langle E_2^a(X'') E_2^b(Y'')\right\rangle\Big)\nonumber\\
	&&-2\partial_{X''1}\partial_{X''1}\partial_{Y''2}\partial_{Y''2}\Big(\left\langle E_1^a(X') E_2^b(Y')\right\rangle \left\langle E_2^a(X'') E_1^b(Y'')\right\rangle\Big)\nonumber\\
	&&-2\partial_{X''1}\partial_{X''2}\partial_{Y''2}\partial_{Y''1}\Big(\left\langle E_1^a(X') E_2^b(Y')\right\rangle\left\langle E_1^a(X'') E_2^b(Y'')\right\rangle\Big)\nonumber\\
	&&-2\partial_{X''1}\partial_{X''2}\partial_{Y''2}\partial_{Y''2}\Big(\left\langle E_2^a(X') E_2^b(Y')\right\rangle \left\langle E_2^a(X'') E_1^b(Y'')\right\rangle\Big)\nonumber\\
	&&+2\partial_{X''1}\partial_{X''2}\partial_{Y''2}\partial_{Y''2}\Big(\left\langle E_1^a(X') E_2^b(Y')\right\rangle \left\langle E_1^a(X'') E_1^b(Y'')\right\rangle\Big)
	\Big],\label{eq:JE}
\end{eqnarray}
\begin{eqnarray}\nonumber
	\mathcal{J}_F&=& \int^{p,X}_{k,X'}\int^{p,X'}_{k',X''}\int^{p,Y}_{\bar{k},Y'}\int^{p,Y'}_{\bar{k}',Y''}(X_0''-X_0')(Y_0''-Y_0')\nonumber\\
	&&\times\Big[\partial_{X''1}\partial_{X''1}\partial_{Y''1}\partial_{Y''1}\Big(\left\langle E_1^a(X')E_2^b (Y'')\right\rangle  \left\langle E_2^a(X'')E_1^b(Y')\right\rangle\Big)\nonumber\\
	&&+\partial_{X''1}\partial_{X''2}\partial_{Y''1}\partial_{Y''2}\Big(\left\langle E_1^a(X')E_1^b (Y'')\right\rangle  \left\langle E_1^a(X'')E_1^b(Y')\right\rangle\Big)\nonumber\\
	&&+\partial_{X''1}\partial_{X''2}\partial_{Y''1}\partial_{Y''2}\Big(\left\langle E_2^a(X')E_2^b (Y'')\right\rangle  \left\langle E_2^a(X'')E_2^b(Y')\right\rangle\Big)\nonumber\\
	&&+\partial_{X''2}\partial_{X''2}\partial_{Y''2}\partial_{Y''2}\Big(\left\langle E_2^a(X')E_1^b (Y'')\right\rangle  \left\langle E_1^a(X'')E_2^b(Y')\right\rangle\Big)\nonumber\\
	&&-2\partial_{X''1}\partial_{X''1}\partial_{Y''1}\partial_{Y''2}\Big(\left\langle E_1^a(X')E_1^b (Y'')\right\rangle  \left\langle E_2^a(X'')E_1^b(Y')\right\rangle\Big)\nonumber\\
	&&+2\partial_{X''1}\partial_{X''1}\partial_{Y''1}\partial_{Y''2}\Big(\left\langle E_1^a(X')E_2^b (Y'')\right\rangle \left\langle E_2^a(X'')E_2^b(Y')\right\rangle\Big)\nonumber\\
	&&-2\partial_{X''1}\partial_{X''1}\partial_{Y''2}\partial_{Y''2}\Big(\left\langle E_1^a(X')E_1^b (Y'')\right\rangle  \left\langle E_2^a(X'')E_2^b(Y')\right\rangle\Big)\nonumber\\
	&&-2\partial_{X''1}\partial_{X''2}\partial_{Y''2}\partial_{Y''1}\Big(\left\langle E_1^a(X')E_2^b (Y'')\right\rangle \left\langle E_1^a(X'')E_2^b(Y')\right\rangle\Big)\nonumber\\
	&&-2\partial_{X''1}\partial_{X''2}\partial_{Y''2}\partial_{Y''2}\Big(\left\langle E_2^a(X')E_1^b (Y'')\right\rangle  \left\langle E_2^a(X'')E_2^b(Y')\right\rangle\Big)\nonumber\\
	&&+2\partial_{X''1}\partial_{X''2}\partial_{Y''2}\partial_{Y''2}\Big(\left\langle E_1^a(X')E_1^b (Y'')\right\rangle  \left\langle E_1^a(X'')E_2^b(Y')\right\rangle\Big)\Big].\label{eq:JF}
\end{eqnarray} 
Now, first using Eqs.~(\ref{eq:bibjcorr_sim}), (\ref{eq:eiejcorr_sim}), (\ref{eq:b3b3corr}), (\ref{eq:e3e3corr_sim}) in Eqs.(\ref{eq:JA}), (\ref{eq:JB}), (\ref{eq:JC}), (\ref{eq:JD}), (\ref{eq:JE}) and  (\ref{eq:JF}) then using the following relations 
\begin{eqnarray}
	\int^{p,X}_{k,X'}G(X,X')\approx\frac{1}{2p_0}\int^{X_0}_{X'_0}G(X,X')|_{X'_{1,2,3}=X_{1,2,3}}, \label{eq:useful_int_2}
\end{eqnarray}
\begin{eqnarray}
	\int^{X_0}_{X'_0}\equiv \int^{\infty}_{-\infty} d X'_0\Big(1+{\rm sgn}(X_0-X_0')\Big),
\end{eqnarray}
\begin{eqnarray}
\partial_{X'_{\perp}j}\int^{X'}_{\perp;q,u}G(u_{\perp})=\int^{X'}_{\perp;q,u}\partial_{u_{\perp}j}G(u_{\perp}),\label{eq:Tr1}
\end{eqnarray}
we can obtain
\begin{eqnarray}
	\mathcal{J}_A(p,X,Y)&=&-\frac{g^4N_c^2\delta^{ab}\delta^{ab}}{64p_0^4}
	\int^{X}_{\perp;q,u}\int^{Y}_{\perp;l,v}\int^{X}_{\perp;q',u'}\int^{Y}_{\perp;l',v'}
	\nonumber\\&&
	\bigg[\bigg(\frac{q'^y l'^y}{q' l'}\partial_{ux}\partial_{vx}\Omega_{-}(u_{\perp},v_{\perp})\Omega_{-}(u'_{\perp},v'_{\perp})\mathcal{\rho}_{Ia}(X_0,Y_0,q,l,q',l')\nonumber\\&&\quad \,+\frac{q^y l^y}{q l}\partial_{ux}\partial_{vx}\Omega_{+}(u_{\perp},v_{\perp})\Omega_{+}(u'_{\perp},v'_{\perp})\mathcal{\rho}_{Ib}(X_0,Y_0,q,l,q',l')\bigg)
	\nonumber\\ &&
	-2\bigg(\frac{q'^y l'^x}{q' l'}\partial_{ux}\partial_{vy}\Omega_{-}(u_{\perp},v_{\perp})\Omega_{-}(u'_{\perp},v'_{\perp})\mathcal{\rho}_{Ia}(X_0,Y_0,q,l,q',l')\nonumber\\&&\quad \,+\frac{q^y l^x}{q l}\partial_{ux}\partial_{vy}\Omega_{+}(u_{\perp},v_{\perp})\Omega_{+}(u'_{\perp},v'_{\perp})\mathcal{\rho}_{Ib}(X_0,Y_0,q,l,q',l')\bigg)
	\nonumber\\ &&
	+\bigg(\frac{q'^x l'^x}{q' l'}\partial_{uy}\partial_{vy}\Omega_{-}(u_{\perp},v_{\perp})\Omega_{-}(u'_{\perp},v'_{\perp})\mathcal{\rho}_{Ia}(X_0,Y_0,q,l,q',l')\nonumber\\&& \quad \,+\frac{q^x l^x}{q l}\partial_{uy}\partial_{vy}\Omega_{+}(u_{\perp},v_{\perp})\Omega_{+}(u'_{\perp},v'_{\perp})\mathcal{\rho}_{Ib}(X_0,Y_0,q,l,q',l')\bigg)
	\bigg],
	\label{eq:JA1}
\end{eqnarray}
where
\begin{eqnarray}\nonumber
	&&\mathcal{\rho}_{Ia}(X_0,Y_0,q,l,q',l')
	\\\nonumber	&&\equiv\int^{X_0}_{X'_0}\int^{Y_0}_{Y'_0}\int^{X'_0}_{X''_0}\int^{Y'_0}_{Y''_0}
	J_0(qX'_0)J_0(lY'_0)\Theta(X'_0)\Theta(Y'_0)
	J_1(q'X''_0)J_1(l'Y''_0)\Theta(X''_0)\Theta(Y''_0)
	\\\nonumber
	&&=16\int^{\infty}_{-\infty} dX'_0\int^{\infty}_{-\infty}dY'_0\int^{\infty}_{-\infty}dX''_0\int^{\infty}_{-\infty}dY''_0 J_0(qX'_0)J_0(lY'_0)
	J_1(q'X''_0)J_1(l'Y''_0)
	\\
	&&\quad\times \Theta(X_0-X'_0)\Theta(Y_0-Y'_0)\Theta(X'_0-X''_0)\Theta(Y'_0-Y''_0)\Theta(X'_0)\Theta(Y'_0)\Theta(X''_0)\Theta(Y''_0),
\end{eqnarray}
\begin{eqnarray}\nonumber
	&&\mathcal{\rho}_{Ib}(X_0,Y_0,q,l,q',l')
	\\\nonumber	&&\equiv\int^{X_0}_{X'_0}\int^{Y_0}_{Y'_0}\int^{X'_0}_{X''_0}\int^{Y'_0}_{Y''_0}
	J_1(qX'_0)J_1(lY'_0)\Theta(X'_0)\Theta(Y'_0)
	J_0(q'X''_0)J_0(l'Y''_0)\Theta(X''_0)\Theta(Y''_0)
	\\\nonumber
	&&=16\int^{\infty}_{-\infty} dX'_0\int^{\infty}_{-\infty}dY'_0\int^{\infty}_{-\infty}dX''_0\int^{\infty}_{-\infty}dY''_0 J_1(qX'_0)J_1(lY'_0)
	J_0(q'X''_0)J_0(l'Y''_0)
	\\
	&&\quad\times \Theta(X_0-X'_0)\Theta(Y_0-Y'_0)\Theta(X'_0-X''_0)\Theta(Y'_0-Y''_0)\Theta(X'_0)\Theta(Y'_0)\Theta(X''_0)\Theta(Y''_0),
\end{eqnarray}
and
\begin{eqnarray}
	\mathcal{J}_{B}(p,X,Y)&=&-\frac{g^4N_c^2\delta^{ab}\delta^{ab}}{64p_0^4}
	\int^{X}_{\perp;q,u}\int^{Y}_{\perp;l,v}\int^{X}_{\perp;q',u'}\int^{Y}_{\perp;l',v'}
	\nonumber\\&&
	\bigg[\bigg(\frac{l^y q'^y}{l q'}\partial_{ux}\Omega_{-}(u_{\perp},v_{\perp})\partial_{vx'}\Omega_{-}(u'_{\perp},v'_{\perp})\mathcal{\rho}_{IIa}(X_0,Y_0,q,l,q',l')\nonumber\\&& \quad \,+\frac{q^y l'^y}{q l'}\partial_{ux}\Omega_{+}(u_{\perp},v_{\perp})\partial_{vx'}\Omega_{+}(u'_{\perp},v'_{\perp})\mathcal{\rho}_{IIb}(X_0,Y_0,q,l,q',l')\bigg)
	\nonumber\\ &&
	-2\bigg(\frac{l^x q'^y}{l q'}\partial_{ux}\Omega_{-}(u_{\perp},v_{\perp})\partial_{vy'}\Omega_{-}(u'_{\perp},v'_{\perp})\mathcal{\rho}_{IIa}(X_0,Y_0,q,l,q',l')\nonumber\\&& \quad \,+\frac{q^y l'^x}{q l'}\partial_{ux}\Omega_{+}(u_{\perp},v_{\perp})\partial_{vy'}\Omega_{+}(u'_{\perp},v'_{\perp})\mathcal{\rho}_{IIb}(X_0,Y_0,q,l,q',l')\bigg)
	\nonumber\\ &&
	+\bigg(\frac{l^x q'^x}{l q'}\partial_{uy}\Omega_{-}(u_{\perp},v_{\perp})\partial_{vy'}\Omega_{-}(u'_{\perp},v'_{\perp})\mathcal{\rho}_{IIa}(X_0,Y_0,q,l,q',l')\nonumber\\&& \quad \,+\frac{q^x l'^x}{q l'}\partial_{uy}\Omega_{+}(u_{\perp},v_{\perp})\partial_{vy'}\Omega_{+}(u'_{\perp},v'_{\perp})\mathcal{\rho}_{IIb}(X_0,Y_0,q,l,q',l')\bigg)
	\bigg],\nonumber\\
	\label{eq:J_Ib.1}
\end{eqnarray}
where
\begin{eqnarray}\nonumber
	&&\mathcal{\rho}_{IIa}(X_0,Y_0,q,l,q',l')
	\\\nonumber	&&\equiv\int^{X_0}_{X'_0}\int^{Y_0}_{Y'_0}\int^{X'_0}_{X''_0}\int^{Y'_0}_{Y''_0}
	J_0(qX'_0)J_1(lY''_0)\Theta(X'_0)\Theta(Y''_0)
	J_1(q'X''_0)J_0(l'Y'_0)\Theta(X''_0)\Theta(Y'_0)
	\\\nonumber
	&&=16\int^{\infty}_{-\infty} dX'_0\int^{\infty}_{-\infty}dY'_0\int^{\infty}_{-\infty}dX''_0\int^{\infty}_{-\infty}dY''_0 J_0(qX'_0)J_1(lY''_0)
	J_1(q'X''_0)J_0(l'Y'_0)
	\\
	&&\quad\times \Theta(X_0-X'_0)\Theta(Y_0-Y'_0)\Theta(X'_0-X''_0)\Theta(Y'_0-Y''_0)\Theta(X'_0)\Theta(Y'_0)\Theta(X''_0)\Theta(Y''_0),
\end{eqnarray}
\begin{eqnarray}\nonumber
	&&\mathcal{\rho}_{IIb}(X_0,Y_0,q,l,q',l')
	\\\nonumber	&&\equiv\int^{X_0}_{X'_0}\int^{Y_0}_{Y'_0}\int^{X'_0}_{X''_0}\int^{Y'_0}_{Y''_0}
	J_1(qX'_0)J_0(lY''_0)\Theta(X'_0)\Theta(Y''_0)
	J_0(q'X''_0)J_1(l'Y'_0)\Theta(X''_0)\Theta(Y'_0)
	\\\nonumber
	&&=16\int^{\infty}_{-\infty} dX'_0\int^{\infty}_{-\infty}dY'_0\int^{\infty}_{-\infty}dX''_0\int^{\infty}_{-\infty}dY''_0 J_1(qX'_0)J_0(lY''_0)
	J_0(q'X''_0)J_1(l'Y'_0)
	\\
	&&\quad\times \Theta(X_0-X'_0)\Theta(Y_0-Y'_0)\Theta(X'_0-X''_0)\Theta(Y'_0-Y''_0)\Theta(X'_0)\Theta(Y'_0)\Theta(X''_0)\Theta(Y''_0),
\end{eqnarray}
and
\begin{eqnarray}\nonumber
	\mathcal{J}_C(p,X,Y)&\approx& +2i\frac{g^4N_c^2\delta^{ab}\delta^{ab}}{64p_0^4}
	\int^{X}_{\perp;q,u}\int^{Y}_{\perp;l,v}\int^{X}_{\perp;q',u'}\int^{Y}_{\perp;l',v'}\nonumber\\
	&&
	\bigg[\frac{l^y}{l}\frac{q'^y l'^x}{q' l'}\partial_{ux}\Omega_{-}(u_{\perp},v_{\perp})\partial_{vx'}\partial_{vx'}\Omega_{-}(u'_{\perp},v'_{\perp})\nonumber\\
	&&\quad\,+\frac{l^y}{l}\frac{q'^y l'^y}{q' l'}\partial_{ux}\Omega_{-}(u_{\perp},v_{\perp})\partial_{vx'}\partial_{vy'}\Omega_{-}(u'_{\perp},v'_{\perp})\nonumber\\
	&&\quad\,-\frac{l^x}{l}\frac{q'^y l'^x}{q' l'}\partial_{ux}\Omega_{-}(u_{\perp},v_{\perp})\partial_{vy'}\partial_{vx'}\Omega_{-}(u'_{\perp},v'_{\perp})\nonumber\\
	&&\quad\,-\frac{l^x}{l}\frac{q'^y l'^y}{q' l'}\partial_{ux}\Omega_{-}(u_{\perp},v_{\perp})\partial_{vy'}\partial_{vy'}\Omega_{-}(u'_{\perp},v'_{\perp})\nonumber\\
	&&\quad\,-\frac{l^y}{l}\frac{q'^x l'^x}{q' l'}\partial_{uy}\Omega_{-}(u_{\perp},v_{\perp})\partial_{vx'}\partial_{vx'}\Omega_{-}(u'_{\perp},v'_{\perp})
	\nonumber\\
	&&\quad\,-\frac{l^y}{l}\frac{q'^x l'^y}{q' l'}\partial_{uy}\Omega_{-}(u_{\perp},v_{\perp})\partial_{vx'}\partial_{vy'}\Omega_{-}(u'_{\perp},v'_{\perp})
	\nonumber\\
	&&\quad\,+\frac{l^x}{l}\frac{q'^x l'^x}{q' l'}\partial_{uy}\Omega_{-}(u_{\perp},v_{\perp})\partial_{vy'}\partial_{vx'}\Omega_{-}(u'_{\perp},v'_{\perp})\nonumber\\
	&&\quad\,+\frac{l^x}{l}\frac{q'^x l'^y}{q' l'}\partial_{uy}\Omega_{-}(u_{\perp},v_{\perp})\partial_{vy'}\partial_{vy'}\Omega_{-}(u'_{\perp},v'_{\perp})
	\bigg]\times
	\mathcal{\rho}_{C}(X_0,Y_0,q,l,q',l'),
	\label{eq:JC1}
\end{eqnarray}
where
\begin{eqnarray}\nonumber
	&&\mathcal{\rho}_{C}(X_0,Y_0,q,l,q',l')
	\\\nonumber	&&\equiv\int^{X_0}_{X'_0}\int^{Y_0}_{Y'_0}\int^{X'_0}_{X''_0}\int^{Y'_0}_{Y''_0}(Y_0''-Y_0')
	J_0(qX'_0)J_1(lY'_0)\Theta(X'_0)\Theta(Y'_0)
	J_1(q'X''_0)J_1(l'Y''_0)\Theta(X''_0)\Theta(Y''_0)
	\\\nonumber
	&&=16\int^{\infty}_{-\infty} dX'_0\int^{\infty}_{-\infty}dY'_0\int^{\infty}_{-\infty}dX''_0\int^{\infty}_{-\infty}dY''_0
	(Y_0''-Y_0')J_0(qX'_0)J_1(lY'_0)
	J_1(q'X''_0)J_1(l'Y''_0)
	\\\nonumber
	&&\quad\times \Theta(X_0-X'_0)\Theta(Y_0-Y'_0)\Theta(X'_0-X''_0)\Theta(Y'_0-Y''_0)\Theta(X'_0)\Theta(Y'_0)\Theta(X''_0)\Theta(Y''_0),
\end{eqnarray}
and
\begin{eqnarray}\nonumber
	\mathcal{J}_D(p,X,Y)&\approx& +2i\frac{g^4N_c^2\delta^{ab}\delta^{ab}}{64p_0^4}
	\int^{X}_{\perp;q,u}\int^{Y}_{\perp;l,v}\int^{X}_{\perp;q',u'}\int^{Y}_{\perp;l',v'}
	\nonumber\\
	&&
	\bigg[\frac{l^x}{l}\frac{q'^y l'^y}{q' l'}\partial_{ux}\partial_{vx}\partial_{vx}\Omega_{-}(u_{\perp},v_{\perp})\Omega_{-}(u'_{\perp},v'_{\perp})\nonumber\\
	&&\quad\,+\frac{l^y}{l}\frac{q'^y l'^y}{q' l'}\partial_{ux}\partial_{vx}\partial_{vy}\Omega_{-}(u_{\perp},v_{\perp})\Omega_{-}(u'_{\perp},v'_{\perp})\nonumber\\
	&&\quad\,-\frac{l^x}{l}\frac{q'^y l'^x}{q' l'}\partial_{ux}\partial_{vy}\partial_{vx}\Omega_{-}(u_{\perp},v_{\perp})\Omega_{-}(u'_{\perp},v'_{\perp})\nonumber\\
	&&\quad\,-\frac{l^y}{l}\frac{q'^y l'^x}{q' l'}\partial_{ux}\partial_{vy}\partial_{vy}\Omega_{-}(u_{\perp},v_{\perp})\Omega_{-}(u'_{\perp},v'_{\perp})\nonumber\\
	&&\quad\,-\frac{l^x}{l}\frac{q'^x l'^y}{q' l'}\partial_{uy}\partial_{vx}\partial_{vx}\Omega_{-}(u_{\perp},v_{\perp})\Omega_{-}(u'_{\perp},v'_{\perp})
	\nonumber\\
	&&\quad\,-\frac{l^y}{l}\frac{q'^x l'^y}{q' l'}\partial_{uy}\partial_{vx}\partial_{vy}\Omega_{-}(u_{\perp},v_{\perp})\Omega_{-}(u'_{\perp},v'_{\perp})
	\nonumber\\
	&&\quad\,+\frac{l^x}{l}\frac{q'^x l'^x}{q' l'}\partial_{uy}\partial_{vy}\partial_{vx}\Omega_{-}(u_{\perp},v_{\perp})\Omega_{-}(u'_{\perp},v'_{\perp})\nonumber\\
	&&\quad\,+\frac{l^y}{l}\frac{q'^x l'^x}{q' l'}\partial_{uy}\partial_{vy}\partial_{vy}\Omega_{-}(u_{\perp},v_{\perp})\Omega_{-}(u'_{\perp},v'_{\perp})
	\bigg]\times
	\mathcal{\rho}_{D}(X_0,Y_0,q,l,q',l'),
	\label{eq:JD1}
\end{eqnarray}
where
\begin{eqnarray}\nonumber
	&&\mathcal{\rho}_{D}(X_0,Y_0,q,l,q',l')
	\\\nonumber	&&\equiv\int^{X_0}_{X'_0}\int^{Y_0}_{Y'_0}\int^{X'_0}_{X''_0}\int^{Y'_0}_{Y''_0}(Y_0''-Y_0')
	J_0(qX'_0)J_1(lY''_0)\Theta(X'_0)\Theta(Y''_0)
	J_1(q'X''_0)J_1(l'Y'_0)\Theta(X''_0)\Theta(Y'_0)
	\\\nonumber
	&&=16\int^{\infty}_{-\infty} dX'_0\int^{\infty}_{-\infty}dY'_0\int^{\infty}_{-\infty}dX''_0\int^{\infty}_{-\infty}dY''_0
	(Y_0''-Y_0')J_0(qX'_0)J_1(lY''_0)
	J_1(q'X''_0)J_1(l'Y'_0)
	\\\nonumber
	&&\quad\times \Theta(X_0-X'_0)\Theta(Y_0-Y'_0)\Theta(X'_0-X''_0)\Theta(Y'_0-Y''_0)\Theta(X'_0)\Theta(Y'_0)\Theta(X''_0)\Theta(Y''_0),
\end{eqnarray}
and
\begin{eqnarray}\nonumber
	\mathcal{J}_E(p,X,Y)&\approx& \frac{g^4N_c^2\delta^{ab}\delta^{ab}}{64p_0^4}
	\int^{X}_{\perp;q,u}\int^{Y}_{\perp;l,v}\int^{X}_{\perp;q',u'}\int^{Y}_{\perp;l',v'}
	\bigg[\frac{q^y l^y}{q l}\frac{q'^x l'^x}{q' l'}\partial^2_{ux'}\partial^2_{vx'}\Omega_{-}(u'_{\perp},v'_{\perp})\Omega_{-}(u_{\perp},v_{\perp})\nonumber\\
	&&\quad\,+\frac{q^y l^y}{q l}\frac{q'^y l'^y}{q' l'}\partial_{ux'}\partial_{uy'}\partial_{vx'}\partial_{vy'}\Omega_{-}(u'_{\perp},v'_{\perp})\Omega_{-}(u_{\perp},v_{\perp})
	\nonumber\\
	&&\quad\,+\frac{q^x l^x}{q l}\frac{q'^x l'^x}{q' l'}\partial_{ux'}\partial_{uy'}\partial_{vx'}\partial_{vy'}\Omega_{-}(u'_{\perp},v'_{\perp})\Omega_{-}(u_{\perp},v_{\perp})\nonumber\\
	&&\quad\,+\frac{q^x l^x}{q l}\frac{q'^y l'^y}{q' l'}\partial^2_{uy'}\partial^2_{vy'}\Omega_{-}(u'_{\perp},v'_{\perp})\Omega_{-}(u_{\perp},v_{\perp})\nonumber\\
	&&\quad\,+2\frac{q^y l^y}{q l}\frac{q'^x l'^y}{q' l'}\partial^2_{ux'}\partial_{vx'}\partial_{vy'}\Omega_{-}(u'_{\perp},v'_{\perp})\Omega_{-}(u_{\perp},v_{\perp})\nonumber\\
	&&\quad\,-2\frac{q^y l^x}{q l}\frac{q'^x l'^x}{q' l'}\partial^2_{ux'}\partial_{vx'}\partial_{vy'}\Omega_{-}(u'_{\perp},v'_{\perp})\Omega_{-}(u_{\perp},v_{\perp})\nonumber\\
	&&\quad\,-2\frac{q^y l^x}{q l}\frac{q'^x l'^y}{q' l'}\partial^2_{ux'}\partial^2_{vy'}\Omega_{-}(u'_{\perp},v'_{\perp})\Omega_{-}(u_{\perp},v_{\perp})\nonumber\\
	&&\quad\,-2\frac{q^y l^x}{q l}\frac{q'^y l'^x}{q' l'}\partial_{ux'}\partial_{uy'}\partial_{vx'}\partial_{vy'}\Omega_{-}(u'_{\perp},v'_{\perp})\Omega_{-}(u_{\perp},v_{\perp})\nonumber\\
	&&\quad\,+2\frac{q^x l^x}{q l}\frac{q'^x l'^y}{q' l'}\partial_{ux'}\partial_{uy'}\partial^2_{vy'}\Omega_{-}(u'_{\perp},v'_{\perp})\Omega_{-}(u_{\perp},v_{\perp})\nonumber\\
	&&\quad\,-2\frac{q^y l^x}{q l}\frac{q'^y l'^y}{q' l'}\partial_{ux'}\partial_{uy'}\partial^2_{vy'}\Omega_{-}(u'_{\perp},v'_{\perp})\Omega_{-}(u_{\perp},v_{\perp})
	\bigg]\times
	\mathcal{\rho}_{E}(X_0,Y_0,q,l,q',l'),
	\label{eq:Jsum1.2}
\end{eqnarray}
where
\begin{eqnarray}\nonumber
	&&\mathcal{\rho}_{E}(X_0,Y_0,q,l,q',l')
	\\\nonumber	&&\equiv\int^{X_0}_{X'_0}\int^{Y_0}_{Y'_0}\int^{X'_0}_{X''_0}\int^{Y'_0}_{Y''_0}(X_0''-X_0')(Y_0''-Y_0')
	J_1(qX'_0)J_1(lY'_0)\Theta(X'_0)\Theta(Y'_0)
	J_1(q'X''_0)J_1(l'Y''_0)\Theta(X''_0)\Theta(Y''_0)
	\\\nonumber
	&&=16\int^{\infty}_{-\infty} dX'_0\int^{\infty}_{-\infty}dY'_0\int^{\infty}_{-\infty}dX''_0\int^{\infty}_{-\infty}dY''_0
	(X_0''-X_0') (Y_0''-Y_0')J_1(qX'_0)J_1(lY'_0)
	J_1(q'X''_0)J_1(l'Y''_0)
	\\
	&&\quad\times \Theta(X_0-X'_0)\Theta(Y_0-Y'_0)\Theta(X'_0-X''_0)\Theta(Y'_0-Y''_0)\Theta(X'_0)\Theta(Y'_0)\Theta(X''_0)\Theta(Y''_0),
\end{eqnarray}
and
\begin{eqnarray}\nonumber
	\mathcal{J}_F(p,X,Y)&\approx& \frac{g^4N_c^2\delta^{ab}\delta^{ab}}{64p_0^4}
	\int^{X}_{\perp;q,u}\int^{Y}_{\perp;l,v}\int^{X}_{\perp;q',u'}\int^{Y}_{\perp;l',v'}
	\bigg[\frac{q^y l^x}{q l}\frac{q'^x l'^y}{q' l'}\partial^2_{vx}\Omega_{-}(u_{\perp},v_{\perp})\partial^2_{ux'}\Omega_{-}(u'_{\perp},v'_{\perp})\nonumber\\
	&&\quad\,+\frac{q^y l^y}{q l}\frac{q'^y l'^y}{q' l'}\partial_{vx}\partial_{vy}\Omega_{-}(u_{\perp},v_{\perp})\partial_{ux'}\partial_{uy'}\Omega_{-}(u'_{\perp},v'_{\perp})
	\nonumber\\
	&&\quad\,+\frac{q^x l^x}{q l}\frac{q'^x l'^x}{q' l'}\partial_{vx}\partial_{vy}\Omega_{-}(u_{\perp},v_{\perp})\partial_{ux'}\partial_{uy'}\Omega_{-}(u'_{\perp},v'_{\perp})\nonumber\\
	&&\quad\,+\frac{q^x l^y}{q l}\frac{q'^y l'^x}{q' l'}\partial^2_{vy}\Omega_{-}(u_{\perp},v_{\perp})\partial^2_{uy'}\Omega_{-}(u'_{\perp},v'_{\perp})\nonumber\\
	&&\quad\,+2\frac{q^y l^y}{q l}\frac{q'^x l'^y}{q' l'}\partial_{vx}\partial_{vy}\Omega_{-}(u_{\perp},v_{\perp})\partial^2_{ux'}\Omega_{-}(u'_{\perp},v'_{\perp})\nonumber\\
	&&\quad\,-2\frac{q^y l^x}{q l}\frac{q'^x l'^x}{q' l'}\partial_{vx}\partial_{vy}\Omega_{-}(u_{\perp},v_{\perp})\partial^2_{ux'}\Omega_{-}(u'_{\perp},v'_{\perp})\nonumber\\
	&&\quad\,-2\frac{q^y l^y}{q l}\frac{q'^x l'^x}{q' l'}\partial^2_{vy}\Omega_{-}(u_{\perp},v_{\perp})\partial^2_{ux'}\Omega_{-}(u'_{\perp},v'_{\perp})\nonumber\\
	&&\quad\,-2\frac{q^y l^x}{q l}\frac{q'^y l'^x}{q' l'}\partial_{vx}\partial_{vy}\Omega_{-}(u_{\perp},v_{\perp})\partial_{ux'}\partial_{uy'}\Omega_{-}(u'_{\perp},v'_{\perp})\nonumber\\
	&&\quad\,+2\frac{q^x l^y}{q l}\frac{q'^x l'^x}{q' l'}\partial^2_{vy}\Omega_{-}(u_{\perp},v_{\perp})\partial_{ux'}\partial_{uy'}\Omega_{-}(u'_{\perp},v'_{\perp})\nonumber\\
	&&\quad\,-2\frac{q^y l^y}{q l}\frac{q'^y l'^x}{q' l'}\partial^2_{vy}\Omega_{-}(u_{\perp},v_{\perp})\partial_{ux'}\partial_{uy'}\Omega_{-}(u'_{\perp},v'_{\perp})
	\bigg]\times
	\mathcal{\rho}_{F}(X_0,Y_0,q,l,q',l'),
	\label{eq:Jsum2.2}
\end{eqnarray}
where
\begin{eqnarray}\nonumber
	&&\mathcal{\rho}_{F}(X_0,Y_0,q,l,q',l')
	\\\nonumber	&&\equiv\int^{X_0}_{X'_0}\int^{Y_0}_{Y'_0}\int^{X'_0}_{X''_0}\int^{Y'_0}_{Y''_0}(X_0''-X_0')(Y_0''-Y_0')
	J_1(qX'_0)J_1(lY''_0)\Theta(X'_0)\Theta(Y''_0)
	J_1(q'X''_0)J_1(l'Y'_0)\Theta(X''_0)\Theta(Y''_0)
	\\\nonumber
	&&=16\int^{\infty}_{-\infty} dX'_0\int^{\infty}_{-\infty}dY'_0\int^{\infty}_{-\infty}dX''_0\int^{\infty}_{-\infty}dY''_0
	(X_0''-X_0') (Y_0''-Y_0')J_1(qX'_0)J_1(lY''_0)
	J_1(q'X''_0)J_1(l'Y'_0)
	\\
	&&\quad\times \Theta(X_0-X'_0)\Theta(Y_0-Y'_0)\Theta(X'_0-X''_0)\Theta(Y'_0-Y''_0)\Theta(X'_0)\Theta(Y'_0)\Theta(X''_0)\Theta(Y''_0).
\end{eqnarray}
We next carry out integration over variables $q_{\perp}$, $l_{\perp}$, $q'_{\perp}$, and 
$l'_{\perp}$ and $X_0'$, $Y_0'$, $X''_0$, and $Y''_0$. For convenience, we make the following decomposition~\cite{Kumar:2022ylt} of $q_{\perp}$, $l_{\perp}$,
\begin{eqnarray}
	q^i=\frac{(X-u)_{\perp}^i}{|X_{\perp}-u_{\perp}|}q\cos\theta_q+\Theta^{ij}_{X-u}q_j\sin\theta_q,\quad
	l^i=\frac{(Y-v)_{\perp}^i}{|Y_{\perp}-v_{\perp}|}l\cos\theta_l+\Theta^{ij}_{Y-v}l_j\sin\theta_l.
\end{eqnarray} 
where $\Theta^{ij}_{V}\equiv \eta^{ij}_{\perp}+\frac{V^{i}_{\perp}V^{j}_{\perp}}{|V_{\perp}|^2}$. A similar decomposition can be taken for $q'_{\perp}$ and 
$l'_{\perp}$.  The angular variables $\theta_q$ and $\theta_l$ will appear in $\int d^2q_{\perp}=\int dqqd\theta_{q}$ and $\int d^2l_{\perp}=\int dlld\theta_{l}$ which can accordingly be evaluated 
by using the formulas
\begin{eqnarray}
	\int^{2\pi}_0 d\theta e^{ia\cos \theta}&=&2\pi J_0(|a|),\quad \int^{2\pi}_0 d\theta e^{ia\cos \theta}\cos\theta=2i\pi J_1(a),
	\\
	\int^{2\pi}_0 d\theta e^{ia\cos \theta}\sin\theta&=&0.
\end{eqnarray}
After carrying out integration over angular variables, we perform integration over variables $q$, $l$, $q'$ and $l'$ using the formula~(\ref{eq:Bessel_rel}). 
Finally, carrying out integration over $X_0'$, $Y_0'$, $X''_0$ and $Y''_0$ and defining new variables $s_{\perp}=X_{\perp}-u_{\perp},\quad 
s'_{\perp}=X_{\perp}-u'_{\perp}, \quad
t_{\perp}=Y_{\perp}-v_{\perp}, \quad
t'_{\perp}=Y_{\perp}-v'_{\perp}$, we can obtain
\begin{eqnarray}
	J_A(p,X,Y)&\approx&+\frac{g^4N_c^2(N_c^2-1)}{4(2\pi)^4p_0^4}
	\int^{\perp}_{u,v,u',v'}\frac{\Theta(X_0-|s_{\perp}|)\Theta(Y_0-|t_{\perp}|)\Theta(|s_{\perp}|-|s'_{\perp}|)\Theta(|t_{\perp}|-|t'_{\perp}|)}{|s_{\perp}||t_{\perp}||s'_{\perp}||t'_{\perp}|}
	\nonumber\\
	&&\times \bigg[\left(\hat{s}_{\perp}^{\prime y}\hat{t}_{\perp}^{\prime y}\partial_{ux}\partial_{vx}\Omega_{-}(u_{\perp},v_{\perp})\Omega_{-}(u'_{\perp},v'_{\perp})+\hat{s}_{\perp}^{ y}\hat{t}_{\perp}^{y}\partial_{ux}\partial_{vx}\Omega_{+}(u_{\perp},v_{\perp})\Omega_{+}(u'_{\perp},v'_{\perp})\right)
	\nonumber\\ &&
	\quad\,-2\left(\hat{s}_{\perp}^{\prime y}\hat{t}_{\perp}^{\prime x} \partial_{ux}\partial_{vy}\Omega_{-}(u_{\perp},v_{\perp})\Omega_{-}(u'_{\perp},v'_{\perp})+\hat{s}_{\perp}^{y}\hat{t}_{\perp}^{x}\partial_{ux}\partial_{vy}\Omega_{+}(u_{\perp},v_{\perp})\Omega_{+}(u'_{\perp},v'_{\perp})\right)
	\nonumber\\ &&
	\quad\,+\left(\hat{s}_{\perp}^{\prime x}\hat{t}_{\perp}^{\prime x}\partial_{uy}\partial_{vy}\Omega_{-}(u_{\perp},v_{\perp})\Omega_{-}(u'_{\perp},v'_{\perp})+\hat{s}_{\perp}^{x}\hat{t}_{\perp}^{x}\partial_{uy}\partial_{vy}\Omega_{+}(u_{\perp},v_{\perp})\Omega_{+}(u'_{\perp},v'_{\perp})\right)
	\bigg],\nonumber\\ \label{eq:JA2}
\end{eqnarray}
where
$\hat{s}^i_{\perp}=s^i_{\perp}/|s_{\perp}|$ and $\int^{\perp}_{u,v,u'v'}\equiv \int \!d^2u_{\perp}\!\!\int \!d^2v_{\perp}
\int \!d^2u'_{\perp}\!\!\int \!d^2v'_{\perp}.$
\begin{eqnarray}
	\mathcal{J}_{B}(p,X,Y)&\approx&+\frac{g^4N_c^2(N_c^2-1)}{4(2\pi)^4p_0^4}
	\int^{\perp}_{u,v,u',v'}\frac{\Theta(X_0-|s_{\perp}|)\Theta(Y_0-|t'_{\perp}|)\Theta(|s_{\perp}|-|s'_{\perp}|)\Theta(|t'_{\perp}|-|t_{\perp}|)}{|s_{\perp}||t_{\perp}||s'_{\perp}||t'_{\perp}|}
	\nonumber\\
	&&\times \bigg[\left(\hat{s}_{\perp}^{\prime y}\hat{t}_{\perp}^{y}\partial_{ux}\Omega_{-}(u_{\perp},v_{\perp})\partial_{vx'}\Omega_{-}(u'_{\perp},v'_{\perp})+\hat{s}_{\perp}^{ y}\hat{t}_{\perp}^{\prime y}\partial_{ux}\Omega_{+}(u_{\perp},v_{\perp})\partial_{vx'}\Omega_{+}(u'_{\perp},v'_{\perp})\right)
	\nonumber\\ &&
	\quad\,-2 \left(\hat{s}_{\perp}^{\prime y}\hat{t}_{\perp}^{x}\partial_{ux}\Omega_{-}(u_{\perp},v_{\perp})\partial_{vy'}\Omega_{-}(u'_{\perp},v'_{\perp})+\hat{s}_{\perp}^{ y}\hat{t}_{\perp}^{\prime x}\partial_{ux}\Omega_{+}(u_{\perp},v_{\perp})\partial_{vy'}\Omega_{+}(u'_{\perp},v'_{\perp})\right)
	\nonumber\\ &&
	\quad\,+\left(\hat{s}_{\perp}^{\prime x}\hat{t}_{\perp}^{x}\partial_{uy}\Omega_{-}(u_{\perp},v_{\perp})\partial_{vy'}\Omega_{-}(u'_{\perp},v'_{\perp})+\hat{s}_{\perp}^{x}\hat{t}_{\perp}^{\prime x}\partial_{uy}\Omega_{+}(u_{\perp},v_{\perp})\partial_{vy'}\Omega_{+}(u'_{\perp},v'_{\perp})\right)
	\bigg], \nonumber\\ \label{eq:JB2}
\end{eqnarray}
\begin{eqnarray}
	\mathcal{J}_C(p,X,Y)&\approx&+2\frac{g^4N_c^2(N_c^2-1)}{4(2\pi)^4p_0^4}
	\int^{\perp}_{u,v,u',v'}\frac{\Theta(X_0-|s_{\perp}|)\Theta(Y_0-|t_{\perp}|)\Theta(|s_{\perp}|-|s'_{\perp}|)\Theta(|t_{\perp}|-|t'_{\perp}|)}{|s_{\perp}||t_{\perp}||s'_{\perp}||t'_{\perp}|}
	\nonumber\\
	&&\times \big(|t'_{\perp}|-|t_{\perp}|\big) \bigg[\hat{t}_{\perp}^{y} \hat{s}_{\perp}^{\prime x} \hat{t}_{\perp}^{\prime x}\partial_{ux}\Omega_{-}(u_{\perp},v_{\perp})\partial_{vx'}\partial_{vx'}\Omega_{-}(u'_{\perp},v'_{\perp})\nonumber\\
	&&\quad\,+\hat{t}_{\perp}^{y} \hat{s}_{\perp}^{\prime y} \hat{t}_{\perp}^{\prime y} \partial_{ux}\Omega_{-}(u_{\perp},v_{\perp})\partial_{vx'}\partial_{vy'}\Omega_{-}(u'_{\perp},v'_{\perp})\nonumber\\
	&&\quad\,-\hat{t}_{\perp}^{x} \hat{s}_{\perp}^{\prime y} \hat{t}_{\perp}^{\prime x} \partial_{ux}\Omega_{-}(u_{\perp},v_{\perp})\partial_{vy'}\partial_{vx'}\Omega_{-}(u'_{\perp},v'_{\perp})\nonumber\\
	&&\quad\,-\hat{t}_{\perp}^{x} \hat{s}_{\perp}^{\prime y} \hat{t}_{\perp}^{\prime y}\partial_{ux}\Omega_{-}(u_{\perp},v_{\perp})\partial_{vy'}\partial_{vy'}\Omega_{-}(u'_{\perp},v'_{\perp})\nonumber\\
	&&\quad\,-\hat{t}_{\perp}^{y} \hat{s}_{\perp}^{\prime x} \hat{t}_{\perp}^{\prime x}\partial_{uy}\Omega_{-}(u_{\perp},v_{\perp})\partial_{vx'}\partial_{vx'}\Omega_{-}(u'_{\perp},v'_{\perp})
	\nonumber\\
	&&\quad\,-\hat{t}_{\perp}^{y} \hat{s}_{\perp}^{\prime x} \hat{t}_{\perp}^{\prime y}\partial_{uy}\Omega_{-}(u_{\perp},v_{\perp})\partial_{vx'}\partial_{vy'}\Omega_{-}(u'_{\perp},v'_{\perp})
	\nonumber\\
	&&\quad\,+\hat{t}_{\perp}^{x} \hat{s}_{\perp}^{\prime x} \hat{t}_{\perp}^{\prime x}\partial_{uy}\Omega_{-}(u_{\perp},v_{\perp})\partial_{vy'}\partial_{vx'}\Omega_{-}(u'_{\perp},v'_{\perp})\nonumber\\
	&&\quad\,+\hat{t}_{\perp}^{x} \hat{s}_{\perp}^{\prime x} \hat{t}_{\perp}^{\prime y}\partial_{uy}\Omega_{-}(u_{\perp},v_{\perp})\partial_{vy'}\partial_{vy'}\Omega_{-}(u'_{\perp},v'_{\perp})
	\bigg], \label{eq:JC2}
\end{eqnarray}
\begin{eqnarray}
	\mathcal{J}_{D}(p,X,Y)&\approx&+2\frac{g^4N_c^2(N_c^2-1)}{4(2\pi)^4p_0^4}
	\int^{\perp}_{u,v,u',v'}\frac{\Theta(X_0-|s_{\perp}|)\Theta(Y_0-|t'_{\perp}|)\Theta(|s_{\perp}|-|s'_{\perp}|)\Theta(|t'_{\perp}|-|t_{\perp}|)}{|s_{\perp}||t_{\perp}||s'_{\perp}||t'_{\perp}|}
	\nonumber\\
	&&\times \big(|t_{\perp}|-|t'_{\perp}|\big) \bigg[\hat{t}_{\perp}^{x} \hat{s}_{\perp}^{\prime y} \hat{t}_{\perp}^{\prime y}\partial_{ux}\partial_{vx}\partial_{vx}\Omega_{-}(u_{\perp},v_{\perp})\Omega_{-}(u'_{\perp},v'_{\perp})\nonumber\\
	&&\quad\,+\hat{t}_{\perp}^{y} \hat{s}_{\perp}^{\prime y} \hat{t}_{\perp}^{\prime y}\partial_{ux}\partial_{vx}\partial_{vy}\Omega_{-}(u_{\perp},v_{\perp})\Omega_{-}(u'_{\perp},v'_{\perp})\nonumber\\
	&&\quad\,-\hat{t}_{\perp}^{x} \hat{s}_{\perp}^{\prime y} \hat{t}_{\perp}^{\prime x}\partial_{ux}\partial_{vy}\partial_{vx}\Omega_{-}(u_{\perp},v_{\perp})\Omega_{-}(u'_{\perp},v'_{\perp})\nonumber\\
	&&\quad\,-\hat{t}_{\perp}^{y} \hat{s}_{\perp}^{\prime y} \hat{t}_{\perp}^{\prime x}\partial_{ux}\partial_{vy}\partial_{vy}\Omega_{-}(u_{\perp},v_{\perp})\Omega_{-}(u'_{\perp},v'_{\perp})\nonumber\\
	&&\quad\,-\hat{t}_{\perp}^{x} \hat{s}_{\perp}^{\prime x} \hat{t}_{\perp}^{\prime y}\partial_{uy}\partial_{vx}\partial_{vx}\Omega_{-}(u_{\perp},v_{\perp})\Omega_{-}(u'_{\perp},v'_{\perp})
	\nonumber\\
	&&\quad\,-\hat{t}_{\perp}^{y} \hat{s}_{\perp}^{\prime x} \hat{t}_{\perp}^{\prime y}\partial_{uy}\partial_{vx}\partial_{vy}\Omega_{-}(u_{\perp},v_{\perp})\Omega_{-}(u'_{\perp},v'_{\perp})
	\nonumber\\
	&&\quad\,+\hat{t}_{\perp}^{x} \hat{s}_{\perp}^{\prime x} \hat{t}_{\perp}^{\prime x}\partial_{uy}\partial_{vy}\partial_{vx}\Omega_{-}(u_{\perp},v_{\perp})\Omega_{-}(u'_{\perp},v'_{\perp})\nonumber\\
	&&\quad\,+\hat{t}_{\perp}^{y} \hat{s}_{\perp}^{\prime x} \hat{t}_{\perp}^{\prime x}\partial_{uy}\partial_{vy}\partial_{vy}\Omega_{-}(u_{\perp},v_{\perp})\Omega_{-}(u'_{\perp},v'_{\perp})
	\bigg], \label{eq:JD2}
\end{eqnarray}
\begin{eqnarray}\nonumber
	\mathcal{J}_{E}(p,X,Y)&\approx& \frac{g^4N_c^2(N_c^2-1)}{4(2\pi)^4p_0^4} 
	\int^{\perp}_{u,v,u',v'}  \bigg[\frac{\left(|s'_{\perp}|-|s_{\perp}|\right)\left(|t'_{\perp}|-|t_{\perp}|\right)}{|s'_{\perp}||s_{\perp}||t'_{\perp}||t_{\perp}|}\Omega_{-}(u_{\perp},v_{\perp})\bigg]
	\nonumber\\
	&&\times
	\bigg[\hat{s}_{\perp}^{y}\hat{s}_{\perp}^{\prime x}\hat{t}_{\perp}^{y}\hat{t}_{\perp}^{\prime x}\partial^2_{ux'}\partial^2_{vx'}\Omega_{-}(u'_{\perp},v'_{\perp})+\hat{s}_{\perp}^{y}\hat{s}_{\perp}^{\prime y}\hat{t}_{\perp}^{y}\hat{t}_{\perp}^{\prime y}\partial_{ux'}\partial_{uy'}\partial_{vx'}\partial_{vy'}\Omega_{-}(u'_{\perp},v'_{\perp})
	\nonumber\\
	&&\quad\,+\hat{s}_{\perp}^{x}\hat{s}_{\perp}^{\prime x}\hat{t}_{\perp}^{x}\hat{t}_{\perp}^{\prime x}\partial_{ux'}\partial_{uy'}\partial_{vx'}\partial_{vy'}\Omega_{-}(u'_{\perp},v'_{\perp})+\hat{s}_{\perp}^{x}\hat{s}_{\perp}^{\prime y}\hat{t}_{\perp}^{x}\hat{t}_{\perp}^{\prime y}\partial^2_{uy'}\partial^2_{vy'}\Omega_{-}(u'_{\perp},v'_{\perp})\nonumber\\
	&&\quad\,+2\hat{s}_{\perp}^{y}\hat{s}_{\perp}^{\prime x}\hat{t}_{\perp}^{y}\hat{t}_{\perp}^{\prime y}\partial^2_{ux'}\partial_{vx'}\partial_{vy'}\Omega_{-}(u'_{\perp},v'_{\perp})-2\hat{s}_{\perp}^{y}\hat{s}_{\perp}^{\prime x}\hat{t}_{\perp}^{x}\hat{t}_{\perp}^{\prime x}\partial^2_{ux'}\partial_{vx'}\partial_{vy'}\Omega_{-}(u'_{\perp},v'_{\perp})\nonumber\\
	&&\quad\,-2\hat{s}_{\perp}^{y}\hat{s}_{\perp}^{\prime x}\hat{t}_{\perp}^{x}\hat{t}_{\perp}^{\prime y}\partial^2_{ux'}\partial^2_{vy'}\Omega_{-}(u'_{\perp},v'_{\perp})-2\hat{s}_{\perp}^{y}\hat{s}_{\perp}^{\prime y}\hat{t}_{\perp}^{x}\hat{t}_{\perp}^{\prime x}\partial_{ux'}\partial_{uy'}\partial_{vx'}\partial_{vy'}\Omega_{-}(u'_{\perp},v'_{\perp})\nonumber\\
	&&\quad\,+2\hat{s}_{\perp}^{x}\hat{s}_{\perp}^{\prime x}\hat{t}_{\perp}^{x}\hat{t}_{\perp}^{\prime y}\partial_{ux'}\partial_{uy'}\partial^2_{vy'}\Omega_{-}(u'_{\perp},v'_{\perp})-2\hat{s}_{\perp}^{y}\hat{s}_{\perp}^{\prime y}\hat{t}_{\perp}^{x}\hat{t}_{\perp}^{\prime y}\partial_{ux'}\partial_{uy'}\partial^2_{vy'}\Omega_{-}(u'_{\perp},v'_{\perp})
	\bigg]
	\nonumber\\
	&&
	\times{\Theta(X_0-|s_{\perp}|)\Theta(Y_0-|t_{\perp}|)\Theta(|s_{\perp}|-|s'_{\perp}|)\Theta(|t_{\perp}|-|t'_{\perp}|)} 
	\nonumber \\
	&&\times{\Theta(|s_{\perp}|)\Theta(|t_{\perp}|)\Theta(|s'_{\perp}|)\Theta(|t'_{\perp}|)},\label{eq:Jsum1.5}
\end{eqnarray}
\begin{eqnarray}\nonumber
	\mathcal{J}_{F}(p,X,Y)&\approx& \frac{g^4N_c^2(N_c^2-1)}{4(2\pi)^4p_0^4} 
	\int^{\perp}_{u,v,u',v'} \bigg[\frac{\left(|s'_{\perp}|-|s_{\perp}|\right)\left(|t_{\perp}|-|t'_{\perp}|\right)}{|s'_{\perp}||t'_{\perp}||s_{\perp}||t_{\perp}|}\bigg]
	\nonumber\\
	&&\times
	\bigg[\hat{s}_{\perp}^{y}\hat{s}_{\perp}^{\prime x}\hat{t}_{\perp}^{x}\hat{t}_{\perp}^{\prime y}\partial^2_{vx}\Omega_{-}(u_{\perp},v_{\perp})\partial^2_{ux'}\Omega_{-}(u'_{\perp},v'_{\perp})
	\nonumber\\
	&&\quad\,+\hat{s}_{\perp}^{y}\hat{s}_{\perp}^{\prime y}\hat{t}_{\perp}^{y}\hat{t}_{\perp}^{\prime y}\partial_{vx}\partial_{vy}\Omega_{-}(u_{\perp},v_{\perp})\partial_{ux'}\partial_{uy'}\Omega_{-}(u'_{\perp},v'_{\perp})
	\nonumber\\
	&&\quad\,+\hat{s}_{\perp}^{x}\hat{s}_{\perp}^{\prime x}\hat{t}_{\perp}^{x}\hat{t}_{\perp}^{\prime x}\partial_{vx}\partial_{vy}\Omega_{-}(u_{\perp},v_{\perp})\partial_{ux'}\partial_{uy'}\Omega_{-}(u'_{\perp},v'_{\perp})\nonumber\\
	&&\quad\,+\hat{s}_{\perp}^{x}\hat{s}_{\perp}^{\prime y}\hat{t}_{\perp}^{y}\hat{t}_{\perp}^{\prime x}\partial^2_{vy}\Omega_{-}(u_{\perp},v_{\perp})\partial^2_{uy'}\Omega_{-}(u'_{\perp},v'_{\perp})\nonumber\\
	&&\quad\,+2\hat{s}_{\perp}^{y}\hat{s}_{\perp}^{\prime x}\hat{t}_{\perp}^{y}\hat{t}_{\perp}^{\prime y}\partial_{vx}\partial_{vy}\Omega_{-}(u_{\perp},v_{\perp})\partial^2_{ux'}\Omega_{-}(u'_{\perp},v'_{\perp})\nonumber\\
	&&\quad\,-2\hat{s}_{\perp}^{y}\hat{s}_{\perp}^{\prime x}\hat{t}_{\perp}^{x}\hat{t}_{\perp}^{\prime x}\partial_{vx}\partial_{vy}\Omega_{-}(u_{\perp},v_{\perp})\partial^2_{ux'}\Omega_{-}(u'_{\perp},v'_{\perp})\nonumber\\
	&&\quad\,-2\hat{s}_{\perp}^{y}\hat{s}_{\perp}^{\prime x}\hat{t}_{\perp}^{y}\hat{t}_{\perp}^{\prime x}\partial^2_{vy}\Omega_{-}(u_{\perp},v_{\perp})\partial^2_{ux'}\Omega_{-}(u'_{\perp},v'_{\perp})\nonumber\\
	&&\quad\,-2\hat{s}_{\perp}^{y}\hat{s}_{\perp}^{\prime y}\hat{t}_{\perp}^{x}\hat{t}_{\perp}^{\prime x}\partial_{vx}\partial_{vy}\Omega_{-}(u_{\perp},v_{\perp})\partial_{ux'}\partial_{uy'}\Omega_{-}(u'_{\perp},v'_{\perp})\nonumber\\
	&&\quad\,+2\hat{s}_{\perp}^{x}\hat{s}_{\perp}^{\prime x}\hat{t}_{\perp}^{y}\hat{t}_{\perp}^{\prime x}\partial^2_{vy}\Omega_{-}(u_{\perp},v_{\perp})\partial_{ux'}\partial_{uy'}\Omega_{-}(u'_{\perp},v'_{\perp})\nonumber\\
	&&\quad\,-2\hat{s}_{\perp}^{y}\hat{s}_{\perp}^{\prime y}\hat{t}_{\perp}^{y}\hat{t}_{\perp}^{\prime x}\partial^2_{vy}\Omega_{-}(u_{\perp},v_{\perp})\partial_{ux'}\partial_{uy'}\Omega_{-}(u'_{\perp},v'_{\perp})
	\bigg]
	\nonumber\\
	&&
	\times{\Theta(X_0-|s_{\perp}|)\Theta(|s_{\perp}|-|s'_{\perp}|)\Theta(Y_0-|t'_{\perp}|)\Theta(|t'_{\perp}|-|t_{\perp}|)} 
	\nonumber \\
	&&\times{\Theta(|s_{\perp}|)\Theta(|t_{\perp}|)\Theta(|s'_{\perp}|)\Theta(|t'_{\perp}|)}.\label{eq:Jsum2.5}
\end{eqnarray}
Now we adopt the GBW type of gluon distribution~\cite{Guerrero-Rodriguez:2021ask},  
\begin{align}
	\Omega_{\pm}(u_{\perp},v_{\perp})=\Omega(u_{\perp},v_{\perp})=\frac{Q_s^4}{g^4N_c^2}\left(\frac{1-e^{-Q_s^2|u_{\perp}-v_{\perp}|^2/4}}{Q_s^2|u_{\perp}-v_{\perp}|^2/4}\right)^2=\frac{Q_s^4}{g^4N_c^2}\left(\frac{1-e^{-Q_s^2|s_{\perp}-t_{\perp}-r_{\perp}|^2/4}}{Q_s^2|s_{\perp}-t_{\perp}-r_{\perp}|^2/4}\right)^2, 
\end{align}
where $Q_s$ is the gluon saturation momentum and $r_{\perp}\equiv X_{\perp}-Y_{\perp}$.

\begin{figure}[ht!]
		\begin{center}
			{\includegraphics[width=0.6\hsize,height=8cm,clip]{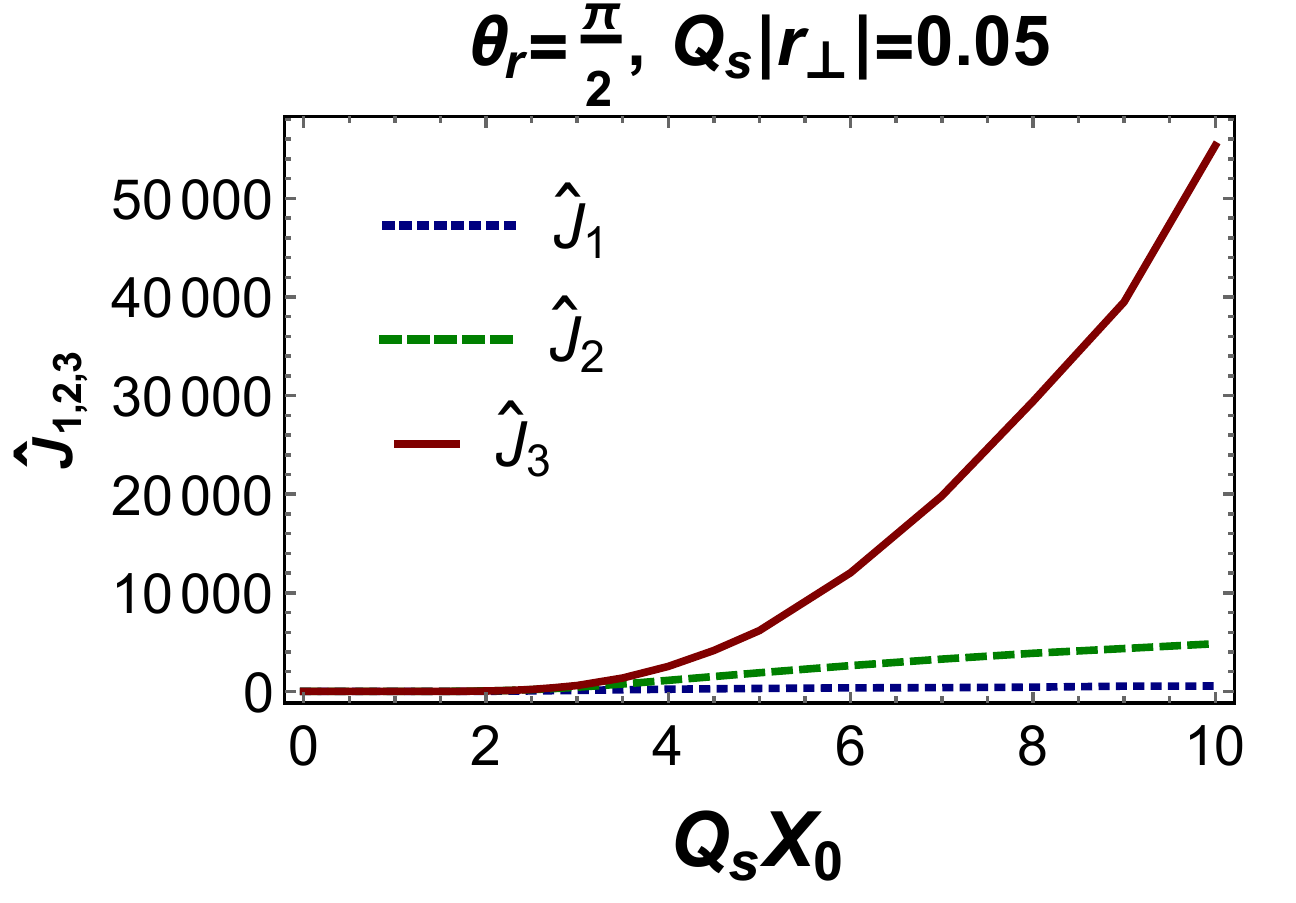}}
			
			\caption{Numerical results for $\hat{\mathcal{J}}_{1}$, $\hat{\mathcal{J}}_{2}$, $\hat{\mathcal{J}}_{3}$ as a function of $Q_sX_0$.
			}
			\label{fig_Jhat123_QX0}
		\end{center}
\end{figure}

Carrying out integration over variables $\bf{u}$, $\bf{v}$, $\bf{u'}$, $\bf{v'}$, we can write $\mathcal{J}_{1}+\mathcal{J}_{2}+\mathcal{J}_{3}=\mathcal{J}$, which depends on $p_0$, $X_0=Y_0$, $r_{\perp}$ and $\theta_r$. It turns out that $\mathcal{J}$ is given by 
\begin{eqnarray}\label{eq:J_estimate}
	\mathcal{J}(p_0,Q_sX_0,Q_s|r_{\perp}|,\theta_{r})=\frac{g^4N_c^2(N_c^2-1)}{4(2\pi)^4p_0^4}\frac{Q_s^8}{g^8N_c^4}\frac{\hat{\mathcal{J}}(Q_sX_0,Q_s|r_{\perp}|,\theta_{r})}{Q_s^2},
\end{eqnarray}
where $\theta_{r}\equiv \cos^{-1}(r^x_{\perp}/|r_{\perp}|)$ and $\hat{\mathcal{J}}(Q_sX_0,Q_s|r_{\perp}|,\theta_r)=\hat{\mathcal{J}}_1(Q_sX_0,Q_s|r_{\perp}|,\theta_r)+\hat{\mathcal{J}}_2(Q_sX_0,Q_s|r_{\perp}|,\theta_r)+\hat{\mathcal{J}}_3(Q_sX_0,Q_s|r_{\perp}|,\theta_r)$ is a dimensionless quantity obtained after carrying out the eight-dimensional integrals. The behavior of $\hat{\mathcal{J}}_1(Q_sX_0,Q_s|r_{\perp}|,\theta_r)+\hat{\mathcal{J}}_2(Q_sX_0,Q_s|r_{\perp}|,\theta_r)+\hat{\mathcal{J}}_3(Q_sX_0,Q_s|r_{\perp}|,\theta_r)$ with respect to $Q_sX_0$ for fixed values $Q_s|r_{\perp}|=0.05$ and $\theta_r=\pi/2$ is shown in Fig.~\ref{fig_Jhat123_QX0}. Since $\hat{\mathcal{J}}_{1,2,3}(Q_sX_0,0.05,\theta_r)\approx \hat{\mathcal{J}}_{1,2,3}(Q_sX_0,0.05,0)$, for numerical efficiency, we approximate $\hat{\mathcal{J}}_{1,2,3}(Q_sX_0,0,0)\approx \hat{\mathcal{J}}_{1,2,3}(Q_sX_0,0.05,\pi/2)$ and also $\hat{\mathcal{J}}(Q_sX_0,0,0)\approx \hat{\mathcal{J}}(Q_sX_0,0.05,\pi/2)$. The same approximation is applied to $\hat{\mathcal{I}}_{1,2,3}$ and $\hat{\mathcal{I}}$ (see Ref.~\cite{Kumar:2022ylt}). For brevity, we denote $\hat{\mathcal{J}}(Q_sX_0)=\hat{\mathcal{J}}(Q_sX_0,0,0)$ and similarly for $\hat{\mathcal{I}}(Q_sX_0)$, $\hat{\mathcal{J}}_{1,2,3}(Q_sX_0)$, and $\hat{\mathcal{I}}_{1,2,3}(Q_sX_0)$.   

\bibliography{reexam_spin_alignmnet_color_fields_v2.bbl}
\end{document}